\documentclass[11pt]{scrartcl}
\usepackage{geometry} % see geometry.pdf on how to lay out the page. There's lots.
\usepackage{booktabs} % for much better looking tables
\usepackage{array} % for better arrays (eg matrices) in maths
\usepackage{paralist} % very flexible & customisable lists (eg. enumerate/itemize, etc.)
\usepackage{verbatim} % adds environment for commenting out blocks of text & for better verbatim
\usepackage{amsmath,amssymb}
\usepackage{tabularx}
\usepackage{multirow}
\usepackage{diagbox}
\usepackage{hyperref}

\usepackage[export]{adjustbox}
\usepackage{xcolor}

\usepackage{tikz}
\usepackage{graphicx}
\graphicspath{{figure/}{../figures/}}

\usepackage{fancyhdr} % This should be set AFTER setting up the page geometry
\usepackage{sectsty}
\allsectionsfont{\sffamily\mdseries\upshape} % (See the fntguide.pdf for font help)
\renewcommand{\vec}[1]	{\pmb{#1}}

\newcommand{\dmin}		{d_\text{min}}
\newcommand{\df}			{d_\text{f}}
\newcommand{\darc}[1]		{\text{dim}_\text{B}^{#1}}
\newcommand{\dw}			{d_\text{w}}

\newcommand{\dminErr}		{\Delta \dmin}
\newcommand{\dminMean}	{\langle{\dmin}\rangle}
\newcommand{\dminVar}		{\sigma^2(\dmin)}
\newcommand{\dminApprox}	{\widetilde{\dmin}}

\newcommand{\mass}		{w}
\newcommand{\linl}			{m}
\newcommand{\iter}			{n}
\newcommand{\repre}		{u}

\newcommand{\ep}			{\gamma}			% exit pair combination

\renewcommand{\path}[2]			{\ell_{#1}^{#2}}	% #1 - iteration stage, #2 - representaion
\newcommand{\pathVec}[2]		{\vec{\ell}_{#1}^{#2}}
\newcommand{\minPath}			{\ell_\text{min}}
\newcommand{\minPathScale}[2]	{\lambda_{#1}^{#2}}

\newcommand{\minPathScaleErr}[1]	{\minPathScale{\text{err}}{#1}}

\newcommand{\minPathScaleMean}	{\langle{\minPathScale{}{}}\rangle}
\newcommand{\minPathScaleVar}		{\sigma^2(\minPathScale{}{})}

\newcommand{\pathMatrix}[1]			{M_{#1}}
\newcommand{\pathMatrixApprox}[1] 	{\widetilde{M}_{#1}}
\newcommand{\specRadius}			{r}

\newcommand{\Sierpinski}	{Sierpi\'nski }

\newcommand{\A}	{\includegraphics{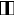}}
\newcommand{\B}	{\includegraphics{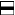}}
\newcommand{\C}	{\includegraphics{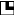}}
\newcommand{\D}	{\includegraphics{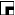}}
\newcommand{\E}	{\includegraphics{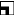}}
\newcommand{\F}	{\includegraphics{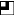}}

\newcommand{\W} {\mathcal W}
\newcommand{\Ac}{\mathcal A}
\newcommand{\G}{\mathcal G}
\newcommand{\linf}{L_{\infty}}
\renewcommand{\P} {\mathcal P}

\newcommand{\gva}[1]		{\textbf{g4a{#1}}}
\newcommand{\gvb}[1]		{\textbf{g4b{#1}}}
\newcommand{\gvc}[1]		{\textbf{g4c{#1}}}
\newcommand{\gfa}[1]		{\textbf{g5a{#1}}}
\newcommand{\gfb}[1]		{\textbf{g5b{#1}}}
\newcommand{\gfc}[1]		{\textbf{g5c{#1}}}
\newcommand{\gsea}[1]		{\textbf{g6a{#1}}}
\newcommand{\gseb}[1]		{\textbf{g6b{#1}}}
\newcommand{\gsia}[1]		{\textbf{g7a{#1}}}
\newcommand{\gsib}[1]		{\textbf{g7b{#1}}}
\newcommand{\gsic}[1]		{\textbf{g7c{#1}}}
\newcommand{\gsid}[1]		{\textbf{g7d{#1}}}
\newcommand{\gsie}[1]		{\textbf{g7e{#1}}}
\newcommand{\gsif}[1]		{\textbf{g7f{#1}}}

\newtheorem{property}{Property}

\title{Randomised mixed labyrinth fractals}
\author{Janett Prehl$^1$, Ligia Loretta Cristea$^2$, and Daniel Dick$^3$ \\
	\small  $^1$ Technische Universität Chemnitz, D-09107, Germany \\[-1ex]
	\small  $^2$ Freelancer, Austria \\[-1ex]
	\small  $^3$ Technische Universtät Chemnitz, D-09107, Germany}
\begin{document}
	
	\maketitle

\date{\today}% It is always \today, today,

\begin{abstract}
In this paper, the class of randomised mixed labyrinth fractals is introduced. It is a class of finitely ramified Sierpinski carpets that generalize mixed labyrinth fractals. The structures are generated by randomly selected labyrinth patterns with fixed selection probabilities at each iteration level, offering a flexible framework to study fractal topology, arc dimensions, and shortest path properties. Here, the focus lies on analysing how the randomised mixing of patterns — specifically their shape, symmetry, and path geometry — effects arc dimensions, path lengths, and isotropy restoration. The study reveals that isotropy, previously shown for self-similar fractals, extends to the randomised mixed class. Various scaling behaviours of shortest path dimensions with respect to the mixing probability are identified, including linear and nonlinear monotonic trends, as well as transitions with maxima. The approximated path matrix is proposed as an efficient alternative to extensive iterative simulations, reliably reproducing statistical results. The findings highlight the relevance of pattern properties in determining fractal structures and dynamics and suggest applications in physical systems such as diffusion, signal processing, and antenna design.
\end{abstract}

%\pacs{}% insert suggested PACS numbers in braces on next line

\maketitle

% SECTION%%SECTION%%SECTION%%SECTION%%SECTION%%SECTION%%SECTION%%SECTION%%SECTION%
\section{Introduction}
% SECTION%%SECTION%%SECTION%%SECTION%%SECTION%%SECTION%%SECTION%%SECTION%%SECTION%

\textit{Labyrinth fractals} are finitely ramified \Sierpinski carpets, 
which were first introduced and studied as self-similar dendrites 
\cite{cristea.l.09.curves.1,%
cristea.l.11.curves.329}, 
with focus on their arcs, 
i.e.,
the length and dimension thereof. 
Passing to more general objects, called \emph{mixed labyrinth fractals} 
\cite{cristea.l.17.mixed.112,%
cristea.l.18.on.575}, 
has lead to new challenges in handling the problems that had been solved in the self-similar case
and was solved only for mixed labyrinth fractals with particular shape 
\cite{cristea.l.18.on.575,%
	cristea.l.20.on.1}.
In the present paper we introduce and study new fractal objects that generalise the self-similar labyrinth fractals: 
\textit{randomised mixed labyrinth fractals}. 
We 
examine
them with respect to properties of their arcs and fractal dimensions of arcs,
and study their isotropy.
Randomised mixed labyrinth fractals offer a new framework for the 
investigation
of mixed labyrinth fractals with respect to fractal 
and shortest path
dimensions. 

In the last decades, \Sierpinski carpets occur in numerous research papers, 
e.g., when studying porous structures of materials 
\cite{franz.a.01.pore.8751,%
blaudeck.p.06.coastline.1609}, 
(anomalous)  diffusion in porous media  
\cite{anh.d.07.anomalous.11453,%
haber.r.13.diffusion.2840},  
and in disordered media 
\cite{havlin.s.02.diffusion.187}. 
 They are also considered in the context of  random walks on fractals 
 \cite{franz.a.00.efficient.155}, and diffusion on so-called ``disordered'' fractals 
 \cite{anh.d.05.diffusion.109}. 
There, finitely ramified \Sierpinski carpets have shown to be convenient tools for implementing various algorithms and 
 computational tools in the study of random walks  
 \cite{schulzky.c.00.resistance.1,%
 seeger.s.01.random.307} 
 or various dimensions of these fractals 
 \cite{franz.a.02.using.18,%
 franz.a.02.diffusion.52}.

Furthermore, 
 dynamical processes on Cantor sets 
 \cite{golmankhaneh.a.18.sub.960}, 
 which are totally disconnected fractals,
and on random Koch curves
 \cite{seeger.s.09.random.225002}, 
have been studied.
In labyrinth fractals, arcs that connect special points (called exits) can be viewed as families of Koch curves. 
In this paper we investigate the dimensions of these arcs in labyrinth fractals
which in general are not self-similar,
based on a new, statistical approach.

In the last decade, the results on  labyrinth fractals
have found applications in physics,  e.g.,   
in the fractal reconstruction of complicated
images
\cite{husain.a.22.fractals.379}, 
signals and radar backgrounds 
\cite{potapov.a.13.nano.941}, 
the construction of fractal (nano-) antennas
\cite{puente.c.96.fractal.1,potapov.a.16.simulation.1, potapov.a.17.fractal.499, anguera.j.20.fractal.4},
or the development of fractal gas sensors 
\cite{tian.f.21.application.14587, yang.t.21.optimizing.16675}.
A review of further applications in engineering, industrial and commercial applications is given in 
\cite{husain.a.22.fractals.379}.

Aspects like restoration of isotropy were studied on  self-similar fractals 
\cite{barlow.m.95.restoration.3042,%
kumagai.t.96.homogenization.375,%
barlow.m.97.weak.1}.  
In the present paper we address the isotropy of labyrinth fractals, passing from the context of self-similar to randomised mixed labyrinth fractals. 
First, we show that restoration of isotropy, proven for the \Sierpinski gasket 
\cite{barlow.m.95.restoration.3042}, also holds for self-similar labyrinth fractals. 
Subsequently, we pass to mixed and randomised mixed labyrinth fractals, which in general 
are only statistically self-similar or non-self-similar.

We recall that while mathematicians usually deal with the fractal constructed 
as the limit of the infinite sequence of prefractals obtained along the iterative construction, 
in physics the focus is on the prefractals, i.e., 
the sets obtained after $\iter$ iterations, where $\iter$ is chosen large enough such that 
the $\iter$-th prefractal displays the 
proper characteristics or 
the features sought for the problem to be solved. 
This is also the case in our approach.

The present paper aims at a novel approach of open problems on labyrinth fractals. 
Our approach is based on new objects called randomised mixed labyrinth fractals, 
which we introduce in order to obtain objects that can be ``tuned'' with respect to their topology or fractal dimensions. 
We showcase two ways of constructing families of dendrites from labyrinth patterns: 
	mixing the rotated images of an initial pattern and 
	mixing two patterns with different shape features, 
	that occur with probabilities $p$ and $1-p$, 
	respectively 
Fig.~\ref{fig:A1A2A3A4} shows four examples of labyrinth patterns. 
We study different dimensions of arcs in the fractals, and, 
correspondingly, the length of paths in the prefractals. 
We investigate their dependence
 on certain features of the patterns that generate the fractal, 
in particular regarding their shape with respect to the property of being non-blocked or totally blocked.

The paper is structured as follows.
In Section \ref{sec:concepts_and_notation} and \ref{sec:StatAna}, the concept of
randomised mixed labyrinth fractals, and  quantities of interest are introduced.
We then 
focus on a selected set of 
labyrinth patterns in Section \ref{sec:analysed_patterns} 
and discuss the resulting path length scaling behaviour 
for increasing iteration depth in Section \ref{sec:ResultsIsotropy} as well as
their structural properties in Section \ref{sec:ResultsDmin}.
Finally, in Section \ref{sec:Sum}, we give a short summary over all outcomes.

% SECTION%%SECTION%%SECTION%%SECTION%%SECTION%%SECTION%%SECTION%%SECTION%%SECTION%
\section{\label{sec:concepts_and_notation}Concepts and notation}
% SECTION%%SECTION%%SECTION%%SECTION%%SECTION%%SECTION%%SECTION%%SECTION%%SECTION%

\subsection{\label{sec:SS-MLF}Self-similar and mixed labyrinth fractals}
%SUBSECTION%%SUBSECTION%%SUBSECTION%%SUBSECTION%%SUBSECTION%%SUBSECTION%%SUBSECTION%

Labyrinth fractals are constructed based on \emph{labyrinth patterns}. 
For details of the mathematical formalism we refer, e.g., to 
\cite{cristea.l.11.curves.329,%
cristea.l.17.mixed.112}. 
 A pattern of width $\linl$, in short $\linl$-pattern (or simply pattern, 
 when $\linl$ is already specified), is obtained by dividing the unit square into $m \times m$ squares of side length $1/\linl$ and colouring $w$ of the $\linl^2$ squares,  
in white, 
and the rest $\linl^2-\mass$ in black,
where $1\leq\mass\leq\linl^2-1$.
 
 Roughly speaking, the black squares correspond to
 what will be ``cut out'' through the iterative construction of fractals based upon such patterns. 
To any $\linl$-pattern $\Ac$ we associate a graph $\G(\Ac)$ defined as follows: 
each vertex in the graph corresponds to a white square in the pattern, and two distinct vertices in the graph are connected by an edge if the corresponding squares share a common edge. A tree is a connected graph that contains no loops.
A pattern $\Ac$ is called  a \emph{labyrinth pattern} if
is has three properties: the tree property, the exits property and the corner property.

\begin{property}\label{tree_prop}[Tree Property]
$\G({\Ac})$ is a tree.
\end{property}
%------------------------------------------------------------------------------------------------------------------------------------
\begin{property}\label{exits_prop}[Exits Property]
${\Ac}$ has exactly one vertical exit pair, and exactly one horizontal exit pair.
\end{property}
%------------------------------------------------------------------------------------------------------------------------------------

The horizontal exit pair in a pattern is a pair of white squares that lie in the same row of the pattern, such that one of them, called the \emph{left exit},  lies in the leftmost and the other one, called the \emph{right exit}, in the rightmost column of the pattern. The vertical exit pair consists of the \emph{top exit} and the \emph{bottom exit}, which lie in the same column of the pattern, in the top and, respectively, bottom row.
\begin{property}\label{corner_prop}[Corner Property]
If there is a white square in ${\Ac}$ at a corner of ${\Ac}$, then there is no white square in ${\Ac}$ at the diagonally opposite corner of ${\Ac}$. 
\end{property}

% FIGURE%%FIGURE%%FIGURE%%FIGURE%%FIGURE%%FIGURE%%FIGURE%%FIGURE%%FIGURE%%FIGURE %
%---------------neu-------------------
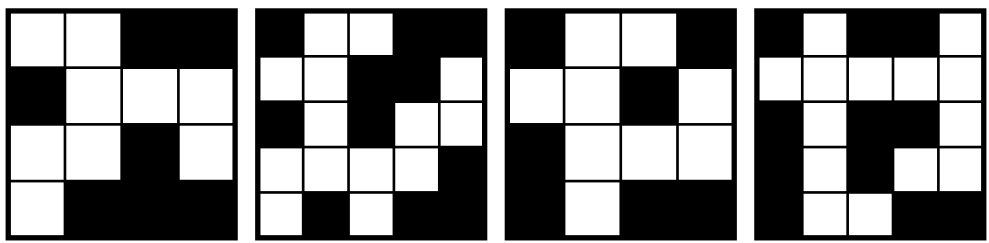
\begin{figure}[hhhh]
\begin{center}
	\begin{tikzpicture}[scale=.30]

\draw[line width=2pt] (0,0) rectangle (10,10);
\draw[line width=1pt] (2.5, 0) -- (2.5,10);
\draw[line width=1pt] (5, 0) -- (5,10);
\draw[line width=1pt] (7.5, 0) -- (7.5,10);
\draw[line width=1pt] (0, 2.5) -- (10,2.5);
\draw[line width=1pt] (0, 5) -- (10,5);
\draw[line width=1pt] (0, 7.5) -- (10,7.5);
\filldraw[fill=black, draw=black] (2.5,0) rectangle (5, 2.5);
\filldraw[fill=black, draw=black] (5,0) rectangle (7.5, 2.5);
\filldraw[fill=black, draw=black] (7.5,0) rectangle (10, 2.5);
\filldraw[fill=black, draw=black] (5,2.5) rectangle (7.5, 5);
\filldraw[fill=black, draw=black] (0,5) rectangle (2.5, 7.5);
\filldraw[fill=black, draw=black] (5,7.5) rectangle (7.5, 10);
\filldraw[fill=black, draw=black] (7.5,7.5) rectangle (10, 10);
\draw[line width=2pt] (11,0) rectangle (21,10);
\draw[line width=1pt] (13, 0) -- (13,10);
\draw[line width=1pt] (15, 0) -- (15,10);
\draw[line width=1pt] (17, 0) -- (17,10);
\draw[line width=1pt] (19, 0) -- (19,10);
\draw[line width=1pt] (11, 2) -- (21,2);
\draw[line width=1pt] (11, 4) -- (21,4);
\draw[line width=1pt] (11, 6) -- (21,6);
\draw[line width=1pt] (11, 8) -- (21,8);
\filldraw[fill=black, draw=black] (13,0) rectangle (15, 2);
\filldraw[fill=black, draw=black] (17,0) rectangle (19, 2);
\filldraw[fill=black, draw=black] (19,0) rectangle (21, 2);
\filldraw[fill=black, draw=black] (19,2) rectangle (21, 4);
\filldraw[fill=black, draw=black] (11,4) rectangle (13, 6);
\filldraw[fill=black, draw=black] (15,4) rectangle (17, 6);
\filldraw[fill=black, draw=black] (15,6) rectangle (17, 8);
\filldraw[fill=black, draw=black] (17,6) rectangle (19, 8);
\filldraw[fill=black, draw=black] (11,8) rectangle (13, 10);
\filldraw[fill=black, draw=black] (17,8) rectangle (19, 10 );
\filldraw[fill=black, draw=black] (19,8) rectangle (21, 10);
%%%%%%%nou: A3
\draw[line width=2pt] (22,0) rectangle (32,10);
\draw[line width=1pt] (24.5, 0) -- (24.5,10);
\draw[line width=1pt] (27,0) -- (27,10);
\draw[line width=1pt] (29.5, 0) -- (29.5,10);
\draw[line width=1pt] (22, 2.5) -- (32,2.5);
\draw[line width=1pt] (22, 5) -- (32,5);
\draw[line width=1pt] (22, 7.5) -- (32,7.5);
\filldraw[fill=black, draw=black] (22,0) rectangle (24.5, 5);
\filldraw[fill=black, draw=black] (22,7.5) rectangle (24.5, 10);
%\filldraw[fill=black, draw=black] (24.5,0) rectangle (27, 2.5);
\filldraw[fill=black, draw=black] (27,0) rectangle (29.5, 2.5);
\filldraw[fill=black, draw=black] (27,5) rectangle (29.5, 7.5);
\filldraw[fill=black, draw=black] (29.5,7.5) rectangle (32, 10);
\filldraw[fill=black, draw=black] (29.5,0) rectangle (32, 2.5);
%%%%%%% neu A4
%Raster
\draw[line width=2pt] (33,0) rectangle (43,10);
\draw[line width=1pt] (35, 0) -- (35,10);
\draw[line width=1pt] (37,0) -- (37,10);
\draw[line width=1pt] (39, 0) -- (39,10);
\draw[line width=1pt] (41, 0) -- (41,10);
\draw[line width=1pt] (33, 2) -- (43,2);
\draw[line width=1pt] (33, 4) -- (43,4);
\draw[line width=1pt] (33, 6) -- (43,6);
\draw[line width=1pt] (33, 8) -- (43,8);
%schwarze Quadrate
\filldraw[fill=black, draw=black] (33,0) rectangle (35, 6);
\filldraw[fill=black, draw=black] (33,8) rectangle (35, 10);
\filldraw[fill=black, draw=black] (37,8) rectangle (41, 10);
\filldraw[fill=black, draw=black] (37,2) rectangle (39, 6);
\filldraw[fill=black, draw=black] (39,0) rectangle (41, 2);
\filldraw[fill=black, draw=black] (39,4) rectangle (41, 6);
\filldraw[fill=black, draw=black] (39,8) rectangle (41, 10);
\filldraw[fill=black, draw=black] (41,0) rectangle (43, 2);

\end{tikzpicture}
\end{center}
\caption{\label{fig:A1A2A3A4}%
	Four labyrinth patterns: 
	${\Ac^{(1)}}$ ($4$-pattern), ${\Ac^{(2)}}$ ($5$-pattern), ${\Ac^{(3)}}$ ($4$-pattern) and $\Ac^{(4)}$ ($5$-pattern)
	}
\end{figure}
% FIGURE%%FIGURE%%FIGURE%%FIGURE%%FIGURE%%FIGURE%%FIGURE%%FIGURE%%FIGURE%%FIGURE %

A labyrinth pattern is called \emph{horizontally blocked}, if the row that contains the left and right exit contains at least one black square, and \emph{vertically blocked} if the row containing the top and bottom exit contains at least one black square. A labyrinth pattern that is both vertically and horizontally blocked is called \emph{totally blocked}. In 
Fig.~\ref{fig:A1A2A3A4}, the labyrinth patterns 
$\Ac^{(1)}$ and 
$\Ac^{(2)}$  are totally blocked, 
$\Ac^{(3)}$ is horizontally but not vertically blocked, and 
$\Ac^{(4)}$  
is neither horizontally nor  vertically blocked, in short \emph{non-blocked}.

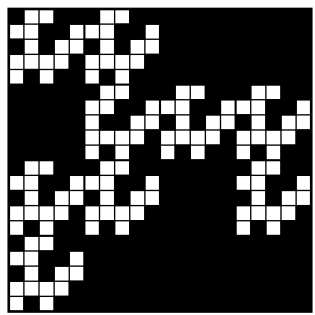
\begin{figure}[hhhh]
\begin{center}
	\begin{tikzpicture}[scale=.2]
\draw[line width=1pt] (0,0) rectangle (20,20);
%------------------------grid of level 1
\draw[line width=0.8pt] (5, 0) -- (5,20);
\draw[line width=0.8pt] (10, 0) -- (10,20);
\draw[line width=0.8pt] (15, 0) -- (15,20);
\draw[line width=0.8pt] (0, 5) -- (20,5);
\draw[line width=0.8pt] (0, 10) -- (20,10);
\draw[line width=0.8pt] (0, 15) -- (20,15);
%---------------------------grid of level 2
%----- vertical
\draw[line width=0.5pt] (1, 0) -- (1,20);
\draw[line width=0.5pt] (2, 0) -- (2,20);
\draw[line width=0.5pt] (3, 0) -- (3,20);
\draw[line width=0.5pt] (4, 0) -- (4,20);
\draw[line width=0.5pt] (6, 0) -- (6,20);
\draw[line width=0.5pt] (7, 0) -- (7,20);
\draw[line width=0.5pt] (8, 0) -- (8,20);
\draw[line width=0.5pt] (9, 0) -- (9,20);
\draw[line width=0.5pt] (11, 0) -- (11,20);
\draw[line width=0.5pt] (12, 0) -- (12,20);
\draw[line width=0.5pt] (13, 0) -- (13,20);
\draw[line width=0.5pt] (14, 0) -- (14,20);
\draw[line width=0.5pt] (16, 0) -- (16,20);
\draw[line width=0.5pt] (17, 0) -- (17,20);
\draw[line width=0.5pt] (18, 0) -- (18,20);
\draw[line width=0.5pt] (19, 0) -- (19,20);
%-------------------horizontal
\draw[line width=0.5pt] (0, 1) -- (20,1);
\draw[line width=0.5pt] (0, 2) -- (20,2);
\draw[line width=0.5pt] (0, 3) -- (20,3);
\draw[line width=0.5pt] (0, 4) -- (20,4);
\draw[line width=0.5pt] (0, 6) -- (20,6);
\draw[line width=0.5pt] (0, 7) -- (20,7);
\draw[line width=0.5pt] (0, 8) -- (20,8);
\draw[line width=0.5pt] (0, 9) -- (20,9);
\draw[line width=0.5pt] (0, 11) -- (20,11);
\draw[line width=0.5pt] (0, 12) -- (20,12);
\draw[line width=0.5pt] (0, 13) -- (20,13);
\draw[line width=0.5pt] (0, 14) -- (20,14);
\draw[line width=0.5pt] (0, 16) -- (20,16);
\draw[line width=0.5pt] (0, 17) -- (20,17);
\draw[line width=0.5pt] (0, 18) -- (20,18);
\draw[line width=0.5pt] (0, 19) -- (20,19);
%--- black quares of level 1
\filldraw[fill=black, draw=black] (5,0) rectangle (10, 5);
\filldraw[fill=black, draw=black] (10,0) rectangle (15, 5);
\filldraw[fill=black, draw=black] (15,0) rectangle (20, 5);
\filldraw[fill=black, draw=black] (10,5) rectangle (15, 10);
\filldraw[fill=black, draw=black] (0,10) rectangle (5, 15);
\filldraw[fill=black, draw=black] (10,15) rectangle (15, 20);
\filldraw[fill=black, draw=black] (15,15) rectangle (20, 20);
%----- black squares of level 2
\filldraw[fill=black, draw=black] (1,0) rectangle (2, 1);
\filldraw[fill=black, draw=black] (3,0) rectangle (4, 1);
\filldraw[fill=black, draw=black] (4,0) rectangle (5, 1);
\filldraw[fill=black, draw=black] (4,1) rectangle (5, 2);
\filldraw[fill=black, draw=black] (0,2) rectangle (1, 3);
\filldraw[fill=black, draw=black] (2,2) rectangle (3, 3);
\filldraw[fill=black, draw=black] (2,3) rectangle (3, 4);
\filldraw[fill=black, draw=black] (3,3) rectangle (4, 4);
\filldraw[fill=black, draw=black] (0,4) rectangle (1, 5);
\filldraw[fill=black, draw=black] (3,4) rectangle (4, 5 );
\filldraw[fill=black, draw=black] (4,4) rectangle (5, 5);
%-----------+(0,5)----------------------------------
\filldraw[fill=black, draw=black] (1,5) rectangle (2, 6);
\filldraw[fill=black, draw=black] (3,5) rectangle (4, 6);
\filldraw[fill=black, draw=black] (4,5) rectangle (5, 6);
\filldraw[fill=black, draw=black] (4,6) rectangle (5, 7);
\filldraw[fill=black, draw=black] (0,7) rectangle (1, 8);
\filldraw[fill=black, draw=black] (2,7) rectangle (3, 8);
\filldraw[fill=black, draw=black] (2,8) rectangle (3, 9);
\filldraw[fill=black, draw=black] (3,8) rectangle (4, 9);
\filldraw[fill=black, draw=black] (0,9) rectangle (1, 10);
\filldraw[fill=black, draw=black] (3,9) rectangle (4, 10);
\filldraw[fill=black, draw=black] (4,9) rectangle (5, 10);
%----------+(5,5)----------------------------------------
\filldraw[fill=black, draw=black] (6,5) rectangle (7, 6);
\filldraw[fill=black, draw=black] (8,5) rectangle (9, 6);
\filldraw[fill=black, draw=black] (9,5) rectangle (10, 6);
\filldraw[fill=black, draw=black] (9,6) rectangle (10, 7);
\filldraw[fill=black, draw=black] (5,7) rectangle (6, 8);
\filldraw[fill=black, draw=black] (7,7) rectangle (8, 8);
\filldraw[fill=black, draw=black] (7,8) rectangle (8, 9);
\filldraw[fill=black, draw=black] (8,8) rectangle (9, 9);
\filldraw[fill=black, draw=black] (5,9) rectangle (6, 10);
\filldraw[fill=black, draw=black] (8,9) rectangle (9, 10);
\filldraw[fill=black, draw=black] (9,9) rectangle (10, 10);
%---- --+(5,10)------------------------------------------
\filldraw[fill=black, draw=black] (6,10) rectangle (7, 11);
\filldraw[fill=black, draw=black] (8,10) rectangle (9, 11);
\filldraw[fill=black, draw=black] (9,10) rectangle (10, 11);
\filldraw[fill=black, draw=black] (9,11) rectangle (10, 12);
\filldraw[fill=black, draw=black] (7,12) rectangle (6, 13);
\filldraw[fill=black, draw=black] (7,12) rectangle (8, 13);
\filldraw[fill=black, draw=black] (7,13) rectangle (8, 14);
\filldraw[fill=black, draw=black] (8,13) rectangle (9, 14);
\filldraw[fill=black, draw=black] (5,14) rectangle (6, 15);
\filldraw[fill=black, draw=black] (8,14) rectangle (9, 15 );
\filldraw[fill=black, draw=black] (9,14) rectangle (10, 15);
%----------+(10,10)-----------------------------------------
\filldraw[fill=black, draw=black] (11,10) rectangle (12, 11);
\filldraw[fill=black, draw=black] (13,10) rectangle (14, 11);
\filldraw[fill=black, draw=black] (14,10) rectangle (15, 11);
\filldraw[fill=black, draw=black] (14,11) rectangle (15, 12);
\filldraw[fill=black, draw=black] (10,12) rectangle (11, 13);
\filldraw[fill=black, draw=black] (12,12) rectangle (13, 13);
\filldraw[fill=black, draw=black] (12,13) rectangle (13, 14);
\filldraw[fill=black, draw=black] (13,13) rectangle (14, 14);
\filldraw[fill=black, draw=black] (10,14) rectangle (11, 15);
\filldraw[fill=black, draw=black] (13,14) rectangle (14, 15 );
\filldraw[fill=black, draw=black] (14,14) rectangle (15, 15);
%----------+(15,5)------------------
\filldraw[fill=black, draw=black] (16,5) rectangle (17, 6);
\filldraw[fill=black, draw=black] (18,5) rectangle (19, 6);
\filldraw[fill=black, draw=black] (19,5) rectangle (20, 6);
\filldraw[fill=black, draw=black] (19,6) rectangle (20, 7);
\filldraw[fill=black, draw=black] (15,7) rectangle (16, 8);
\filldraw[fill=black, draw=black] (17,7) rectangle (18, 8);
\filldraw[fill=black, draw=black] (17,8) rectangle (18, 9);
\filldraw[fill=black, draw=black] (18,8) rectangle (19, 9);
\filldraw[fill=black, draw=black] (15,9) rectangle (16, 10);
\filldraw[fill=black, draw=black] (18,9) rectangle (19, 10 );
\filldraw[fill=black, draw=black] (19,9) rectangle (20, 10);
%---------+(15,10)-----------------
\filldraw[fill=black, draw=black] (16,10) rectangle (17, 11);
\filldraw[fill=black, draw=black] (18,10) rectangle (19, 11);
\filldraw[fill=black, draw=black] (19,10) rectangle (20, 11);
\filldraw[fill=black, draw=black] (19,11) rectangle (20, 12);
\filldraw[fill=black, draw=black] (15,12) rectangle (16, 13);
\filldraw[fill=black, draw=black] (17,12) rectangle (18, 13);
\filldraw[fill=black, draw=black] (17,13) rectangle (18, 14);
\filldraw[fill=black, draw=black] (18,13) rectangle (19, 14);
\filldraw[fill=black, draw=black] (15,14) rectangle (16, 15);
\filldraw[fill=black, draw=black] (18,14) rectangle (19, 15);
\filldraw[fill=black, draw=black] (19,14) rectangle (20, 15);
%-----------+(0,15)------------
\filldraw[fill=black, draw=black] (1,15) rectangle (2, 16);
\filldraw[fill=black, draw=black] (3,15) rectangle (4, 16);
\filldraw[fill=black, draw=black] (4,15) rectangle (5, 16);
\filldraw[fill=black, draw=black] (4,16) rectangle (5, 17);
\filldraw[fill=black, draw=black] (0,17) rectangle (1, 18);
\filldraw[fill=black, draw=black] (2,17) rectangle (3, 18);
\filldraw[fill=black, draw=black] (2,18) rectangle (3, 19);
\filldraw[fill=black, draw=black] (3,18) rectangle (4, 19);
\filldraw[fill=black, draw=black] (0,19) rectangle (1, 20);
\filldraw[fill=black, draw=black] (3,19) rectangle (4, 20);
\filldraw[fill=black, draw=black] (4,19) rectangle (5, 20);
%----------+(5,15)--------------------
\filldraw[fill=black, draw=black] (6,15) rectangle (7, 16);
\filldraw[fill=black, draw=black] (8,15) rectangle (9, 16);
\filldraw[fill=black, draw=black] (9,15) rectangle (10, 16);
\filldraw[fill=black, draw=black] (9,16) rectangle (10, 17);
\filldraw[fill=black, draw=black] (5,17) rectangle (6, 18);
\filldraw[fill=black, draw=black] (7,17) rectangle (8, 18);
\filldraw[fill=black, draw=black] (7,18) rectangle (8, 19);
\filldraw[fill=black, draw=black] (8,18) rectangle (9, 19);
\filldraw[fill=black, draw=black] (5,19) rectangle (6, 20);
\filldraw[fill=black, draw=black] (8,19) rectangle (9, 20);
\filldraw[fill=black, draw=black] (9,19) rectangle (10, 20);
%--------+(15,15)-----------------
\filldraw[fill=black, draw=black] (16,15) rectangle (17, 16);
\filldraw[fill=black, draw=black] (18,15) rectangle (19, 16);
\filldraw[fill=black, draw=black] (19,15) rectangle (20, 16);
\filldraw[fill=black, draw=black] (19,16) rectangle (20, 17);
\filldraw[fill=black, draw=black] (15,17) rectangle (16, 18);
\filldraw[fill=black, draw=black] (17,17) rectangle (18, 18);
\filldraw[fill=black, draw=black] (17,18) rectangle (18, 19);
\filldraw[fill=black, draw=black] (18,18) rectangle (19, 19);
\filldraw[fill=black, draw=black] (15,19) rectangle (16, 20);
\filldraw[fill=black, draw=black] (18,19) rectangle (19, 20);
\filldraw[fill=black, draw=black] (19,19) rectangle (20, 20);
\end{tikzpicture}
\end{center}
\caption{\label{fig:W2}%
	The mixed labyrinth set ${\W}_2$ of level 2, constructed based on the above patterns 
	$\Ac_1=\Ac^{(1)}$  and 
	$\Ac_2=\Ac^{(2)}$ (see Fig.~\ref{fig:A1A2A3A4})
	can also be viewed as a $20$-pattern} 
\end{figure}

 Let $\{\Ac_k\}_{k\ge 1}$ be a sequence of labyrinth patterns. 
 A \emph{labyrinth set of level} $\iter$, 
 denoted by $\W_\iter$, for $n \ge 1,$ is constructed as follows: 
 We start with a labyrinth pattern $\Ac_1$, 
 and define the corresponding labyrinth set of level $1$, $\W_1:=\Ac_1$. 
 Given $\W_{\iter-1}$, $\iter \ge 2$, 
 we obtain $\W_\iter$ by replacing every white square in $\W_{\iter-1}$ 
 by the scaled image of the pattern $\Ac_\iter$. Then $\W_\iter$ has the tree, 
 the exits 
 and the corner property 
 \cite{cristea.l.17.mixed.112}, 
 and can therefore be viewed as an $\linl(\iter)$-labyrinth pattern, 
 where $\linl(\iter)=\prod_{k=1}^\iter \linl_k$.  
 The limit set $\linf$  of a sequence of labyrinth sets of $\{\W_\iter \}_{\iter\ge 1}$  
 constructed as above is called a \emph{labyrinth fractal}. 
In particular, if all patterns $\Ac_k$ are identical,
then we obtain a self-similar labyrinth fractal $\linf$, 
otherwise  $\linf$ is a \emph{mixed} labyrinth fractal.
Correspondingly, we distinguish between  self-similar and mixed prefractals:  
For $n \ge 2$ call the labyrinth set $\W_n$ generated by the finite sequence $\{ \Ac_k \}_{k=1}^n$ 
a self-similar labyrinth set of level $n$ if $\Ac_1=\Ac_2=\dots\Ac_n,$ 
and a mixed labyrinth set of level $n$ if $\Ac_1, \dots, \Ac_n$ do not all coincide. 
Fig.~\ref{fig:W2} shows an example of a mixed labyrinth set of level 2: 
here $\W_2$ is obtained by starting with 
$\Ac_1=\Ac^{(1)}$  and replacing  all white squares in 
$\W_1=\Ac_1$
with
the pattern 
$\Ac_2=\Ac^{(2)}$, scaled by the factor $1/4$.

% FIGURE%%FIGURE%%FIGURE%%FIGURE%%FIGURE%%FIGURE%%FIGURE%%FIGURE%%FIGURE%%FIGURE %
% FIGURE%%FIGURE%%FIGURE%%FIGURE%%FIGURE%%FIGURE%%FIGURE%%FIGURE%%FIGURE%%FIGURE %

The fact that any labyrinth pattern and any labyrinth set of level $\iter\ge 1$ 
has the Exits Property naturally yields (in the limit for $\iter \to \infty$)
four special points: the exits of the fractal, one on each side of the unit square \cite{cristea.l.11.curves.329, cristea.l.17.mixed.112}.
Such fractals are also called finitely ramified.

The path matrix of a labyrinth pattern or labyrinth set is a convenient tool
 when studying arcs in the fractal or lengths of paths in (the graph of) a labyrinth set of level $\iter$, both in the self-similar and in the mixed case.
For $\iter\ge 1$, let $\W_1, \W_2, \W_3 \dots$ denote the labyrinth sets of level $\iter\ge 1$ occurring in the construction of a labyrinth fractal. 
 We call a path in $\G({\W}_{\iter})$, $\iter\ge 1$, the $\A$\emph{-path} if it connects the top and 
the bottom exit of $\W_\iter$. 
The $\B,\C,\D,\E$, and $\F$\emph{-paths} connect the other pairs of exits, respectively, as indicated by the symbols.
Let $\P=\{\A, \B, \C, \D,\E, \F \}$ be an ordered set of indices, 
which represent the types of paths between exits in a labyrinth pattern or 
labyrinth set of level $\iter\ge 1$.
We denote by $a(n;\ep)$, with $\ep \in \P$, the path of type $\ep$ in $\G(\W_n)$ and by 
 $\ell_n(\gamma)=\ell(a(\iter;\ep))$
 the length of $a(\iter;\ep)$, i.e., the number of squares in the path, for $\ep \in \P$.

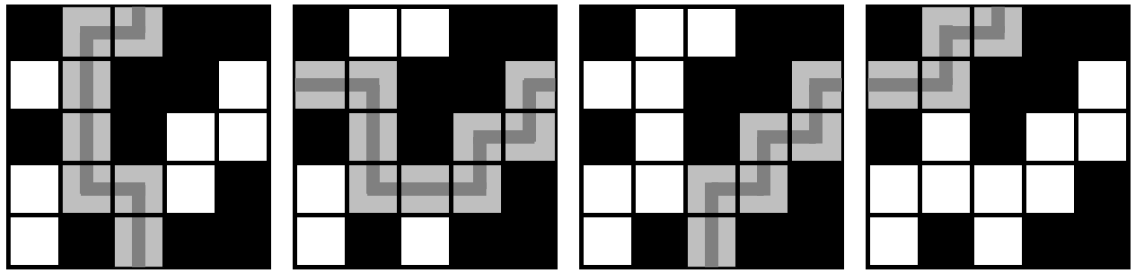
\begin{figure}[h]
\begin{center}
	\scalebox{.84}{\begin{tikzpicture}[scale=.41]
%path from bottom to top exit
\draw[line width=2pt] (0,0) rectangle (10,10);

%------ die grauen quadrate durch die der weg geht ----
\filldraw[fill=gray!50, draw= black] (4,0) rectangle (6,2);
\filldraw[fill=gray!50, draw= black] (4,2) rectangle (6,4);
\filldraw[fill=gray!50, draw= black] (2,2) rectangle (4,4);
\filldraw[fill=gray!50, draw= black] (2,4) rectangle (4,6);
\filldraw[fill=gray!50, draw= black] (2,6) rectangle (4,8);
\filldraw[fill=gray!50, draw= black] (2,8) rectangle (4,10);
\filldraw[fill=gray!50, draw= black] (4,8) rectangle (6,10);
%------ der weg
\draw[line width=6 pt, color=gray] (5, 0) -- (5, 3.15);
\draw[line width=6 pt, color=gray] (2.78,3) -- (5.25,3);
\draw[line width=6 pt, color=gray] (3, 2.82) -- (3, 9);
\draw[line width=6 pt, color=gray] (2.75, 9) -- (5.25, 9);
\draw[line width=6 pt, color=gray] (5, 9) -- (5, 10);
%---- jetzt das dicke raster--------
\draw[line width=2pt] (2, 0) -- (2,10);
\draw[line width=2pt] (4, 0) -- (4,10);
\draw[line width=2pt] (6, 0) -- (6,10);
\draw[line width=2pt] (8, 0) -- (8,10);
\draw[line width=2pt] (0, 2) -- (10,2);
\draw[line width=2pt] (0, 4) -- (10,4);
\draw[line width=2pt] (0, 6) -- (10,6);
\draw[line width=2pt] (0, 8) -- (10,8);
%--- jetzt die schwarzen quadrate-------------
\filldraw[fill=black, draw=black] (2,0) rectangle (4, 2);
\filldraw[fill=black, draw=black] (6,0) rectangle (8, 2);
\filldraw[fill=black, draw=black] (8,0) rectangle (10, 2);
\filldraw[fill=black, draw=black] (8,2) rectangle (10, 4);
\filldraw[fill=black, draw=black] (0,4) rectangle (2, 6);
\filldraw[fill=black, draw=black] (4,4) rectangle (6, 6);
\filldraw[fill=black, draw=black] (4,6) rectangle (6, 8);
\filldraw[fill=black, draw=black] (6,6) rectangle (8, 8);
\filldraw[fill=black, draw=black] (0,8) rectangle (2, 10);
\filldraw[fill=black, draw=black] (6,8) rectangle (8, 10 );
\filldraw[fill=black, draw=black] (8,8) rectangle (10, 10);

% path from left to right exit
\draw[line width=2pt] (11,0) rectangle (21,10);

%------ die grauen quadrate durch die der weg geht ----
\filldraw[fill=gray!50, draw= black] (11,6) rectangle (13,8);
\filldraw[fill=gray!50, draw= black] (13,6) rectangle (15,8);
\filldraw[fill=gray!50, draw= black] (13,4) rectangle (15,6);
\filldraw[fill=gray!50, draw= black] (13,2) rectangle (15,4);
\filldraw[fill=gray!50, draw= black] (15,2) rectangle (17,4);
\filldraw[fill=gray!50, draw= black] (17,2) rectangle (19,4);
\filldraw[fill=gray!50, draw= black] (17,4) rectangle (19,6);
\filldraw[fill=gray!50, draw= black] (19,4) rectangle (21,6);
\filldraw[fill=gray!50, draw= black] (19,6) rectangle (21,8);
%------ der weg
\draw[line width=6 pt, color=gray] (11,7) -- (14.24,7 );
\draw[line width=6 pt, color=gray] (14, 2.82) -- (14, 7);
\draw[line width=6 pt, color=gray] (13.78,3) -- (18.35,3);
\draw[line width=6 pt, color=gray] (18.1,3 ) -- (18.1,5 );
\draw[line width=6 pt, color=gray] (17.85,5 ) -- (20.2,5 );
\draw[line width=6 pt, color=gray] (20,4.82 ) -- (20,7.18 );
\draw[line width=6 pt, color=gray] (19.8,7 ) -- (21,7 );
%---- jetzt das dicke raster--------
\draw[line width=2pt] (13, 0) -- (13,10);
\draw[line width=2pt] (15, 0) -- (15,10);
\draw[line width=2pt] (17, 0) -- (17,10);
\draw[line width=2pt] (19, 0) -- (19,10);
\draw[line width=2pt] (11, 2) -- (21,2);
\draw[line width=2pt] (11, 4) -- (21,4);
\draw[line width=2pt] (11, 6) -- (21,6);
\draw[line width=2pt] (11, 8) -- (21,8);
%--- jetzt die schwarzen quadrate-------------
\filldraw[fill=black, draw=black] (13,0) rectangle (15, 2);
\filldraw[fill=black, draw=black] (17,0) rectangle (19, 2);
\filldraw[fill=black, draw=black] (19,0) rectangle (21, 2);
\filldraw[fill=black, draw=black] (19,2) rectangle (21, 4);
\filldraw[fill=black, draw=black] (11,4) rectangle (13, 6);
\filldraw[fill=black, draw=black] (15,4) rectangle (17, 6);
\filldraw[fill=black, draw=black] (15,6) rectangle (17, 8);
\filldraw[fill=black, draw=black] (17,6) rectangle (19, 8);
\filldraw[fill=black, draw=black] (11,8) rectangle (13, 10);
\filldraw[fill=black, draw=black] (17,8) rectangle (19, 10 );
\filldraw[fill=black, draw=black] (19,8) rectangle (21, 10);

% path from bottom to right exit
\draw[line width=2pt] (22,0) rectangle (32,10);

%------ die grauen quadrate durch die der weg geht ----
\filldraw[fill=gray!50, draw= black] (26,0) rectangle (28,2);
\filldraw[fill=gray!50, draw= black] (26,2) rectangle (28,4);
\filldraw[fill=gray!50, draw= black] (28,2) rectangle (30,4);
\filldraw[fill=gray!50, draw= black] (28,4) rectangle (30,6);
\filldraw[fill=gray!50, draw= black] (30,4) rectangle (32,6);
\filldraw[fill=gray!50, draw= black] (30,6) rectangle (32,8);
%------ der weg
\draw[line width=6 pt, color=gray] (27,0) -- (27, 3.25 );
\draw[line width=6 pt, color=gray] (27, 3) -- (29, 3);
\draw[line width=6 pt, color=gray] (29, 2.75) -- (29, 5);
\draw[line width=6 pt, color=gray] (28.75, 5 ) -- (31, 5 );
\draw[line width=6 pt, color=gray] (31, 4.75 ) -- (31, 7 );
\draw[line width=6 pt, color=gray] (30.75, 7 ) -- (32, 7);
%---- jetzt das dicke raster--------
\draw[line width=2pt] (24, 0) -- (24,10);
\draw[line width=2pt] (26, 0) -- (26,10);
\draw[line width=2pt] (28, 0) -- (28,10);
\draw[line width=2pt] (30, 0) -- (30,10);
\draw[line width=2pt] (22, 2) -- (32,2);
\draw[line width=2pt] (22, 4) -- (32,4);
\draw[line width=2pt] (22, 6) -- (32,6);
\draw[line width=2pt] (22, 8) -- (32,8);
%--- jetzt die schwarzen quadrate-------------
\filldraw[fill=black, draw=black] (24,0) rectangle (26, 2);
\filldraw[fill=black, draw=black] (28,0) rectangle (30, 2);
\filldraw[fill=black, draw=black] (30,0) rectangle (32, 2);
\filldraw[fill=black, draw=black] (30,2) rectangle (32, 4);
\filldraw[fill=black, draw=black] (22,4) rectangle (24, 6);
\filldraw[fill=black, draw=black] (26,4) rectangle (28, 6);
\filldraw[fill=black, draw=black] (26,6) rectangle (28, 8);
\filldraw[fill=black, draw=black] (28,6) rectangle (30, 8);
\filldraw[fill=black, draw=black] (22,8) rectangle (24, 10);
\filldraw[fill=black, draw=black] (28,8) rectangle (30, 10 );
\filldraw[fill=black, draw=black] (30,8) rectangle (32, 10);

% path from left to top exit
\draw[line width=2pt] (33,0) rectangle (43,10);

%------ die grauen quadrate durch die der weg geht ----
\filldraw[fill=gray!50, draw= black] (33,6) rectangle (35,8);
\filldraw[fill=gray!50, draw= black] (35,6) rectangle (37,8);
\filldraw[fill=gray!50, draw= black] (35,8) rectangle (39,10);
\filldraw[fill=gray!50, draw= black] (37,8) rectangle (39,10);
%------ der weg
\draw[line width=6 pt, color=gray] (33, 7) -- (36.25, 7 );
\draw[line width=6 pt, color=gray] (36, 7) -- (36, 9.29);
\draw[line width=6 pt, color=gray] (36, 9) -- (38.25, 9);
\draw[line width=6 pt, color=gray] (38, 9 ) -- (38,10);
%---- jetzt das dicke raster--------
\draw[line width=2pt] (35, 0) -- (35,10);
\draw[line width=2pt] (37, 0) -- (37,10);
\draw[line width=2pt] (39, 0) -- (39,10);
\draw[line width=2pt] (41, 0) -- (41,10);
\draw[line width=2pt] (33, 2) -- (43,2);
\draw[line width=2pt] (33, 4) -- (43,4);
\draw[line width=2pt] (33, 6) -- (43,6);
\draw[line width=2pt] (33, 8) -- (43,8);
%--- jetzt die schwarzen quadrate-------------
\filldraw[fill=black, draw=black] (35,0) rectangle (37, 2);
\filldraw[fill=black, draw=black] (39,0) rectangle (41, 2);
\filldraw[fill=black, draw=black] (41,0) rectangle (43, 2);
\filldraw[fill=black, draw=black] (41,2) rectangle (43, 4);
\filldraw[fill=black, draw=black] (33,4) rectangle (35, 6);
\filldraw[fill=black, draw=black] (37,4) rectangle (39, 6);
\filldraw[fill=black, draw=black] (37,6) rectangle (39, 8);
\filldraw[fill=black, draw=black] (39,6) rectangle (41, 8);
\filldraw[fill=black, draw=black] (33,8) rectangle (35, 10);
\filldraw[fill=black, draw=black] (39,8) rectangle (41, 10 );
\filldraw[fill=black, draw=black] (41,8) rectangle (43, 10);

\end{tikzpicture}}
\end{center}
\caption{\label{fig:A2_paths}%
	Paths from bottom to top (\A), from left to right (\B), from bottom to right (\D), and from left to top (\F) exit 
	of %$\mathcal{A}_2$
	the labyrinth pattern $\mathcal{A}_2$}
\end{figure}

\begin{figure}[h!]
\begin{center}
	\begin{tikzpicture}[scale=.25]
\draw[line width=2pt] (0,0) rectangle (20,20);
%---color the squares of level 1 that contain the path 
\filldraw[fill=gray!95, draw= black] (0,0) rectangle (5,5);
\filldraw[fill=gray!95, draw= black] (0,5) rectangle (5,10);
\filldraw[fill=gray!95, draw= black] (5,5) rectangle (10,10);
\filldraw[fill=gray!95, draw= black] (5,10) rectangle (10,15);
\filldraw[fill=gray!95, draw= black] (10,10) rectangle (15,15);
\filldraw[fill=gray!95, draw= black] (15,10) rectangle (20,15);
\filldraw[fill=gray!95, draw= black] (15,5) rectangle (20,10);
\draw[line width=1pt] (5, 0) -- (5,20);
\draw[line width=1pt] (10, 0) -- (10,20);
\draw[line width=1pt] (15, 0) -- (15,20);
\draw[line width=1pt] (0, 5) -- (20,5);
\draw[line width=1pt] (0, 10) -- (20,10);
\draw[line width=1pt] (0, 15) -- (20,15);
%%---------------------------grid of level 2
%%----- vertical
%\draw[line width=0.5pt] (1, 0) -- (1,20);
%\draw[line width=0.5pt] (2, 0) -- (2,20);
%\draw[line width=0.5pt] (3, 0) -- (3,20);
%\draw[line width=0.5pt] (4, 0) -- (4,20);
%\draw[line width=0.5pt] (6, 0) -- (6,20);
%\draw[line width=0.5pt] (7, 0) -- (7,20);
%\draw[line width=0.5pt] (8, 0) -- (8,20);
%\draw[line width=0.5pt] (9, 0) -- (9,20);
%\draw[line width=0.5pt] (11, 0) -- (11,20);
%\draw[line width=0.5pt] (12, 0) -- (12,20);
%\draw[line width=0.5pt] (13, 0) -- (13,20);
%\draw[line width=0.5pt] (14, 0) -- (14,20);
%\draw[line width=0.5pt] (16, 0) -- (16,20);
%\draw[line width=0.5pt] (17, 0) -- (17,20);
%\draw[line width=0.5pt] (18, 0) -- (18,20);
%\draw[line width=0.5pt] (19, 0) -- (19,20);
%%-------------------horizontal
%\draw[line width=0.5pt] (0, 1) -- (20,1);
%\draw[line width=0.5pt] (0, 2) -- (20,2);
%\draw[line width=0.5pt] (0, 3) -- (20,3);
%\draw[line width=0.5pt] (0, 4) -- (20,4);
%\draw[line width=0.5pt] (0, 6) -- (20,6);
%\draw[line width=0.5pt] (0, 7) -- (20,7);
%\draw[line width=0.5pt] (0, 8) -- (20,8);
%\draw[line width=0.5pt] (0, 9) -- (20,9);
%\draw[line width=0.5pt] (0, 11) -- (20,11);
%\draw[line width=0.5pt] (0, 12) -- (20,12);
%\draw[line width=0.5pt] (0, 13) -- (20,13);
%\draw[line width=0.5pt] (0, 14) -- (20,14);
%\draw[line width=0.5pt] (0, 16) -- (20,16);
%\draw[line width=0.5pt] (0, 17) -- (20,17);
%\draw[line width=0.5pt] (0, 18) -- (20,18);
%\draw[line width=0.5pt] (0, 19) -- (20,19);
%--- black quares of level 1
\filldraw[fill=black, draw=black] (5,0) rectangle (10, 5);
\filldraw[fill=black, draw=black] (10,0) rectangle (15, 5);
\filldraw[fill=black, draw=black] (15,0) rectangle (20, 5);
\filldraw[fill=black, draw=black] (10,5) rectangle (15, 10);
\filldraw[fill=black, draw=black] (0,10) rectangle (5, 15);
\filldraw[fill=black, draw=black] (10,15) rectangle (15, 20);
\filldraw[fill=black, draw=black] (15,15) rectangle (20, 20);

\end{tikzpicture}\hspace{1.5ex}\input{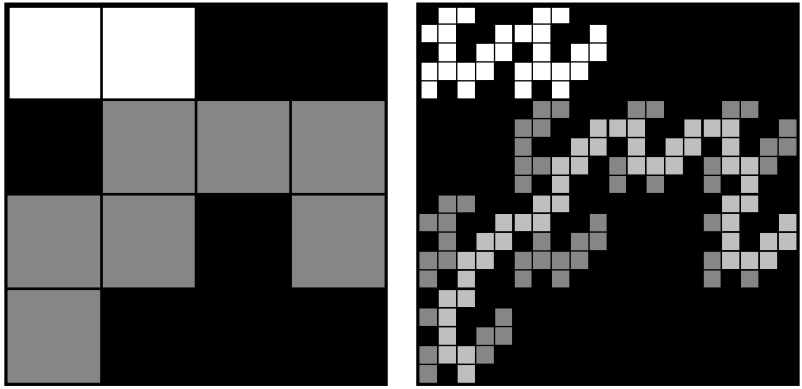}
\end{center}
\caption{\label{fig:W2_path}%
	On the left, the set $\W_1=\Ac_1$ is given, where the dark grey squares
	represent the \D-path. 
	The set $\W_2$ constructed with the patterns 
	$\Ac_1$ and $\Ac_2$ shown in Fig.~\ref{fig:A1A2A3A4} is shown on the right.
	The \D-path of $ \W_2$ is indicated in light grey.
	} 
\end{figure}

In order to construct the (unique) path in the graph $\G(\W_\iter)$ that connects, e.g., the bottom and the right exit of $\W_\iter$, for some $\iter\ge 1$,  we proceed as follows. 
First, we find the path between the right and the bottom exit of ${\W}_{1}$ 
(or, equivalently ${\Ac}_{1}$). Then, we assign a type of square to each white square in the path 
according to its neighbours within the path: 
if it has a top and a bottom neighbour it is called a $\A$-\emph{square}  
(with respect to the path), and
it is called a $\B,\C,\D,\E$, and $\F$-\emph{square}, correspondingly to the position of its neighbours, as indicated by the symbols. 
If a white square is an exit of $\W_1$, it is supposed to have a neighbour outside the 
side of the unit square on which the exit lies. For example, the bottom exit is supposed to have a neighbour below, outside the bottom side of the unit square, additionally to its neighbour that lies inside the unit square.
 This procedure is then repeated for all possible paths in $\G({\W}_{1})$ 
 between two exits of $\W_1$. 
 Fig.~\ref{fig:A2_paths} illustrates this procedure in a labyrinth pattern. 
In order to obtain the $\D$-path in $\G({\W}_{2})$ shown in Fig.~\ref{fig:W2_path},
 we replace each $\A$-square of 
the $\D$-path in $\G({\W}_{1})$ 
with the $\A$-path in $\G(\Ac_{2})$, 
shown in Fig.~\ref{fig:A2_paths}.
Then, we do this analogously for the other marked white squares
of the $\D$-path in $\G({\W}_{1})$, and thus get the
$\D$-path in ${\W}_{2}$ (s.~Fig.~\ref{fig:W2_path}). 
In the $\iter$-th iteration, with $\iter\ge 2$,
in order to obtain the $\gamma$-path of $\G({\W}_{\iter})$, 
for any pair of exits we replace each marked white 
square of type $\gamma'$ in the $\gamma$-path of $\G({\W}_{\iter-1})$ by the $\gamma'$-path in 
$\G({\Ac}_{\iter}),\gamma, \gamma' \in \P$. 
Thus $\P$ is also the set of all  possible types of squares with respect to a path in the graph of a labyrinth pattern or labyrinth set of 
level $n\ge 1$. 
In particular, each element of $\P$ corresponds to  exactly one pair of distinct exits and vice-versa.

We define the \emph{path matrix} $\pathMatrix{}=\left(\linl_{\nu,\pi}\right)_{\nu,\pi \in \P}$, of a labyrinth pattern  or labyrinth set of level $n\ge 1$ in the following way: 
the columns of $\pathMatrix{}$ from left to right and the rows of $\pathMatrix{}$ from top to bottom are indexed by $\A,\B,\C,\D,\E$, and $\F$, respectively,
and the element in row 
{$\nu$ and column $\pi$ of $\pathMatrix{}$ is the number of $\pi$-squares in the $\nu$-path} 
in the graph of the labyrinth pattern or set. 
It was proven 
\cite{cristea.l.17.mixed.112} 
that the path matrix $M(\iter)$ of a mixed labyrinth set of level $\iter$ is 
\begin{equation}
	\label{eq:pathmatrix_mixed}
	\pathMatrix{}(\iter)=\prod_{k=1}^\iter \pathMatrix{k},  \text{\,\, for\ all \,\,} n\ge1, 
\end{equation}
where $\pathMatrix{k}$ is the path matrix of the pattern $\Ac_k$.
Thus the length of all connected pairs of exits  $\pathVec{k}{}=(\ell_k(\A),\ell_k(\B),\ell_k(\C),\ell_k(\D),\ell_k(\E),\ell_k(\F))$ 
	can be determined by
\begin{align}
	\pathVec{k}{} &= \pathMatrix{k}\cdot\pathVec{k-1}{} \;\;, & k &\geq 2 ,\text{ with } \pathVec{1}{}=(1,1,1,1,1,1)^T \;.
\end{align}

% FIGURE%%FIGURE%%FIGURE%%FIGURE%%FIGURE%%FIGURE%%FIGURE%%FIGURE%%FIGURE %
% FIGURE%%FIGURE%%FIGURE%%FIGURE%%FIGURE%%FIGURE%%FIGURE%%FIGURE%%FIGURE %

% FIGURE%%FIGURE%%FIGURE%%FIGURE%%FIGURE%%FIGURE%%FIGURE%%FIGURE%%FIGURE %
% FIGURE%%FIGURE%%FIGURE%%FIGURE%%FIGURE%%FIGURE%%FIGURE%%FIGURE%%FIGURE %

It was shown 
\cite{cristea.l.11.curves.329} 
that the path matrix of a pattern is a primitive matrix 
(i.e., there exists a finite number $\iter$ such that $\pathMatrix{}^\iter$ has only strictly positive entries) 
if and only if the pattern is totally blocked. In this case the spectral radius $r$ of the path matrix 
(the maximum absolute value among its eigenvalues) is strictly greater than the pattern's width $\linl$.

In order to study the dimension of arcs in mixed labyrinth fractals let us introduce here the definitions of the following four types of dimension.
First of all we recall that
the box counting dimension $\darc{}(F)$ of a set $F$ can be obtained from the formula
\begin{equation}
	\label{eq:def_boxcountingdim}
	\darc{}(F)= \lim_{\delta \to 0} \frac{\log N_{\delta}(F)}{-\log \delta} \;,
\end{equation}
where $N_{\delta}(F)$ is the number of $\delta$-mesh squares that cover $F$.
If the limit in (\ref{eq:def_boxcountingdim}) does not exist,
then the upper and lower box counting dimensions are defined analogously,
using $\lim\sup$ and $\liminf$, respectively.
We refer to 
\cite{falconer.k.14.fractal.book} 
for more details and equivalent definitions of this fractal dimension.

The fractal dimension $\df$ of a 
mixed fractal
expresses the scaling of the amount of white squares,
i.e., the so-called mass $\mass$  versus the width $\linl$ of 
the labyrinth patterns that generates 
mixed labyrinth fractal, 
$\mass \sim \linl^{\df}$.
It can be determined as
\begin{align}
	\df &= \lim_{n\to\infty} \frac{ \log\left( \mass(n) %\frac{\mass(n)}{\mass(n-1)} 
		\right) }{ \log\left( \linl(n) %\frac{\linl(n)}{\linl(n-1)} 
		\right) } \;.
\label{eq:df}
\end{align}
In the case of a self-similar labyrinth fractal, the dimension defined by Eq.~\eqref{eq:df} 
reduces to $\df =  \frac{\log{\mass}}{\log{\linl}}$ and
coincides with the Hausdorff dimension of the fractal.
An overview on various fractal dimensions and their properties is given in
\cite{falconer.k.14.fractal.book}.

The shortest path dimension $\dmin$
reflects how the length $\minPath$ of the shortest path between any two points
in 
a mixed fractal 
scales with 
width
$\linl$ of the generating pattern,
i.e., $\minPath \sim \linl^{\dmin}$, and is given as 
\begin{align}
	\dmin &= \lim_{n\to\infty} \frac{ \log\left( \minPath(n) %\frac{\minPath(n)}{\minPath(n-1)} 
		\right) }{ \log\left( \linl(n) %\frac{\linl(n)}{\linl(n-1)} 
		\right) }  \,\, .
\label{eq:dmin}
\end{align}

The dynamical behavior of randomly moving particles on structures 
	is described by the random walk dimension $\dw$ 
	that satisfies the Einstein relation
\cite{bunde.a.96.fractals.book,franz.a.01.einstein.1411}
as
\begin{align}
	\label{eq:einstein}
	\dw = \df + \xi \,\,,
\end{align}
where $\xi$ is the resistance scaling exponent.
The resistance scaling exponent describes how resistance %$R$ 
of the underlying resistance network 
representing the structure  
scales with the linear length %$m$ 
of the structure.
For  finitely ramified \Sierpinski carpets 
 	\cite{franz.a.01.einstein.1411,%
 		franz.a.02.using.18}
 	and objects related to the paths in (mixed)
 	labyrinth fractals, e.g.~linear networks  
 	\cite{juhasz.r.08.superdiffusion.066106},
 	it was shown that $\xi = \dmin$.
 	Since mixed labyrinth fractals can then be represented by a corresponding resistor network,
	the shortest path dimension $\dmin$ can also be described the scaling of the shortest path scaling factor $\lambda$, i.e., the ratio $\lambda_n = \minPath(n)/\minPath(n-1)$, as
 	\begin{align}
 		\label{eq:dmin-lambda}
 		\dmin = \lim_{n\to \infty} \frac{\log \lambda_n}{ \log\linl(n)} \;\; .
 	\end{align}
 	This will be utilised subsequently to determine $\dmin$.
 	Details are given in section \ref{sec:StatAna}.

The box counting dimension of arcs between exits in a self-similar or mixed labyrinth fractal {is given}.
Let $\linf$ be a labyrinth fractal, and $a(\ep)$ an arc  that connects two exits (indicated by $\ep \in \P$ ) in $\linf$. 
The box-counting dimension of the arc $a(\ep)$ is 
\cite{cristea.l.17.mixed.112}
\begin{align}
 	\darc{}(a(\ep)) &= \lim_{\iter \to \infty} \frac{\log{\path{\iter}{}(\ep){}}}{\log{\linl(\iter)}},
\label{eq:darc-math}
\end{align}
where $\ell_{\iter}(\ep)$ 
is the length of the path $a(\iter;\ep)$
in $\G(\W_\iter)$ that connects the pair of exits  of $\W_\iter$ that correspond to the endpoints of $a$ in $\linf$. 
Note that the shortest path $\minPath(n)$ is $\min\{ \ell_{\iter}(\ep), \gamma \in \mathcal{P}\}$.

For non-blocked labyrinth fractals one can show that all arcs have box-counting dimension $1$.
Further 
results on partially or totally blocked labyrinth fractals
were given for
special cases 
\cite{cristea.l.09.curves.1,%
cristea.l.11.curves.329,%
cristea.l.17.mixed.112,%
cristea.l.18.on.575,%
cristea.l.20.on.1}.
It was proven 
\cite{cristea.l.09.curves.1,%
cristea.l.11.curves.329} 
that in any self-similar labyrinth fractal generated by a totally blocked pattern 
the dimension of any arc connecting two distinct points in the fractal (in particular, two exits) is 
strictly greater than one and is
\begin{align}
	\darc{}(a(\ep)) &= \frac{\log \specRadius}{\log \linl},
	\label{eq:darc-tb}
\end{align}
where $\specRadius$ is the spectral radius of the pattern's path matrix, and $\linl$ its width.

While the main results proven on arcs in mixed labyrinth fractals refer to their length 
\cite{cristea.l.18.on.575,%
cristea.l.20.supermixed.183}, 
only basic estimations of the upper and lower box counting dimension of arcs between exits in the fractal were given 
\cite{cristea.l.17.mixed.112}, 
and in recent research 
\cite{cristea.l.20.on.1} results on the box counting dimension of arcs were proven for families of mixed labyrinth fractals generated by special families of patterns with rotational (and path-length-) symmetry.

\subsection{\label{sec:RMLF}Randomised mixed labyrinth fractals}
%SUBSECTION%%SUBSECTION%%SUBSECTION%%SUBSECTION%%SUBSECTION%%SUBSECTION%

In this article we pass, in the quest for more insights on the properties and dimensions of arcs in mixed labyrinth fractals,  to a next 
generalisation step for these objects: we introduce new objects called \emph{randomised mixed labyrinth fractals}, which offer a new framework for the study of the mixed labyrinth fractals, in the following way.

Let $s\ge 2$,  $\{\Ac^{(1)}, \Ac^{(2)}, \dots, \Ac^{(s)}\}$ be a collection of (distinct) labyrinth patterns,
 possibly with different widths, and $0\le p_1,p_2, \dots,p_{s}\le 1$ 
with $\sum_{i=1}^{s} p_i=1$. 
We assume that the pattern $\Ac^{(i)}$ is chosen with the selection probability $p_i$, 
for $i=1,\dots,s$, to construct 
a 
\emph{randomised mixed labyrinth set of level}  $n\ge2$: 
at each iteration step $k\in\{ 1,\dots,n\}$ we select the generating pattern 
(as described in the definition of a mixed labyrinth set) 
from the collection $\Ac^{(1)}, \Ac^{(2)}, \dots, \Ac^{(s)}$ with the corresponding probability, 
as mentioned above. 
The limit set obtained from this construction is then called a 
\emph{randomised mixed labyrinth fractal}.

Throughout this paper we study randomised mixed labyrinth fractals generated 
by two (distinct) labyrinth patterns
 $\Ac$ and $\Ac'$ of the same width $m=m'$. 
Let $p\in [0,1]$ be the selection probability of $\Ac$ and $1-p$ that of $\Ac'$. 
We denote by $(\Ac:p\,,\,\Ac':1-p)$ a system of two labyrinth patterns and their selection probabilities.
A realisation $\repre$ of a randomised mixed labyrinth set of level $n$, for $n\geq1$, denoted by 
$\W^\repre_n(\Ac:p\,,\,\Ac':1-p)$,
is constructed as a mixed labyrinth set, 
with the special feature that in each step $k$, 
with $1\leq k\leq n$, 
the selected labyrinth pattern is chosen randomly. %and uniformly distributed from $(\Ac:p\,,\,\Ac':1-p)$.
The limit set $L_{\infty}(\Ac:p\,,\,\Ac':1-p)$ of the sequence
$\{\W^\repre_n(\Ac:p\,,\,\Ac':1-p)\}_{n\ge 1}$
is called the randomised mixed labyrinth fractal generated by the randomised labyrinth patterns system $(\Ac:p\,,\,\Ac':1-p)$.
The cases $s=2$ with $m\ne m'$ and $s\ge 3$ go beyond the scope of this paper and thus will be studied later.

As each set $\W^\repre_n(\Ac:p\,,\,\Ac':1-p)$ is associated with a unique sequence of labyrinth patterns, 
the corresponding (exact) path matrix $\pathMatrix{}(n)=\pathMatrix{}(n; \Ac:p\,,\, \Ac':1-p)$ 
can be determined via Eq.~(\ref{eq:pathmatrix_mixed}).
By basic combinatorial arguments, 
there are $2^n$ different possible realisations $\W^\repre_n(\Ac:p\,,\,\Ac':1-p)$.
To get an idea of the statistical behaviour of$\W^\repre_n(\Ac:p\,,\,\Ac':1-p)$
we average over an ensemble of $\repre_\text{max}$ different realisations of $\W^\repre_n(\Ac:p\,,\,\Ac':1-p)$.
However, this is numerically very time consuming.
So, our aim is also to provide an alternative approach.
For products of random matrices a variant of the 
	Gelfand's theorem states
	that for a sequence of i.i.d.~random matrices $X_i$ such that $E(||X_i||)<\infty$, where $||\cdot||$ is any submultiplicative norm (like the sup norm) almost surely 
	$\lim_{n\to \infty} \frac{1}{n} \log(||X_1 X_2 \dots X_n||) \to c \leq E(\log(s(X)))$, where $s$ is the spectral radius of $X$. This is a consequence of Kingman's subadditive ergodic theorem.
	This idea has been used amongst others in the context of \Sierpinski gaskets 
\cite{lau.k.12.martin.475,kessebohmer.m.20.sierpinski.113}  
	and random self-similar graphs 
\cite{troscheit.s.17.on.257}.
Here, we use this idea to introduce a 
\emph{combined path matrix} $\pathMatrixApprox{}$ 
which involves the corresponding selection probabilities of the two labyrinth patterns as
\begin{align}
	\pathMatrixApprox{}(\Ac:p\,,\,\Ac':1-p) = p\,\pathMatrix{} + (1-p)\,\pathMatrix{}' \,\, ,
\label{eq:MixPathMat}
\end{align}
where $\pathMatrix{}$ is the path matrix of 
the pattern
$\Ac$ and $\pathMatrix{}'$ that of $\Ac'$. 
In analogy to 
the arc dimension $\darc{}(a(\ep))$ (see Eq.~(\ref{eq:darc-tb})),
we determine the spectral radius $\tilde{r}$ of $\pathMatrixApprox{}$ and the width $m$
in order to analyse
the arc dimension of the combined path matrix, denoted as  $\darc{}(\widetilde{a(\ep))}$, where
\begin{align}
	\darc{}(\widetilde{a(\ep))} = \frac{\log \tilde{r}}{\log m}
	\; .
\label{eq:MixArcDim}
\end{align} 
Then we compare $\darc{}(\widetilde{a(\ep))}$ with the averaged arc dimension $\langle \darc{}(a(\ep)) \rangle$ %
obtained by statistical analysis of 
$\repre$ realisations 
of 
$\W^\repre_n(\Ac:p\,,\,\Ac':1-p)$,
with $1\le \repre \le  2^n$.
The details of the analysis are given in following sections.

% SECTION%%SECTION%%SECTION%%SECTION%%SECTION%%SECTION%%SECTION%%SECTION%%SECTION%
\section{\label{sec:StatAna}Statistical analysis}
% SECTION%%SECTION%%SECTION%%SECTION%%SECTION%%SECTION%%SECTION%%SECTION%%SECTION%

We investigate structural properties
of randomised mixed labyrinth sets
by a statistical analysis of an ensemble of 
different realisations of these sets.
We show
that the averaged ensemble behaviour of the properties under investigation
converge for large enough level $\iter$ of the labyrinth fractal set
to a constant value, 
that approximates the limit values for randomised mixed labyrinth fractal $\linf$.

First, 
	for each realisation $\W_\iter^\repre(\Ac:p\,, \Ac':1-p)$ we 
determine the path length between any pair of exit points.
Due to the tree property, there is exactly one path between a pair of exit points 
and the path lengths  $ \path{\iter}{\repre}(\ep)=\ell(a(\iter, \repre; \ep))$ of all paths 
$a(\iter,\repre;\ep)$ of type $\ep\in \P$
in $\G(\W_\iter^\repre)$
can be calculated via Eq.~(\ref{eq:pathmatrix_mixed}) 
as
\begin{align}
	\pathVec{\iter}{\repre} = \pathMatrix{}(n;\Ac:p\,,\,\Ac':1-p) \cdot (1,1,1,1,1,1)^T \,\,,
	\label{eq:pathLengths}
\end{align}
with $\pathVec{\iter}{\repre}=
(\path{n}{\repre}({\A}),\path{n}{\repre}({\B}),\path{\iter}{\repre}({\C}),\path{n}{\repre}({\D}),\path{n}{\repre}({\E}),\path{n}{\repre}({\F}))$.
Then, the corresponding path scaling factor $\minPathScale{\iter}{\repre}({\ep})$ of the path $a(\iter,\repre;\ep)$ is, 
in analogy to 
\cite{franz.a.02.using.18}, 
obtained as
\begin{align}
	\minPathScale{\iter}{\repre}({\ep}) 
	&= 
	\frac{\path{\iter}{\repre}({\ep})}{\path{\iter-1}{\repre}({\ep})} 
	\hspace{3ex}\text{ with }\hspace{3ex} 
	\path{0}{\repre}({\ep}):=1 \,\, .
\label{eq:PathScale}
\end{align}
Here, one utilises the possibility to represent a labyrinth fractal by a resistor network, as shown in 
	\cite{franz.a.01.einstein.1411, franz.a.02.using.18, juhasz.r.08.superdiffusion.066106},
	that gives (\ref{eq:dmin-lambda}).
In the subsequent section we show that
$\minPathScale{\iter}{\repre}({\ep})$ 
approaches a unique value $\minPathScale{}{\repre}$   
for all paths $a(\iter,\repre;\ep)$ of $\G\{\W_\iter^\repre\}$ 
for large enough $n$, representing the shortest path scaling factor.

The arithmetic mean $\minPathScaleMean$ 
of the ensemble-average $\overline{\minPathScale{k}{}}$
of the path scaling factor is  
\begin{align}
	\minPathScaleMean &= \frac{1}{\iter-k_0+1}\,\sum_{k=k_0}^{\iter}  \overline{\minPathScale{k}{}} \,\, 
	\hspace{3ex}\text{with}\hspace{3ex} 
	\overline{\minPathScale{k}{}} = \frac{1}{\repre_\text{max}}\sum_{\repre=1}^{\repre_\text{max}} \minPathScale{k}{\repre} \,\, ,
\label{eq:MixedLambdaMean}
\end{align}
where 
$k_0$ is the   
first level $k$ for which the the relative deviation $\minPathScaleErr{\iter}$ 
\begin{align}
	\minPathScaleErr{\iter} 
	&= 
	\frac{1}{\repre_\text{max}}\,\sum_{\repre=1}^{\repre_\text{max}} 
		\frac{\max \minPathScale{\iter}{\repre}({\ep})-\min \minPathScale{\iter}{\repre}({\ep})}{\min \minPathScale{\iter}{\repre}({\ep})}
\label{eq:ErrorLambdas}
\end{align}
for all  $\repre$ of $\W_\iter^\repre(\Ac:p\,,\,\Ac':1-p)$ can be neglected.
This implies that from level $k=k_0$ onward the unique scaling factor $\lambda$ is reached.

By the definition of the shortest path dimension (\ref{eq:dmin-lambda}) 
we can identify the scaling of the shortest path $\minPath$ with the minimal value of $\minPathScale{\iter}{\repre}({\ep})$
for the corresponding labyrinth set.
The \emph{approximative shortest path dimension} $\dmin^{\iter,\repre}$  
of the randomised mixed laby\-rinth set $\W_\iter^\repre(\Ac:p\,,\,\Ac':1-p)$ 
is 
\begin{align}
	\dmin^{\iter,\repre} 
	&= \frac{ \log \min \minPathScale{\iter}{\repre}(\ep) } {\log \linl} %\log{ \min \minPathScale{\iter}{\repre}(\ep) } } 
\label{eq:dminApprox0} 
\end{align}
Note that in the case  $\minPathScaleErr{\iter} <10^{-15}$, i.e., $n \geq k_0$, we have
\begin{align}
	\dmin^{\iter,\repre} 
	&= \frac{ \log \minPathScale{\iter}{\repre} } {\log \linl} \,\,.
\label{eq:dminApprox} 
\end{align}  
The corresponding arithmetic mean $\dminMean$ of the ensemble-average $\overline{\dmin^{\iter}}$ 
of the approximative shortest path dimension 
of $\W_n(\Ac:p\,,\,\Ac':1-p)$ is
\begin{align}
	\dminMean &= \frac{1}{\iter-k_0+1}\,\sum_{k=k_0}^\iter \overline{\dmin^{k}} \,\, 
	\hspace{3ex}\text{with}\hspace{3ex} 
	\overline{\dmin^{k}} = \frac{1}{\repre_\text{max}}\sum_{\repre=1}^{\repre_\text{max}} \dmin^{k, \repre} \,\,
\label{eq:dminMean} 
\end{align}
Additionally, we estimate the shortest path dimension error  $\dminErr^\iter$ via
\begin{align}
	\dminErr^\iter &= \frac{\max_\repre\{\minPathScaleErr{\iter,\repre}\}}{\log \linl} \,\, 
\label{eq:dmin-num}
\end{align}
and we determine 
the unbiased estimate of the variances $\minPathScaleVar$ and $\dminVar$, 
\begin{align}
	\minPathScaleVar &= \frac{1}{\iter-k_0}\,\sum_{k=k_0}^{\iter} (\overline{\minPathScale{k}{}} -  \minPathScaleMean)^2  \hspace{3ex}\text{ and }\\
	\dminVar &= \frac{1}{\iter-k_0}\,\sum_{k=k_0}^{\iter} (\overline{\dmin^{k}} - \dminMean)^2 \,\, ,
\label{eq:MixedDminVar} 
\end{align}
as well as
the minimum and maximum among all obtained values of $\minPathScale{\iter}{\repre}$ and $\dmin^{\iter,\repre}$ per level $\iter$.
In order to give an indication of the accuracy of the obtained ensemble-average of $\dminMean$ of the randomised mixed labyrinth sets,
we plot, next to $\dminMean$, 
the variance $\dminVar$ as error bars 
and the min-max-values as shaded area.

Furthermore, we introduce the approximative box-counting dimension 
$\darc{\iter,\repre}(\ep)=\darc{}(a(\iter,\repre; \ep))$ 
for every path $a(\iter, \repre;\ep)$ of type $\ep \in \P$ in $\G(\W_\iter^\repre)$
\begin{align}
	\darc{\iter,\repre}(\ep) &:=\frac{\log \path{\iter}{\repre}(\ep)}{\log \linl(\iter)} \,, ,
	\label{eqn:ApproxDArc}
\end{align}
as well as the ensemble-average of the approximative box-counting dimension $\darc{\iter}(\ep)=\frac{1}{\repre_\text{max}}\sum_{u=1}^{\repre_\text{max}}\darc{\iter,\repre}(\ep)$.
Here we investigate the relative error $\Delta \darc{\iter}({\ep})$ of $\darc{\iter}(\ep)$,
\begin{align}
	\Delta \darc{\iter}({\ep}) &= \frac{\darc{\iter-1}({\ep}) - \darc{\iter}({\ep})}{\darc{\iter-1}({\ep})}
\label{eq:darc-math-err}
\end{align}
over the level $\iter$ as a convergence criterion. 
In all cases we find that  
$\displaystyle \darc{}(\ep) \approx \darc{\iter}(\ep)$
for large enough $\iter$.

% SECTION%%SECTION%%SECTION%%SECTION%%SECTION%%SECTION%%SECTION%%SECTION%%SECTION%
\section{\label{sec:analysed_patterns}Analysed labyrinth patterns}
% SECTION%%SECTION%%SECTION%%SECTION%%SECTION%%SECTION%%SECTION%%SECTION%%SECTION%

We are interested in the  
general 
structural behaviour of randomised mixed labyrinth sets.
Therefore, we analyse a broad variety of labyrinth patterns of width 
$\linl \in\{4,5,6,7\}$ 
to investigate the effect of
structural features of a single pattern on the randomised
mixed labyrinth sets
in comparison to the corresponding self-similar labyrinth sets. 
For all patterns of width $\linl$ we choose the same number of white squares $\mass$ 
in order to avoid additional effects due to the mass scaling while mixing different labyrinth patterns.

\begin{figure}[h!]
	\begin{tabular}{ccccc}
		\hline \hline
		\gvb{1}{} & \gfb{1}{} & \gfc{1}{} & \gseb{1}{} & \gsib{1}{} \\ \hline 
		\includegraphics[width=.17\textwidth]{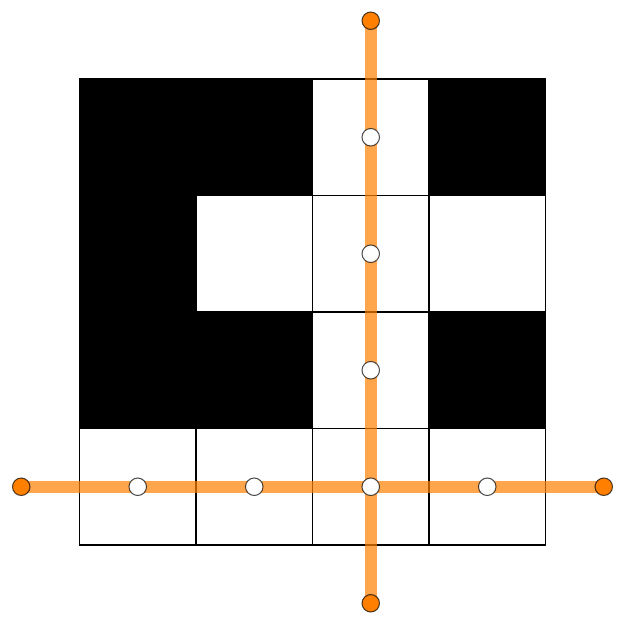} & 
		\includegraphics[width=.17\textwidth]{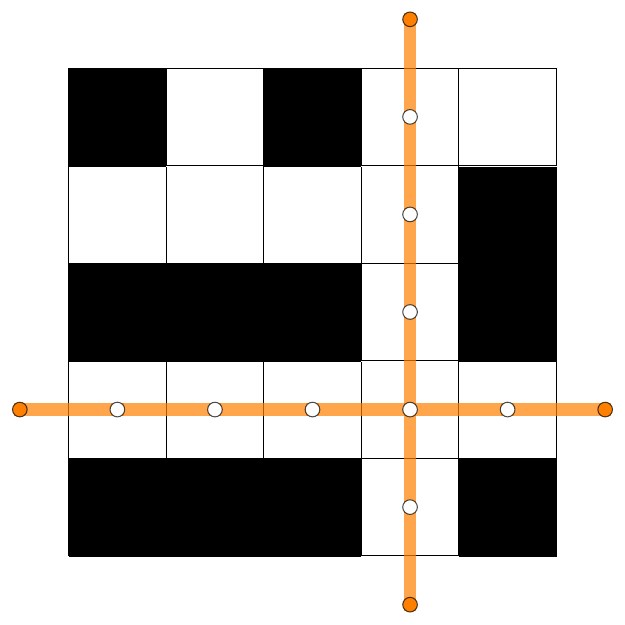} &
		\includegraphics[width=.17\textwidth]{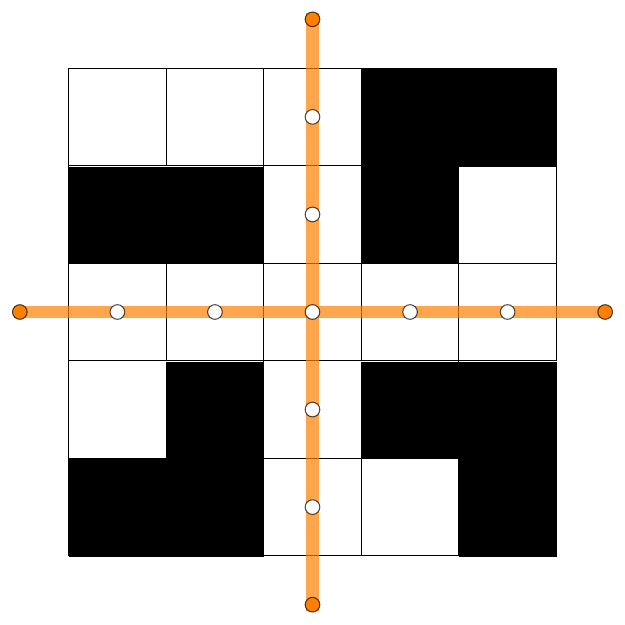} &
		\includegraphics[width=.17\textwidth]{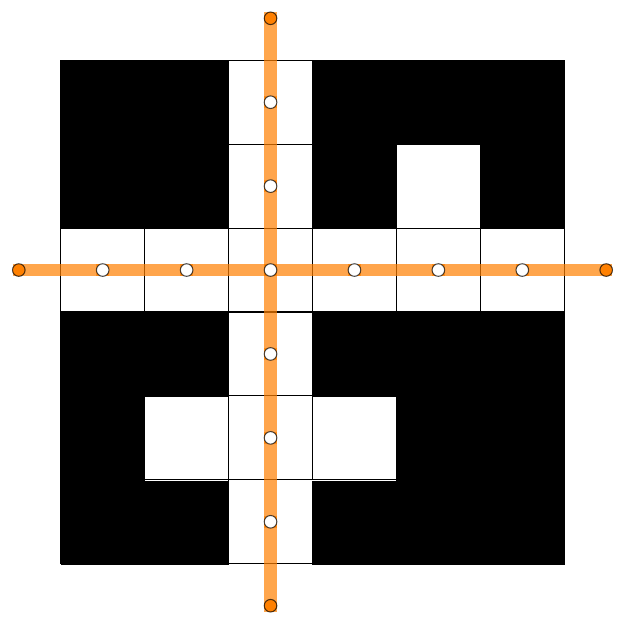} &
		\includegraphics[width=.17\textwidth]{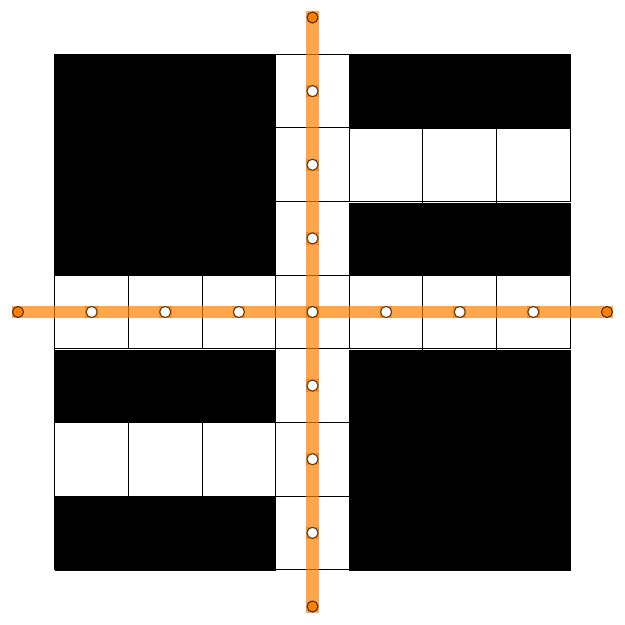}  \\ \hline \hline 
	\end{tabular}
	\caption{\label{fig:nonBlockedPattern}%
		The analysed non-blocked labyrinth patterns.
		The orange lines indicate the shortest paths of type  $\A,\B,\C,\D,\E$, and $\F$, respectively.
	}
\end{figure}

\begin{figure}[h!]
	\begin{tabular}{ccccc}
		\hline \hline
		\gva{1}{} & \gvc{1}{} & \gfa{1}{} & \gsea{1}{} \\ \hline 
		\includegraphics[width=.17\textwidth]{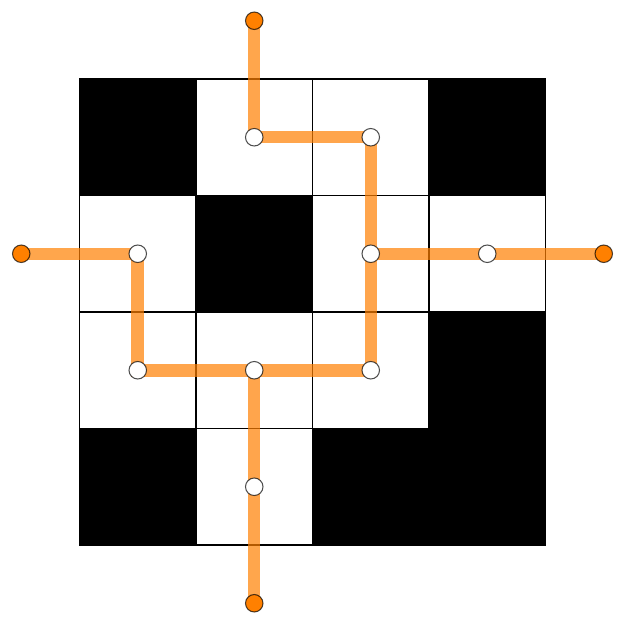} & 
		\includegraphics[width=.17\textwidth]{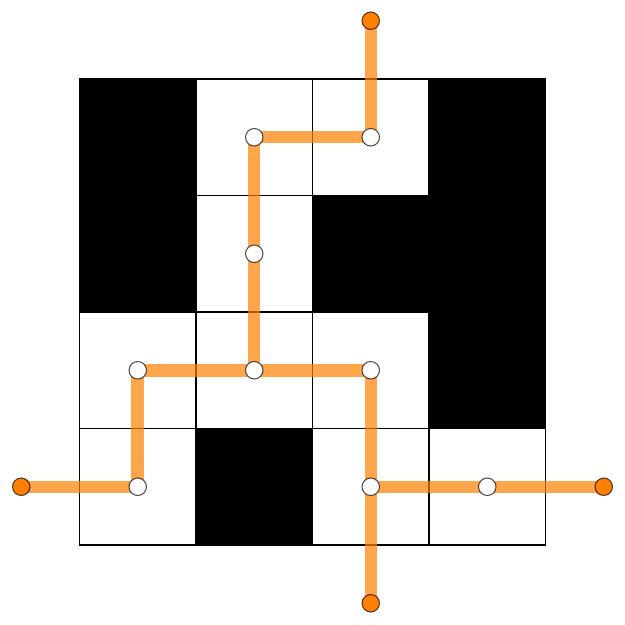} &
		\includegraphics[width=.17\textwidth]{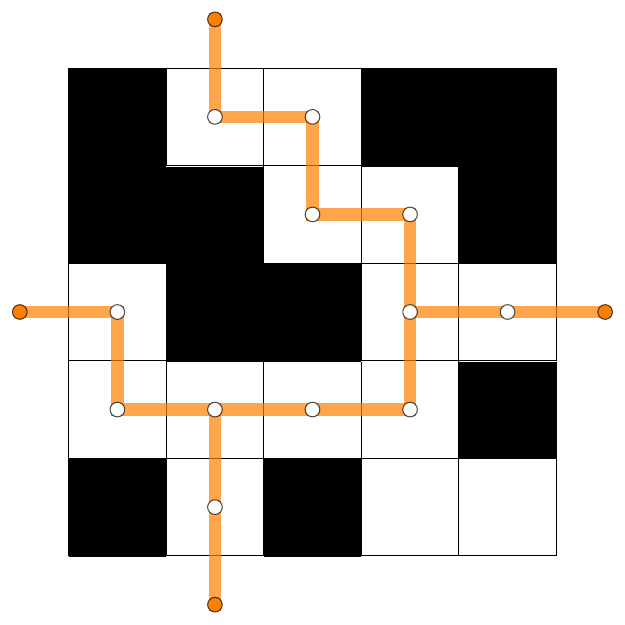} &
		\includegraphics[width=.17\textwidth]{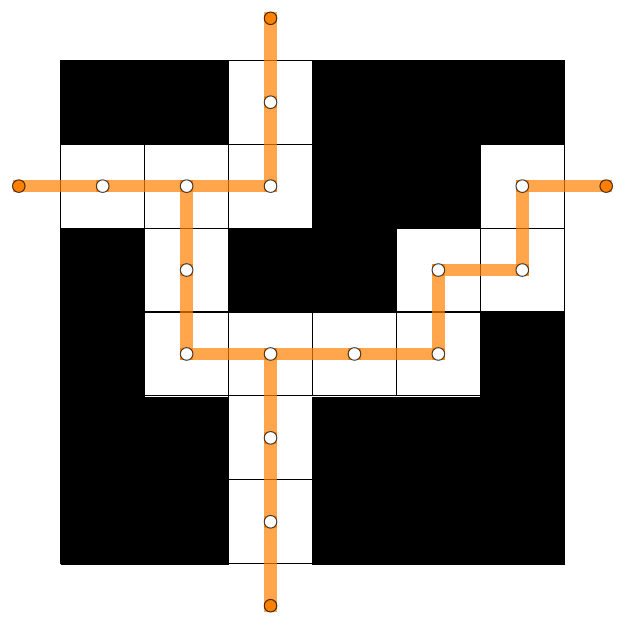} &
		 \\ \hline
		\gsia{1}{} & \gsic{1}{} & \gsid{1}{} & \gsie{1}{} & \gsif{1}{}\\ \hline 
		\includegraphics[width=.17\textwidth]{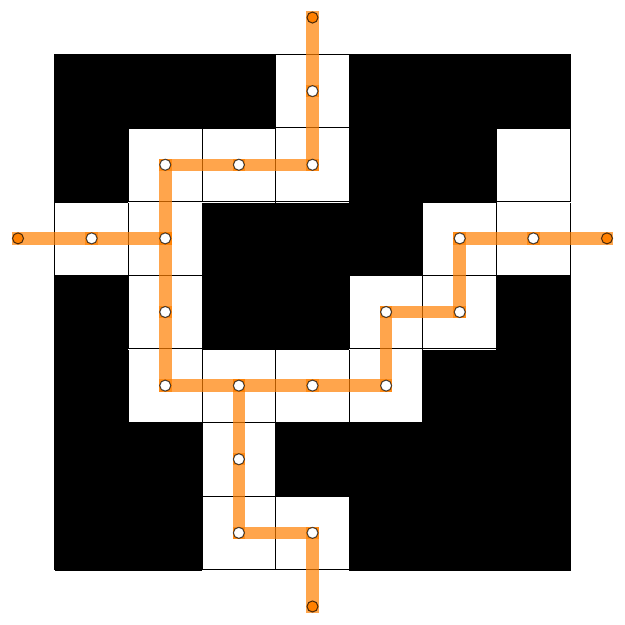} &
		\includegraphics[width=.17\textwidth]{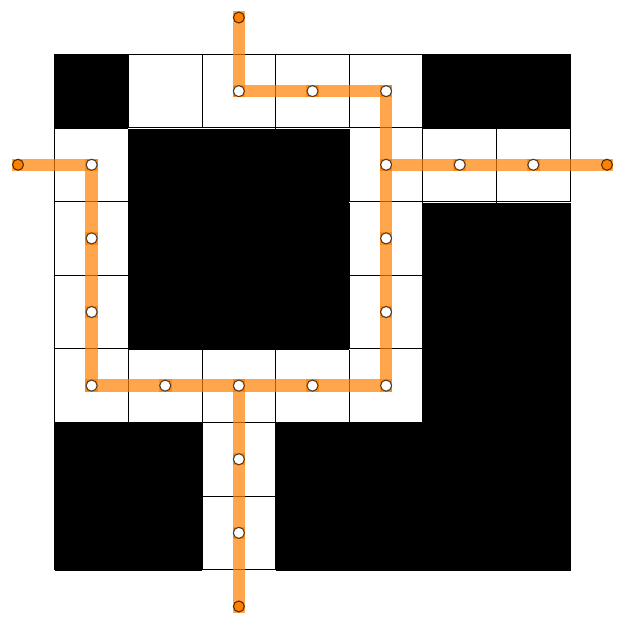} & 
		\includegraphics[width=.17\textwidth]{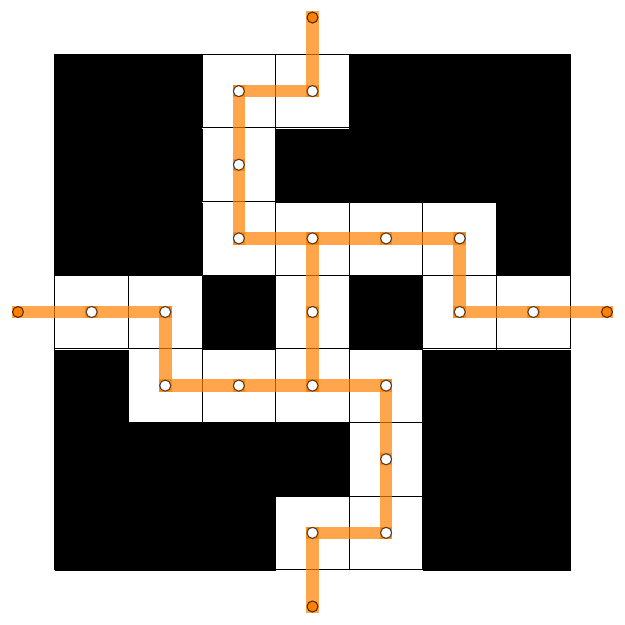} &
		\includegraphics[width=.17\textwidth]{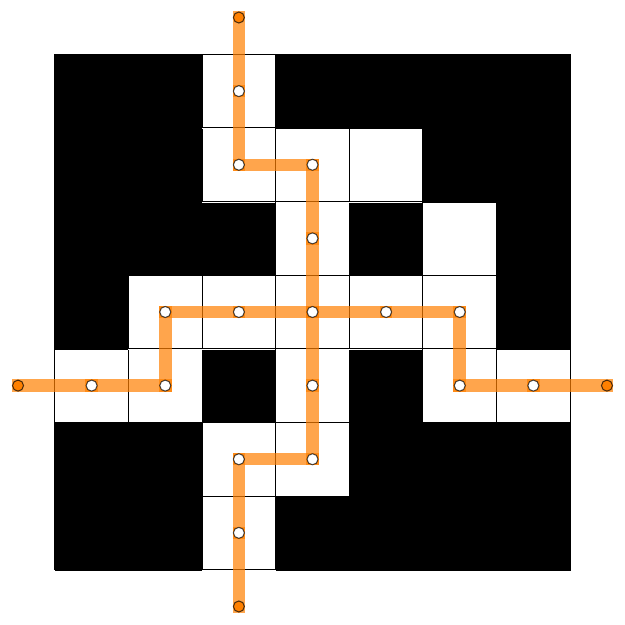} &
		\includegraphics[width=.17\textwidth]{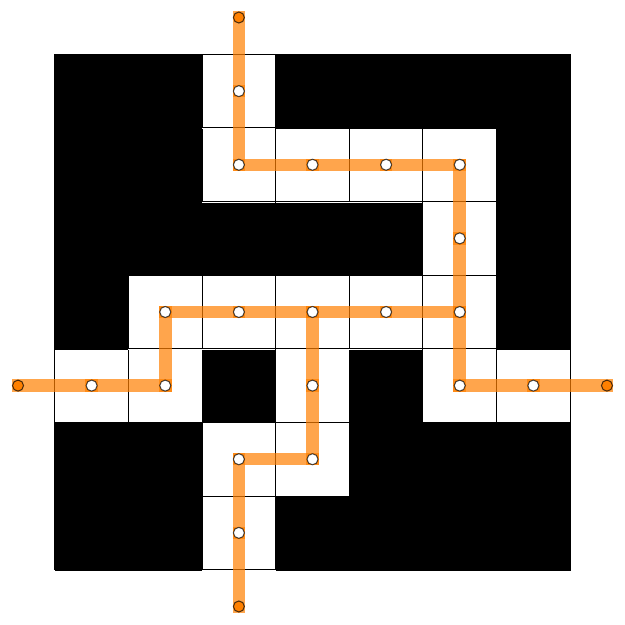} 
		\\ \hline \hline
	\end{tabular}
	\caption{\label{fig:fullBlockedPattern}%
		The analysed  totally blocked labyrinth patterns. 
		The orange lines indicate the shortest paths 
		of type $\A,\B,\C,\D,\E$, and $\F$, respectively.
	}
\end{figure}

\begin{figure}[h!]
	\begin{tabular}{cccc}
		\hline \hline
		\gva{1}{} & \gva{2}{} & \gva{3}{} & \gva{4}{} \\ \hline 
		\includegraphics[width=.17\textwidth]{pattern_g4a1.pdf} & 
		\includegraphics[width=.17\textwidth]{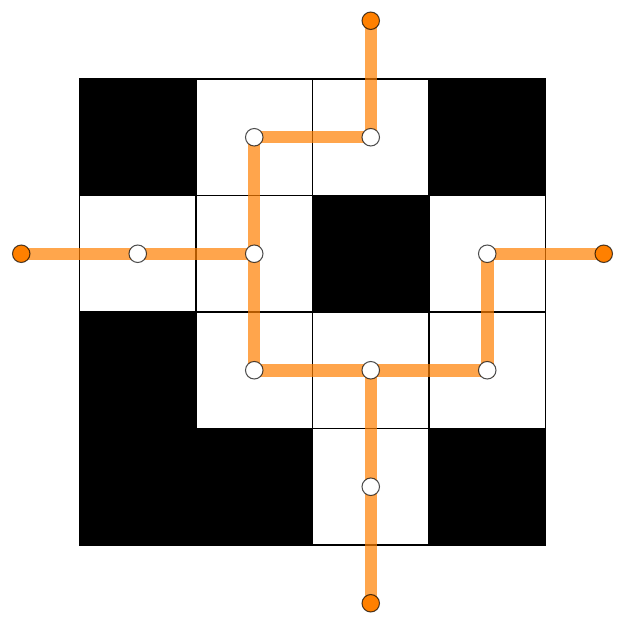} &
		\includegraphics[width=.17\textwidth]{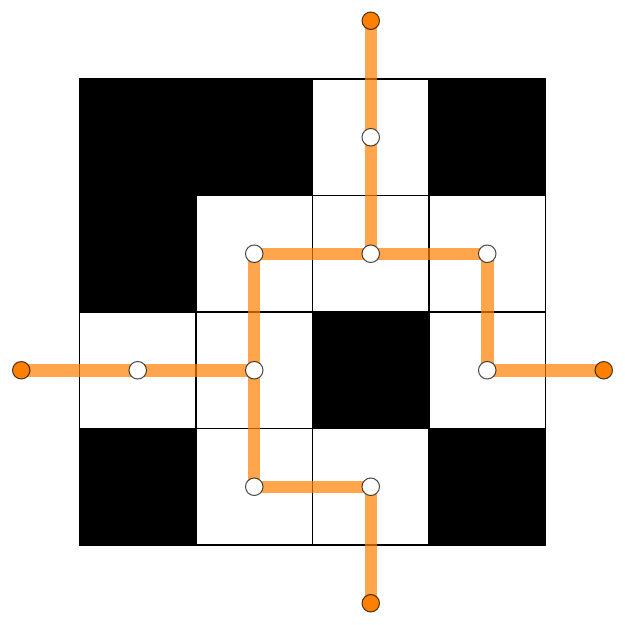} &
		\includegraphics[width=.17\textwidth]{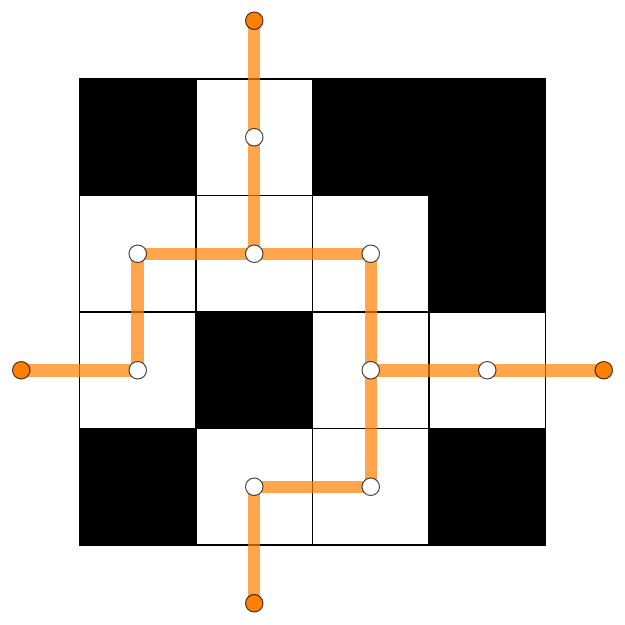} \\ \hline
		$\circlearrowright 0^{\circ}$ & $\circlearrowright 90^{\circ}$ & $\circlearrowright 180^{\circ}$ & $\circlearrowright 270^{\circ}$ 
		\\ \hline \hline
	\end{tabular}
	\caption{\label{fig:rotatedPattern}%
		Example (\gva) for notation of labyrinth patterns obtained by rotation.
		The orange lines indicate the shortest paths 
		of type $\A,\B,\C,\D,\E,$ and $\F$, respectively. 
	}
\end{figure}

There are three structural aspects of the patterns that are taken into account:
the symmetry of a pattern (mirror symmetric, point symmetric, and asymmetric patterns), 
the length of the individual paths of type $\gamma$ of the pattern 
(the paths have all equal, partially equal or different lengths),
and the property of being blocked.
The totally blocked labyrinth patterns are given in Fig.~\ref{fig:fullBlockedPattern}.
The non-blocked ones are shown in Fig.~\ref{fig:nonBlockedPattern}.
For a detailed understanding of the influence of symmetry on the final structural composition,
we also mix patterns that are rotated versions of each other.
This is shown exemplary in Fig.~\ref{fig:rotatedPattern}.
There are four different possible rotations of the given pattern \gva{}, 
clockwise rotated by $i\cdot90^\circ$, 
where $i=0,\dots,3$ is the rotation index.

\begin{table}[h]
\begin{tabular}{cp{2.6cm}llp{1.9cm}lp{1.9cm}}
	\hline \hline
	pattern & $\df$ & $\dmin$ %& $\dchem$ 
		& $\dw$	%\newline=\df+\dmin$ 
		& $\lambda$ & $r$ \\ \hline
	\gva{}{} & $\frac{\ln{9}}{\ln{4}}\approx1.58496$ & 1.16096 %& 1.36521 
		& 2.74593  
		& 5.0 & 5.0  \\
	\gvb{}{} & $\frac{\ln{9}}{\ln{4}}$ & 1.0 %& 1.58496 
	& 2.58496 
		& 4.0 & 4.0 \\
	\gvc{}{} & $\frac{\ln{9}}{\ln{4}}$ & 1.16096 %& 1.36521 
	& 2.74593
		& 5.0 & 5.0  \\ \hline
	\gfa{}{} & $\frac{\ln{14}}{\ln{5}}\approx1.63974$ & 1.13726 %& 1.44183 
	& 2.7770 
		& 6.2361 & $4+\sqrt{5}\newline=6.2361$  \\
	\gfb{}{} & $\frac{\ln{14}}{\ln{5}}$ & 1.0 %& 1.63974 
	& 2.63974 
		& 5.0 & 5.0  \\
	\gfc{}{} & $\frac{\ln{14}}{\ln{5}}$ & 1.0 %& 1.63974 
	& 2.63974 
		& 5.0 & 5.0  \\ \hline
	\gsea{}{} & $\frac{\ln{14}}{\ln{6}}\approx1.47289$ & 1.14347 %& 1.28808 
	& 2.61636 
		& 7.7588 & 7.75877 \\
	\gseb{}{} & $\frac{\ln{14}}{\ln{6}}$ & 1.0 %& 1.47289 
	& 2.47289 
		& 6.0 & 6.0 \\ \hline
	\gsia{}{} & $\frac{\ln{19}}{\ln{7}}\approx1.51314$ & 1.18235 %& 1.27977 
	& 2.6955 
		& 9.98173 & 9.98173\\% $6+\sqrt{13}=9.6056$ \\
	\gsib{}{} & $\frac{\ln{19}}{\ln{7}}$ & 1.0 %& 1.51314 
	& 2.51314 
		& 7.0 & 7.0  \\
	\gsic{}{} & $\frac{\ln{19}}{\ln{7}}$ & 1.22759 %& 1.23261 
	& 2.74074 
		& 10.9003 & 10.9003  \\
	\gsid{}{} & $\frac{\ln{19}}{\ln{7}}$ & 1.19806 %& 1.26299 
	& 2.7112 
		& 10.2915 & $5+2\sqrt{7}\newline=10.2915$  \\
	\gsie{}{} & $\frac{\ln{19}}{\ln{7}}$ & 1.12915 %& 1.34007 
	& 2.64229 
		& 9.0 & 9.0  \\
	\gsif{}{} & $\frac{\ln{19}}{\ln{7}}$ & 1.19132 %& 1.27014 
	& 2.70446 
		& 10.1574 & 10.1574 \\ 	\hline \hline 
\end{tabular}
	\caption{\label{tab:LabPattern}%
		The shortest path scaling factor $\minPathScale{}{}$, spectral radius $\specRadius$ and
		fractal-, shortest path-, and random walk dimensions $\df, \dmin, \dw$ 
		for the self-similar sets of the analysed labyrinth patterns.
	}
\end{table}

We investigate the self-similar labyrinth sets of level $\iter =1000$ for each pattern 
as reference cases. 
The corresponding values for $\df$ (see Eq.~(\ref{eq:df})), 
$\dmin$ (see Eq.~(\ref{eq:dmin})), 
$r$ (spectral radius of $M$), 
and the shortest path scaling factor $\lambda$ (see Eq.~(\ref{eq:PathScale}))
are given in Tab.~\ref{tab:LabPattern}.
Note that in the self-similar case there is only one possible realisation, 
i.e.,~$\repre_\text{max}=1$,
and that all rotated versions of a pattern have equal values.

For the analysis of the randomised mixed labyrinth sets
we generate $\repre_\text{max}=2000$ realisations of $\W_n(\Ac:p\,,\,\Ac':1-p)$ with $\iter=1000$.
If not mentioned otherwise, we vary $p \in [0.0,1.0]$ with step size $\Delta p = \{0.1,0.05\}$. 
$\lambda^\repre$ is considered to be converging once Eq.~(\ref{eq:ErrorLambdas}) holds for at least 5 subsequent levels $k$ for all realisations $\repre$ at the same time. 
We denote the fifth level by $k_0$.
Furthermore, we determine 
the spectral radius $\tilde{r}$ of the combined matrix $\widetilde{M}(\Ac :p\,, \Ac':1-p)$ 
given in Eq.~(\ref{eq:MixPathMat}) and analyse it with respect to the above mentioned values.

%SECTION%%SECTION%%SECTION%%SECTION%%SECTION%%SECTION%%SECTION%%SECTION%%SECTION%
\section{\label{sec:ResultsIsotropy}Restoration of isotropy}
%SECTION%%SECTION%%SECTION%%SECTION%%SECTION%%SECTION%%SECTION%%SECTION%%SECTION%

For \Sierpinski gaskets
\cite{barlow.m.95.restoration.3042}
as well as 
for a broad class of self-similar \Sierpinski carpets
\cite{barlow.m.95.restoration.3042%
,barlow.m.97.weak.1%
,franz.a.02.using.18}
it is known
that the path scaling factors $\minPathScale{\iter}{}(\ep)$ of 
all paths 
$a(\iter; \ep)$ of type $\ep\in\P$ in the corresponding graph
converge  
to a unique shortest path scaling factor 
$\minPathScale{\iter}{}(\ep)\to\minPathScale{}{}$
for $\iter\to \infty$.
Even in cases where 
the initial path lengths $\path{1}{}(\ep)$ 
of the
labyrinth pattern $\Ac$ 
are different, 
the scaling of all paths shows the same behaviour for large $\iter$.
In terms of physics this means that these structures are isotropic on large length scales 
although they are locally anisotropic.

This
effect 
of restoration of isotropy has also been observed for all self-similar and randomised mixed labyrinth sets 
that are 
investigated here,
independently of
the patterns' symmetry, 
the initial path lengths $\path{1}{}(\ep)$,
the blocking type,  
or the width $\linl$.

%FIGURE%%FIGURE%%FIGURE%%FIGURE%%FIGURE%%FIGURE%%FIGURE%%FIGURE%%FIGURE%
%FIGURE%%FIGURE%%FIGURE%%FIGURE%%FIGURE%%FIGURE%%FIGURE%%FIGURE%%FIGURE%
\begin{figure}[h!]
	\begin{tabular}{ll}
	a) & b) \\
	\includegraphics[width = 0.45\textwidth]{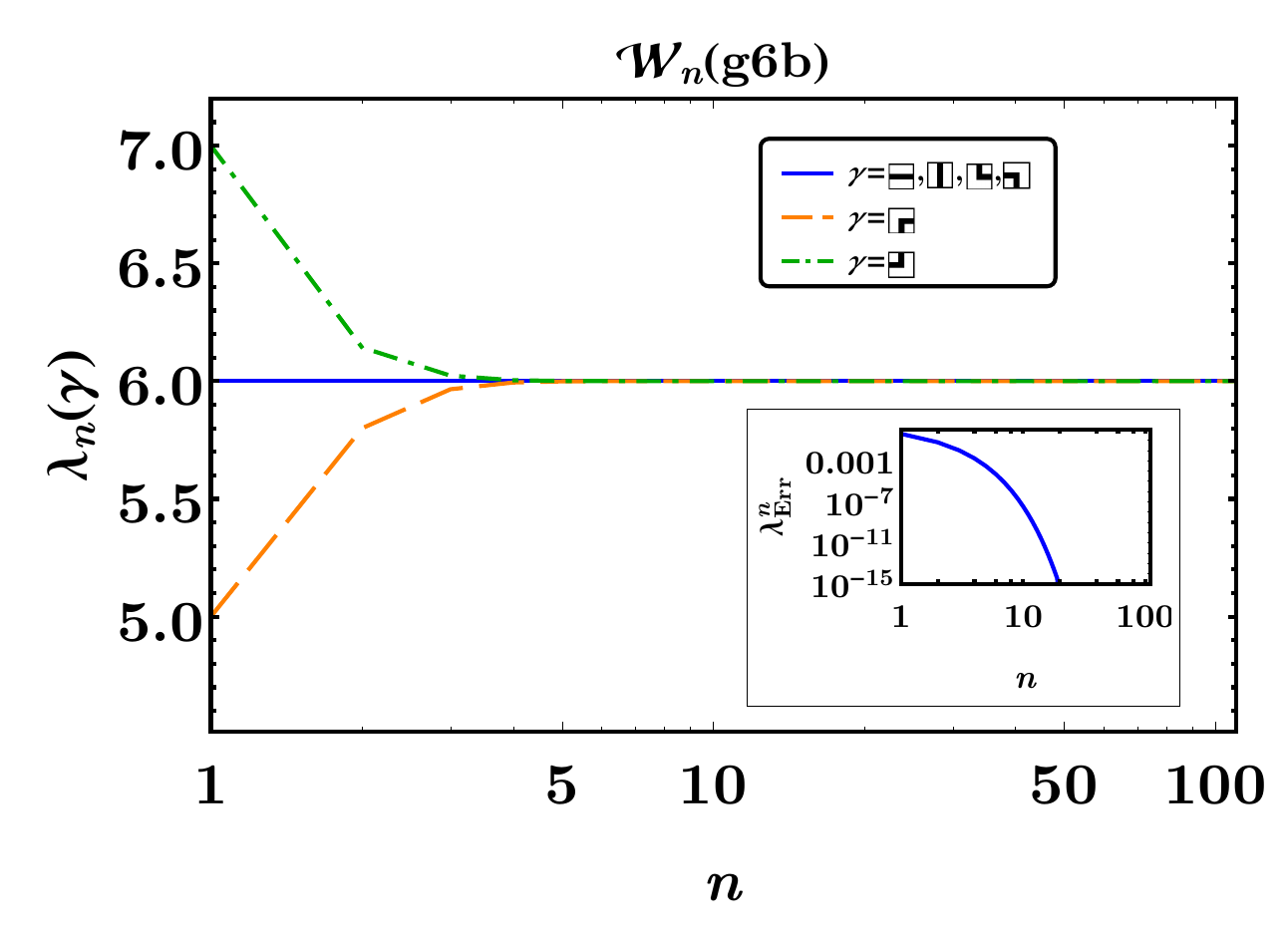} &
	\includegraphics[width = 0.45\textwidth]{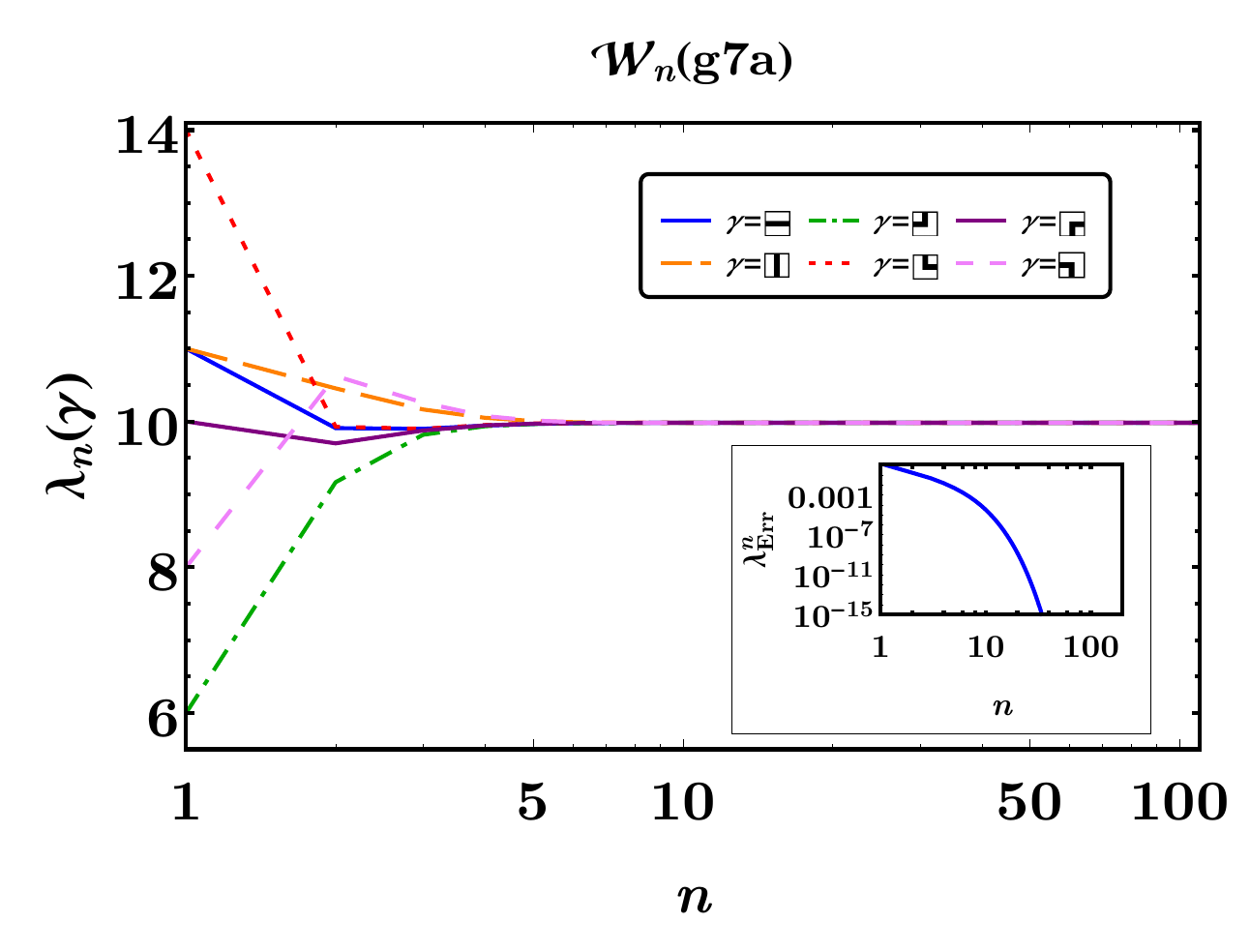} \\
	\end{tabular}
	\caption{\label{fig:lambdaConv1}
		The path scaling factor $\minPathScale{\iter}{}({\ep})$ is shown 
		for the sets a) $\W_\iter(\gseb{})$ and b) $\W_\iter(\gsia{})$
		for each $\ep$-path, for $\ep \in \P$. 
		In the insets the relative deviation $\minPathScaleErr{\iter}$ is plotted over $\iter$ in a log-log-plot. 
	}
\end{figure}
%FIGURE%%FIGURE%%FIGURE%%FIGURE%%FIGURE%%FIGURE%%FIGURE%%FIGURE%%FIGURE%%FIGURE%%FIGURE%%FIGURE%%FIGURE%%FIGURE%%FIGURE%%FIGURE%%FIGURE%%FIGURE%%

The convergence behaviour of the path scaling factor $\minPathScale{\iter}{}(\ep)$ 
for two self-similar labyrinth sets 
is shown in Fig.~\ref{fig:lambdaConv1}
for a) $\W_\iter(\gseb{1})$ (non-blocked) and 
b) $\W_\iter(\gsia{1})$ (totally blocked).
One finds a fast convergence of $\minPathScale{\iter}{}(\ep)$ to the unique shortest path scaling factor 
$\minPathScale{}{}=6.0$ and $\minPathScale{}{}=9.6056$ (cf.~Tab.~\ref{tab:LabPattern}), respectively.
The quick convergence becomes also apparent 
by the fast decrease of the relative deviation error 
$\minPathScaleErr{\iter}$ 
as shown in the insets of the diagrams for both cases.
Although the exact value of the iteration depth necessary for convergence, $k_0$, 
depends on the underlying labyrinth pattern
$\Ac$.
In the self-similar case $k_0$ is typically below 50.
Note, that for a completely symmetric pattern, where all exit pair path lengths $\path{1}{}(\ep)$ are identical,
$k_0$ equals 1.

%FIGURE%%FIGURE%%FIGURE%%FIGURE%%FIGURE%%FIGURE%%FIGURE%%FIGURE%%FIGURE%
%FIGURE%%FIGURE%%FIGURE%%FIGURE%%FIGURE%%FIGURE%%FIGURE%%FIGURE%%FIGURE%
\begin{figure}[h!]
	\begin{tabular}{ll}
	a) & b) \\
	\includegraphics[width = 0.45\textwidth]{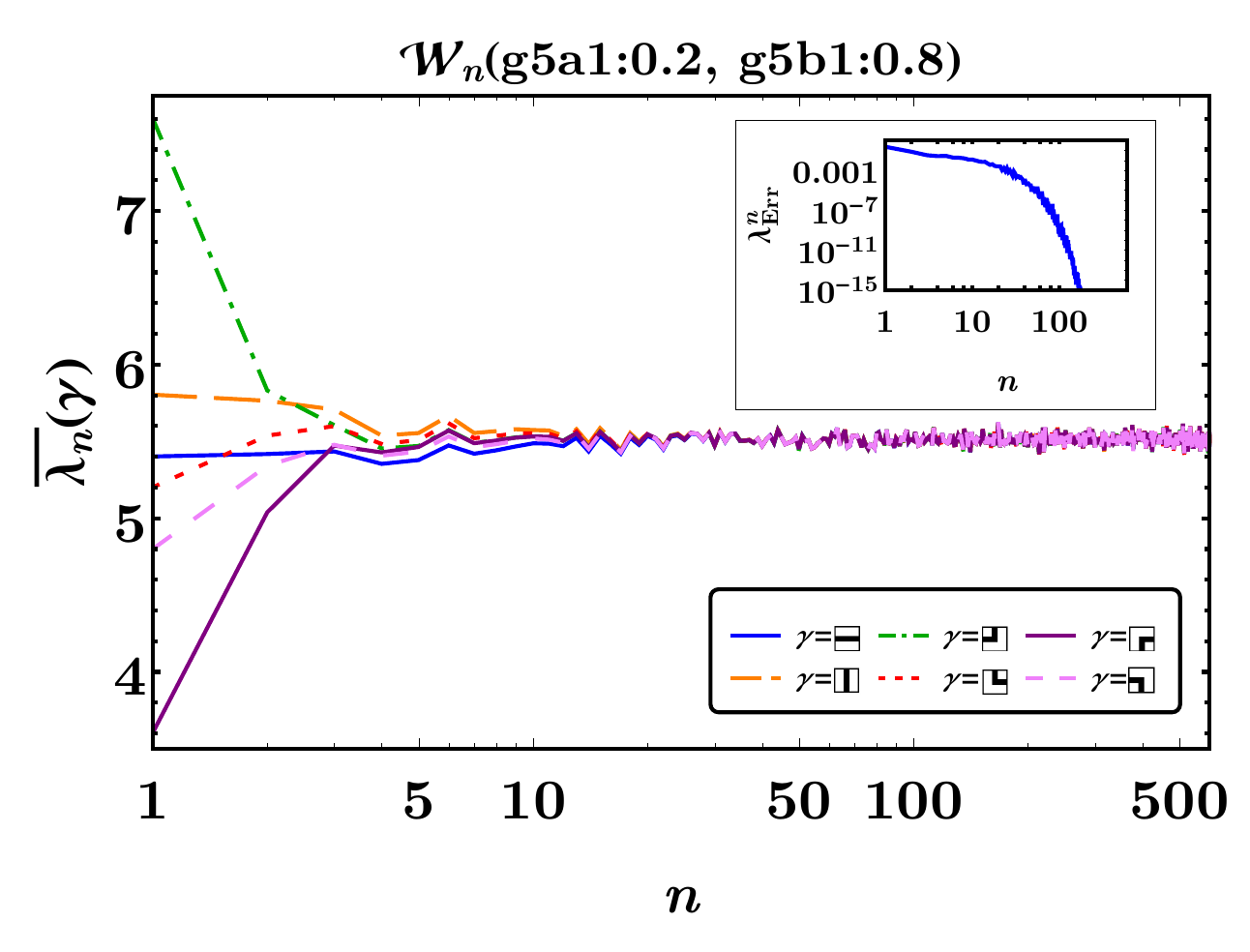} &
	\includegraphics[width = 0.45\textwidth]{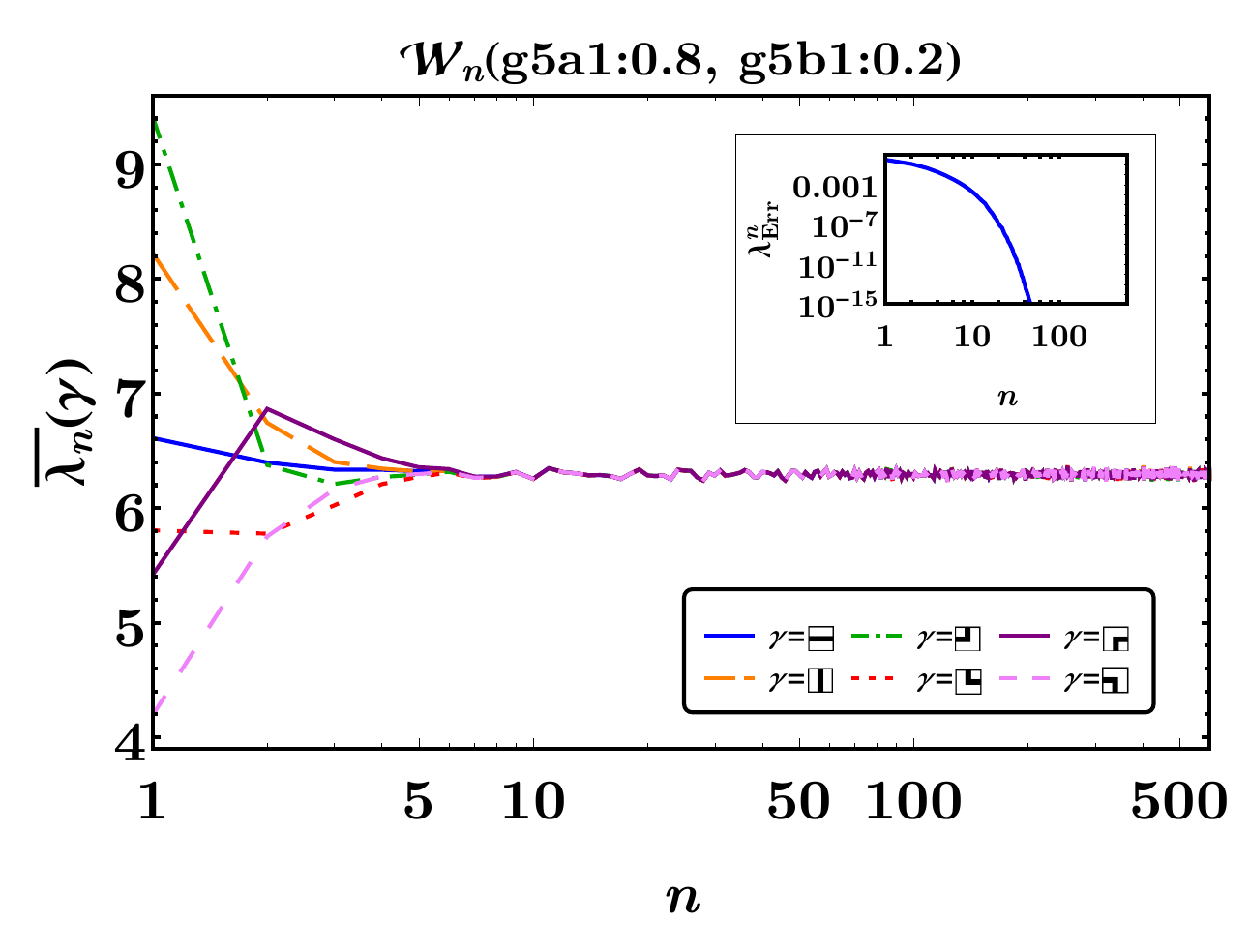} \\
	\end{tabular}
	\caption{\label{fig:lambdaConv2}
		The ensemble-averaged path scaling factor $\overline{\minPathScale{\iter}{}}({\ep})$ is shown 
		for the randomised mixed sets a) $\W_\iter(\gfa{1}:{0.2}\,,\,\gfb{1}:{0.8})$\ and 
		b) $\W_\iter(\gfa{1}:{0.8}\,,\,\gfb{1}:{0.2})$ for every $\ep$-path, for $\ep \in \P$. 
		In the insets the relative deviation $\minPathScaleErr{\iter}$ is plotted over $\iter$ in a log-log-plot. 
	}
\end{figure}
%FIGURE%%FIGURE%%FIGURE%%FIGURE%%FIGURE%%FIGURE%%FIGURE%%FIGURE%%FIGURE%
%FIGURE%%FIGURE%%FIGURE%%FIGURE%%FIGURE%%FIGURE%%FIGURE%%FIGURE%%FIGURE%

For randomised mixed labyrinth sets, 
the ensemble-averaged path scaling factors $\overline{\minPathScale{\iter}{}}(\ep)$ 
are shown
in Fig.~\ref{fig:lambdaConv2} for $\W_\iter(\gfa{1}:{p}\,,\,\gfb{1}:{1-p})$ 
with a) $p=0.2$ 
and b) $p=0.8$.
The convergence of $\overline{\minPathScale{\iter}{}}(\ep)$ takes longer,
as the decrease of the corresponding relative deviation error $\minPathScaleErr{\iter}$ is slower 
than in the self-similar case.
However, $\minPathScaleErr{\iter}<10^{-15}$ is typically reached for $k_0\approx 200$.
For $p=0.2$ one finds $\overline{\minPathScale{\iter}{}}\approx5.5$ and 
for $p=0.8$, $\overline{\minPathScale{\iter}{}}\approx6.3$.
We point out that
$\overline{\minPathScale{\iter}{}}$ fluctuates around $\minPathScaleMean$ due to the 
random generation of $\W_n(\Ac:p\,,\,\Ac':1-p)$ over $\iter$,
i.e.,~the scaling factor changes depending on the selected labyrinth patterns
for
each level.

%FIGURE%%FIGURE%%FIGURE%%FIGURE%%FIGURE%%FIGURE%%FIGURE%%FIGURE%%FIGURE%%FIGURE%%FIGURE%%FIGURE%%FIGURE%%FIGURE%%FIGURE%%FIGURE%%FIGURE%%FIGURE%
\begin{figure}[h!]
	\begin{tabular}{ll}
	a) & b) \\
	\includegraphics[width = 0.45\textwidth]{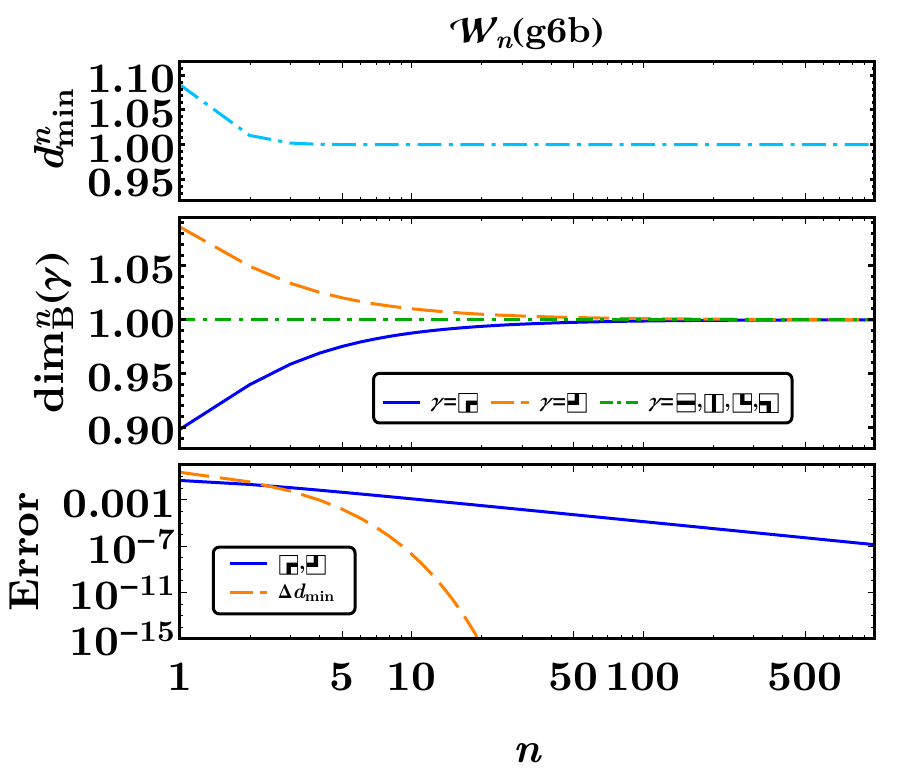} &
	\includegraphics[width = 0.45\textwidth]{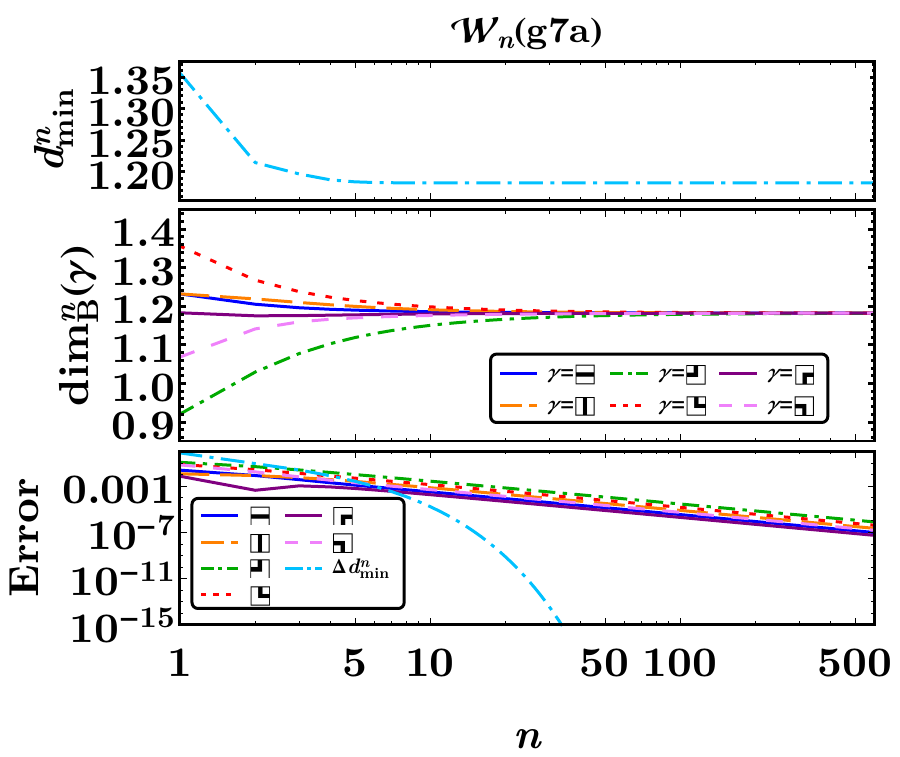} \\
	\end{tabular}
	\caption{\label{fig:dminConv1}%{fig:IsotropMixed}%
		The approximative shortest path dimension $\dmin^\iter$ and 
		box-counting dimension $\darc{\iter}({\ep})$ of the paths $\ep \in \{\A, \B, \C, \D,\E,\F\}$
		as well as the corresponding errors 
		are given over iteration level $\iter$
		for the labyrinth sets a) $\W_\iter(\gseb{})$ and b)  $\W_\iter(\gsia{})$.
		The behaviour  of $\dmin^\iter$ and $\darc{\iter}{}({\ep})$ is shown in linear-log plot and the errors in log-log-plot.
		Note that the error of $\darc{\iter}(\ep)$ is 0 for $\ep=\{\A, \B, \C,\E\}$ for all $n$.
	}
\end{figure}
%FIGURE%%FIGURE%%FIGURE%%FIGURE%%FIGURE%%FIGURE%%FIGURE%%FIGURE%%FIGURE%
%FIGURE%%FIGURE%%FIGURE%%FIGURE%%FIGURE%%FIGURE%%FIGURE%%FIGURE%%FIGURE%

In Fig.~\ref{fig:dminConv1} the 
approximative shortest path dimension $\dmin^{\iter}$, 
approximative arc dimension $\darc{\iter}(\ep)$ 
and the corresponding errors $\dminErr^{\iter}$ and $\Delta\darc{\iter}(\ep)$
are presented for the self-similar labyrinth sets a) $\W_\iter(\gseb{1})$ and b) $\W_\iter(\gsia{1})$.
One observes a fast convergence of $\dmin^{\iter}$ and 
$\darc{\iter}(\ep)$
 to their final values.
Due to the direct relation of $\minPathScale{\iter}{}$ and $\dmin^{\iter}$
given by
Eq.~(\ref{eq:dminApprox}) and Eq.~(\ref{eq:dminMean}),
the convergence occurs at the same iteration level $n=k_0$.

%FIGURE%%FIGURE%%FIGURE%%FIGURE%%FIGURE%%FIGURE%%FIGURE%%FIGURE%%FIGURE%
%FIGURE%%FIGURE%%FIGURE%%FIGURE%%FIGURE%%FIGURE%%FIGURE%%FIGURE%%FIGURE%
\begin{figure}[h!]
	\begin{tabular}{ll}
	a) & b) \\
	\includegraphics[width = 0.45\textwidth]{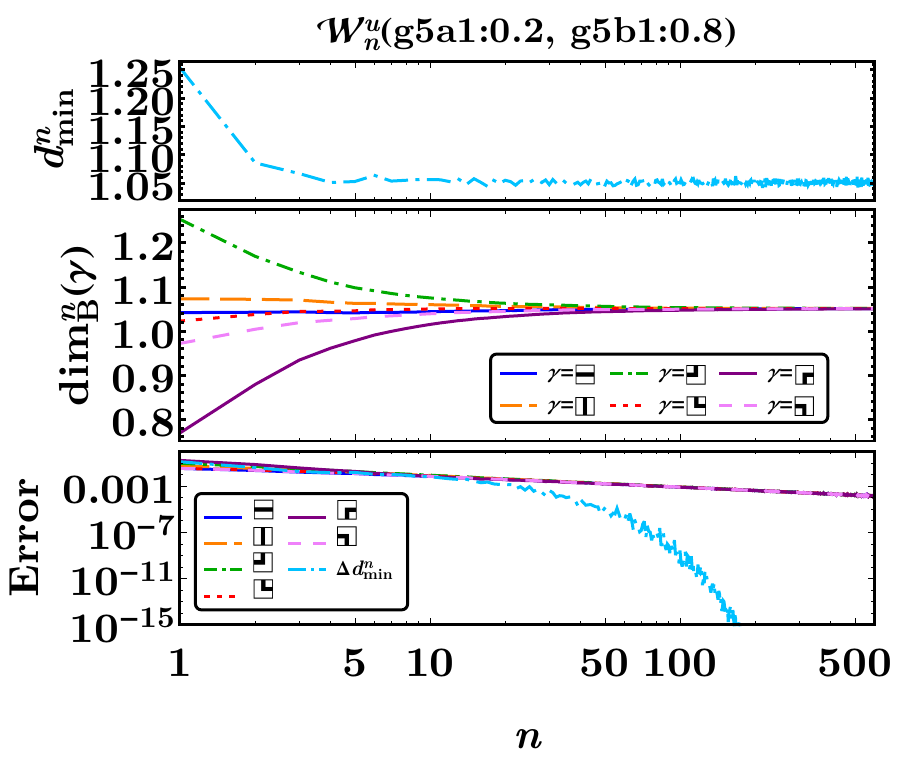} &
	\includegraphics[width = 0.45\textwidth]{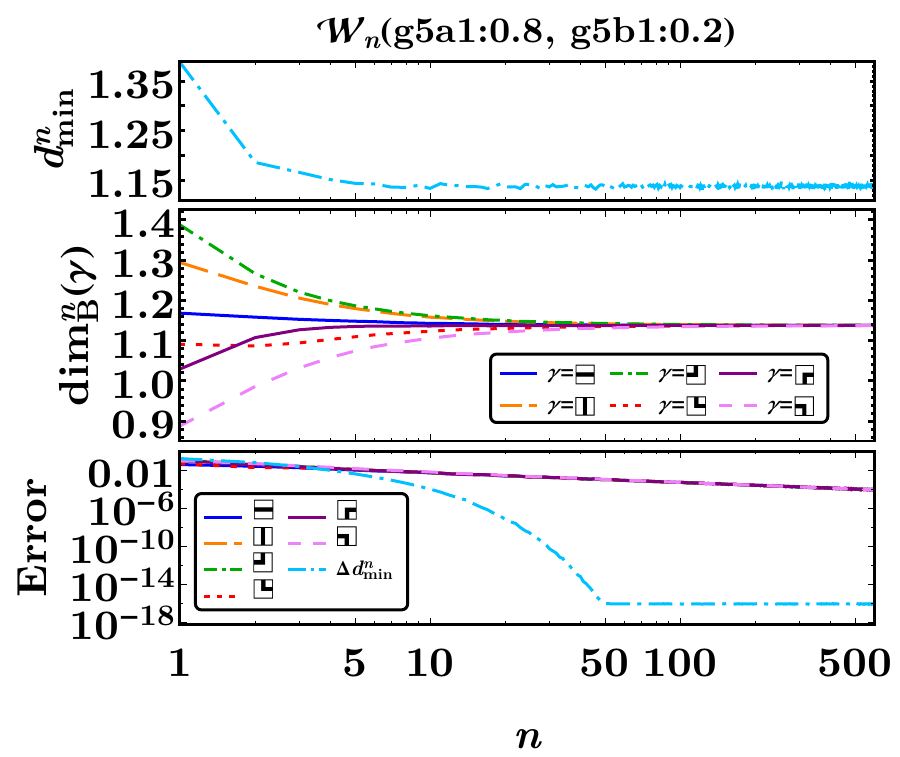} \\
	\end{tabular}
	\caption{\label{fig:dminConv2}%{fig:IsotropMixed}%
		The approximative shortest path dimension $\dmin^\iter$, 
		the arc dimension $\darc{\iter}({\ep})$ of the paths $\ep \in \{\A, \B, \C, \D,\E,\F\}$,
		and the corresponding errors 
		are given over iteration depth $\iter$
		for the randomised mixed labyrinth sets 
		a) $\W_\iter(\gfa{1}:{0.2};\gfb{1}:{0.8})$ and b)  $\W_\iter(\gfa{1}:{0.8};\gfb{1}:{0.2})$.
		$\dmin^\iter$ and $\darc{\iter}{}({\ep})$ are shown in linear-log plot and the errors in log-log-plot.
	}
\end{figure}
%FIGURE%%FIGURE%%FIGURE%%FIGURE%%FIGURE%%FIGURE%%FIGURE%%FIGURE%%FIGURE%%FIGURE%%FIGURE%%FIGURE%%FIGURE%%FIGURE%%FIGURE%%FIGURE%%FIGURE%%FIGURE%

A similar behaviour occurs for the randomised mixed labyrinth sets $\W_\iter(\gfa{1}:{p}\,,\,\gfb{1}:{1-p})$
with $p=0.2$ and $p=0.8$ shown in Fig.~\ref{fig:dminConv2}a) and b).
The approximative ensemble-averaged shortest path dimension $\overline{\dmin^{\iter}}$
and approximative shortest path dimension  $\darc{\iter}(\ep)$ converge after 
50 to 200 iterations.
One observes a fast
decrease of the relative deviation error of $\overline{\dmin^{\iter}}$ and a slower but monotonic decrease of
$\dminErr^{\iter}$.

A closer comparison of the values of the shortest path- and the box counting dimension 
shows that   
both 
lead to approximately the same values. 
In the self-similar cases
$\dmin^{\iter} \approx \darc{\iter}(\ep) = 1.0$ for $\W_\iter(\gseb{1})$
and $\dmin^{\iter} \approx \darc{\iter}(\ep) = 1.18$ for $\W_\iter(\gsia{1})$.
In the randomised mixed cases
$\dmin^{\iter} \approx \darc{\iter}(\ep) = 1.05$ for $\W_\iter(\gfa{1}:{0.2}\,,\,\gfb{1}:{0.8})$
and $\dmin^{\iter} \approx \darc{\iter}(\ep) = 1.14$ for $\W_\iter(\gfa{1}:{0.8}\,,\,\gfb{1}:{0.2})$.
One can see that $\darc{\iter}(\ep)$ converges slower than $\dmin^{\iter}$ and $\overline{\dmin^{\iter}}$
to a unique value $\darc{\iter}(\gamma)$ for all $\ep$,
as the relative error decreases slower than $\dminErr^{\iter}$.
How can this be understood? Is there a direct relation between these quantities?

We recapture the definitions of  $\darc{}({\ep})$ and $\dmin$.
In the case of labyrinth patterns with equal width $\linl$
we can rewrite Eq.~(\ref{eq:darc-math}) as
\begin{align}
	\darc{}({\ep}) 
	&=\lim_{n\to \infty} \frac{\log{\path{\iter}{}({\ep})}}{\log{\linl(\iter)}} 
	=\lim_{n\to \infty} \sum_{k=1}^{\iter} \frac{ \log{ \frac{\path{k}{}({\ep})}{\path{k-1}{}({\ep})} } }{ \iter\,\log{\linl} }
	=\lim_{n\to \infty} \frac{1}{\iter} \,\sum_{k=1}^{\iter} \frac{\log{\minPathScale{k}{}({\ep})}}{\log{\linl}} 
\nonumber
\\
	\darc{}({\ep}) & \approx
	\frac{1}{\iter} \,\sum_{k=1}^{\iter} \dmin
	  \label{eq:darc-reform} \,\, ,
\end{align}
with $\path{0}{}(\ep)=1$, 
and taking into account 
that 
for large enough $n$, i.e., $n>k_0$, 
the initially different $\minPathScale{n}{}({\ep})$ converge to 
a unique value
$\minPathScale{}{}$.
Thus, we find a direct relation between 
$\darc{n}({\ep})$ and $\dmin^{\iter}$.
The quantity $\darc{}({\ep})$
 can be interpreted as the averaged 
shortest path dimension
$\dmin^{n}$ over all preceding levels $n$
and for $\iter\to\infty$  we get  $\darc{}({\ep})= \dmin$.
The error $\Delta\darc{n}({\ep})$ gives the averaged $\Delta\dmin^{n}$ over all preceding levels $n$, 
leading to
a slower decrease than for $\dmin$.

Note, that Eq.~(\ref{eq:darc-reform}) holds
	for a single representation $\repre$ of 
	a randomised mixed labyrinth set.
	In this case $\darc{}({\ep})$ can be also determined by Eq.~(\ref{eq:darc-tb}), i.e.~the spectral radius of the path matrix $M$. 
	As we are interested in the averaged shortest path dimension $\dminMean$ of randomised mixed labyrinth fractals,
	in the following we will investigate to what extend 
	$\dminMean$ 
	is captured by $\darc{}{(\widetilde{a(\ep)})}$ (see Eq.~(\ref{eq:MixArcDim})), which is determined via the approximative path matrix $\widetilde{M}$.
Based on (\ref{eq:darc-reform}), we introduce the short notation $\dminApprox=\darc{}{(\widetilde{a(\ep)})}$
for given $\widetilde{a(\ep)}$.

From the above results, one can conclude three points for the further analysis.
First, a maximum iteration level of $n=1000$ is a convenient choice 
for the analysis of the approximative values for the 
shortest path- and arc dimension, as the values exhibit a 
fast 
convergence behaviour.
Second, as 
for $\iter \to \infty$, $\dmin$ equals $\darc{}({\ep})$,
we will focus on $\dmin$ as it shows faster convergence 
but 
the same asymptotic behaviour as $\darc{}({\ep})$.
Third, we will analyse how this direct relation between $\dmin$ and $\darc{}({\ep})$ 
can be extended to $\dminMean$ and $\dminApprox$ 
for randomised mixed labyrinth sets, respectively.

%SECTION%%SECTION%%SECTION%%SECTION%%SECTION%%SECTION%%SECTION%%SECTION%%SECTION%
\section{\label{sec:ResultsDmin} Shortest path dimension}
%SECTION%%SECTION%%SECTION%%SECTION%%SECTION%%SECTION%%SECTION%%SECTION%%SECTION%

In the following we discuss the results 
of the ensemble-averaged shortest path dimension $\dminMean$ as well as the 
approximated shortest path dimension
$\dminApprox$ determined from $\widetilde{M}$.
Both values are analysed depending on the selection probability $p$ 
of 
the investigated randomised mixed labyrinth sets $\W_n(\Ac:p\,,\,\Ac':1-p)$.
We remind the reader 
that for $p=0$ and $p=1$ $\dminMean$ equals 
$\dmin$ of the corresponding pattern $\Ac'$ and $\Ac$, respectively.

\begin{figure}[tb!]
	\begin{tabular}{ll}
	a) & b) \\
	\includegraphics[width=0.475\textwidth]{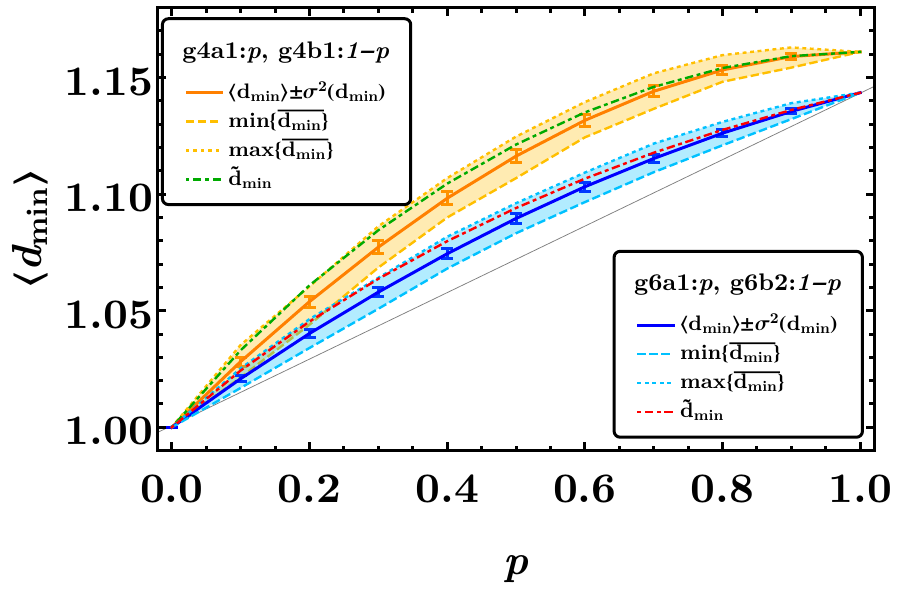} %}
	&
	\includegraphics[width=0.475\textwidth]{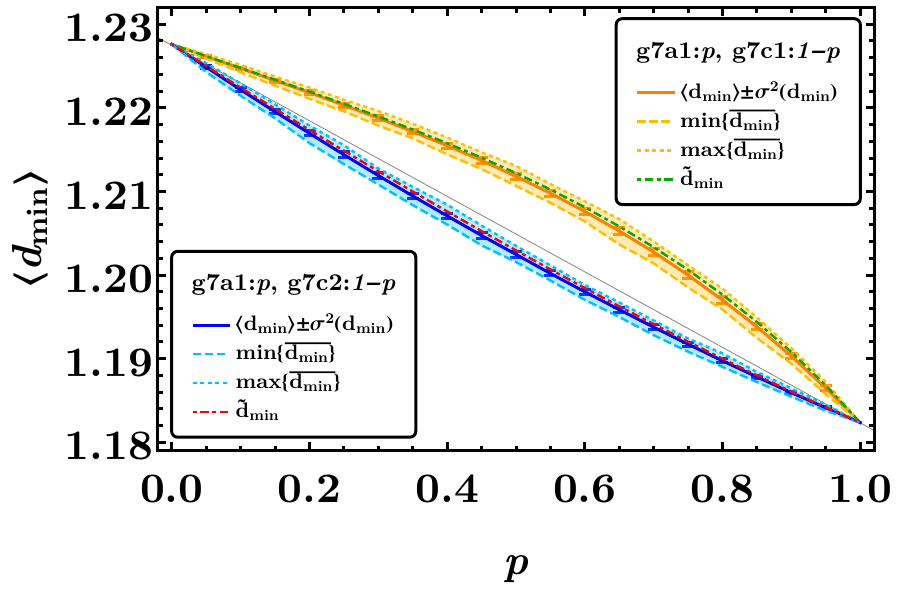}
	\end{tabular}
	\caption{\label{fig:dmin_monotonic}%
		Examples for nonlinear monotonic behaviours of $\dminMean$ and $\dminApprox$. 
		The straight line (grey line) indicates linear behaviour. 
	}
\end{figure}

\begin{figure}[tb!]
	\begin{tabular}{ll}
	a) & b) \\
	\includegraphics[width=0.475\textwidth]{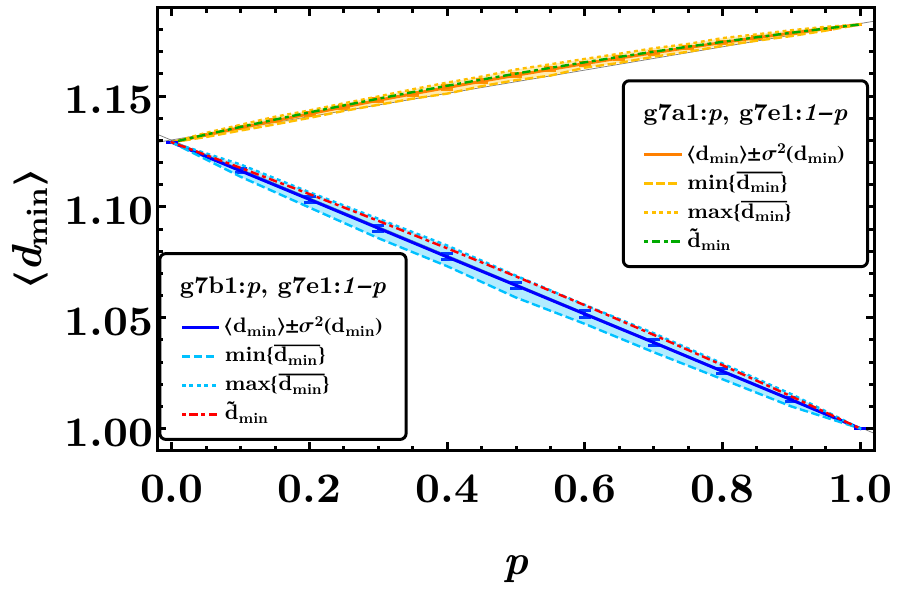}
	&
	\includegraphics[width=0.475\textwidth]{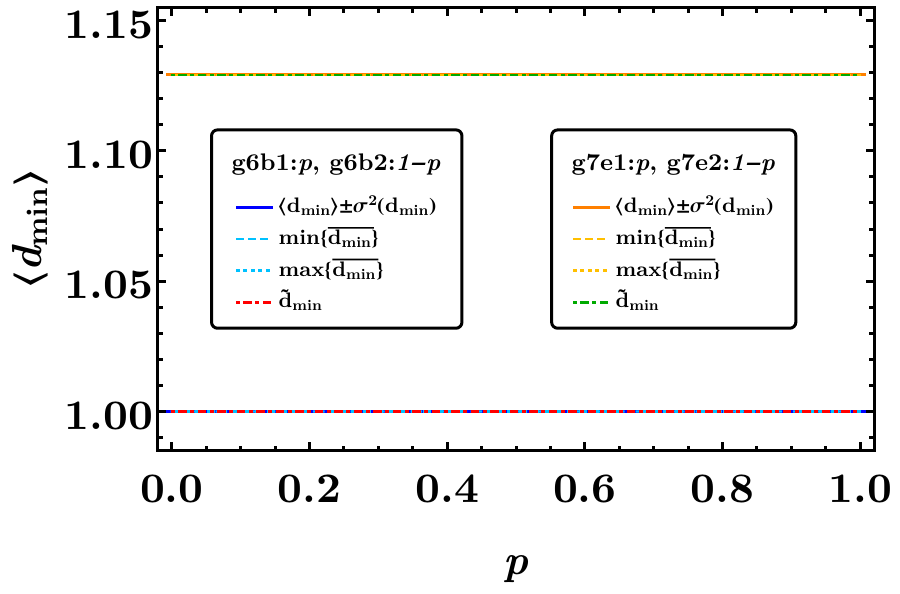}
	\end{tabular}
	\caption{\label{fig:dmin_LinConst}%
		Examples for linear and constant behaviours of $\dminMean$ and $\dminApprox$.
	}
\end{figure}

\begin{figure}[tb!]
	\begin{tabular}{ll}
	a) & b) \\
	\includegraphics[width=0.475\textwidth]{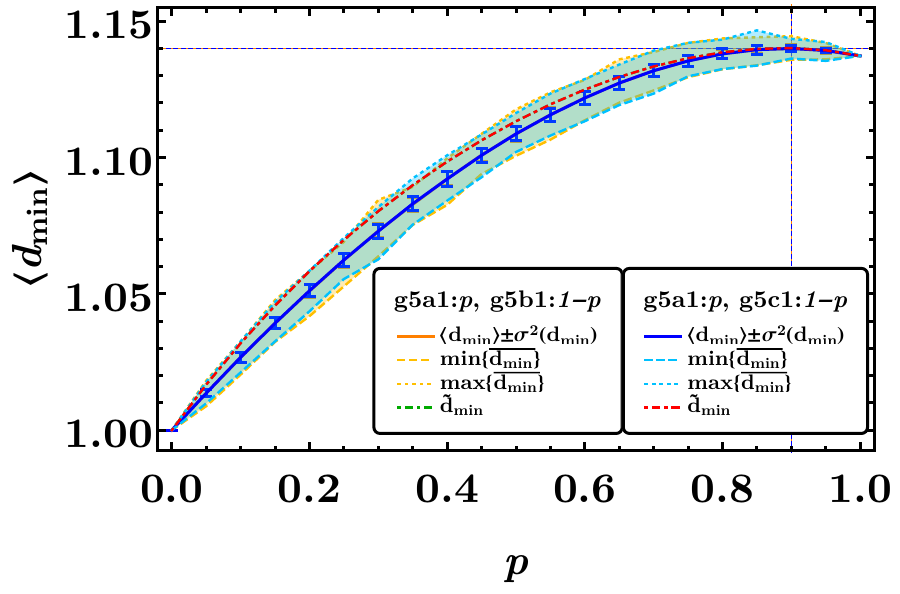} %\\
	&
	\includegraphics[width=0.475\textwidth]{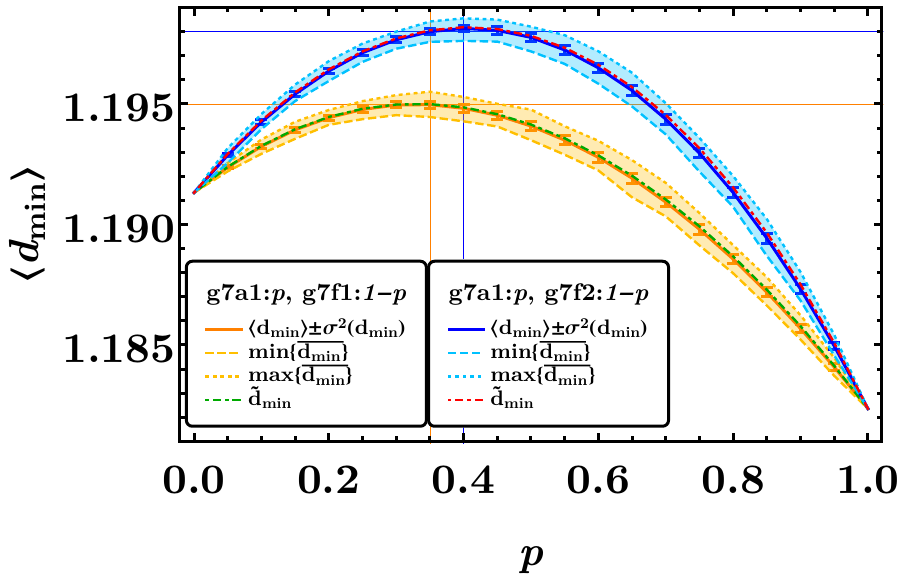}
	\end{tabular}
	\caption{\label{fig:dmin_max}%
		Examples for pattern combinations that lead to a maximum of $\dmin$ for
		$p\in(0,1)$. The maximum positions are indicated by the 
		horizontal and vertical lines.
	}
\end{figure}

\begin{figure}[tb!]%[ht]
	\begin{tabular}{ll}
	a) & b) \\
	\includegraphics[width=0.475\textwidth]{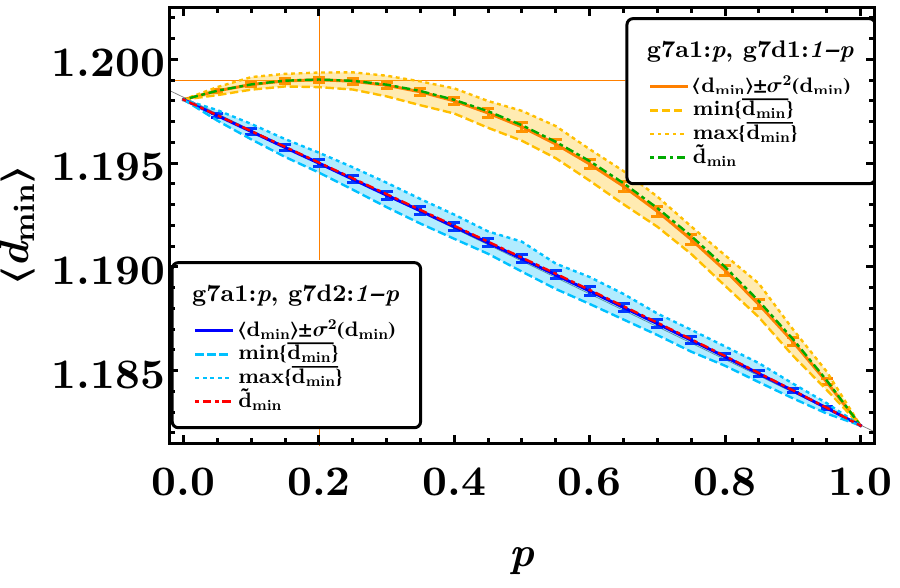}
	&
	\includegraphics[width=0.475\textwidth]{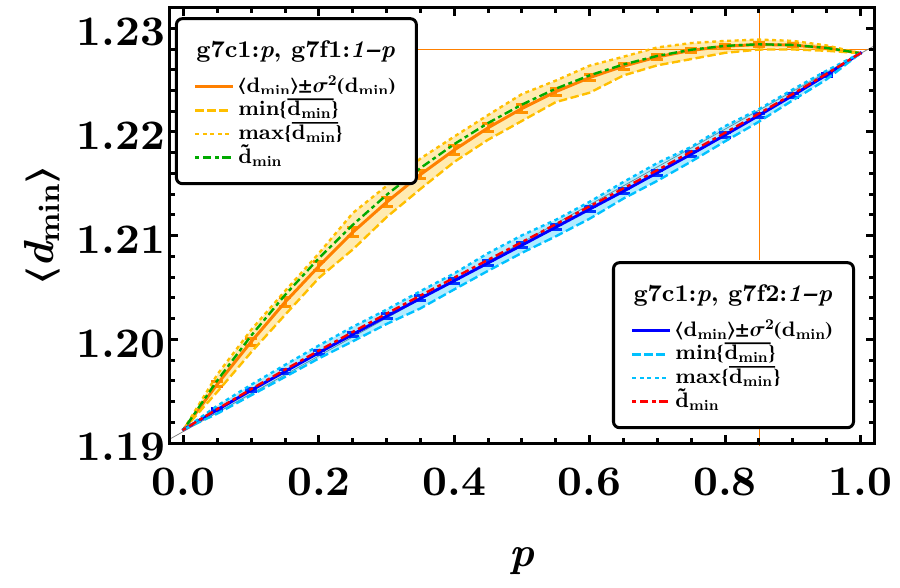} \\
	c) \\
	\multicolumn{2}{c}{\includegraphics[width=0.475\textwidth]{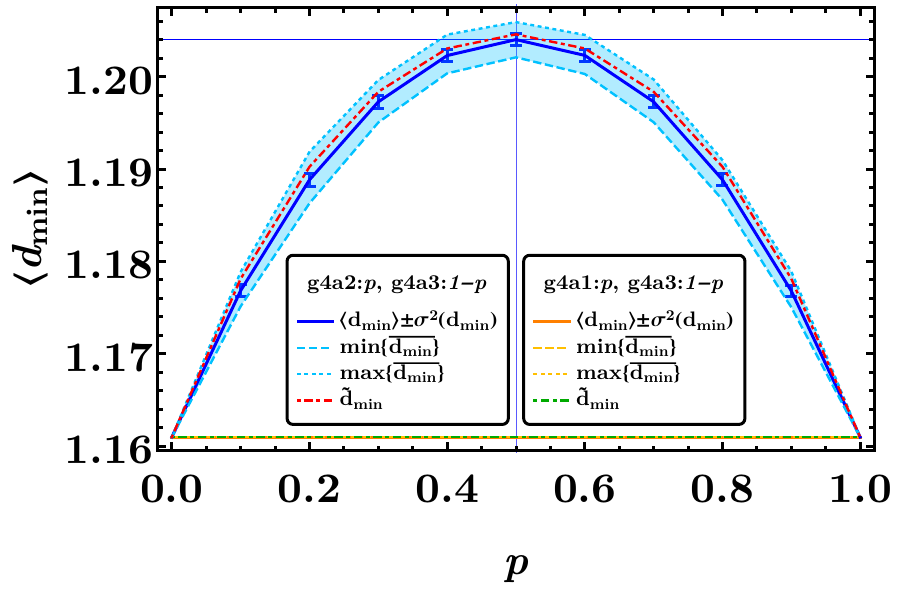}}
	\end{tabular}
	\caption{\label{fig:dmin_switching}%
		Three examples where mixing different labyrinth patterns or their rotated versions lead to a 
		switch of the behaviour of $\dmin$ from a constant/ nonlinear monotonic behaviour to the occurrence of a maximum for $p\in(0,1)$. 
	}
\end{figure}

Depending on the pattern combination,
the observed values of $\dminMean$ and $\dminApprox$
are either constant, monotonically increasing/decreasing over $p$ or 
exhibit a maximum at $p$, for $0<p<1$,
as shown in Fig.~\ref{fig:dmin_monotonic} - \ref{fig:dmin_switching}.
In all cases the approximated shortest path dimension
$\dminApprox$ captures the behaviour of $\dminMean$ within the min-max-errors.
Moreover, we find that the rotation of the patterns with respect to each other can also have an effect
on the resulting $\dminMean$ and $\dminApprox$.
An overview of the discussed results, i.e., the observed behaviours and the path scaling factors, is listed in Tab.~\ref{tab:pathLength}.
The complete list of all combinations $\W_n(\Ac:p\,,\,\Ac':1-p)$ and 
the $\dminMean$-$p$-diagrams are given
in the Supplementary Material..

%VERSION2: %TABLE%%TABLE%%TABLE%%TABLE%%TABLE%%TABLE%%TABLE%%TABLE%%TABLE%%TABLE%
\begin{table}[tb]
\small
\begin{tabular}{cccc@{\hspace{3ex}}c@{\hspace{3ex}}c}
	Fig. & Behaviour & $\Ac$ & $\Ac'$ & $\pathVec{1}{}$ of $\Ac$ &  $\pathVec{1}{}$ of  $\Ac'$ \\ \hline \hline
%\label{eqn:g7a1g7f1}
	\ref{fig:dmin_monotonic} a) & monotonic & $\gva{1}$ & $\gvb{1}$ & $(6,6,4,5,4,7)$ & $(4,4,5,2,3,6)$ \\  \hline
%\label{eqn:g7a1g7c1}
	\ref{fig:dmin_monotonic} a) & monotonic & $\gsea{1}$ & $\gseb{1}$ & $(8,10,11,8,7,4)$ & $(6,6,6,7,6,5)$ \\  \hline
	\ref{fig:dmin_monotonic} b) & monotonic & $\gsia{1}$ & $\gsic{1}$ & $(11,11,14,10,8,6)$ & $(11,13,6,10,8,14)$ \\  \hline
%\label{eqn:g7a1g7c2}
	\ref{fig:dmin_monotonic} b) & monotonic & $\gsia{1}$ & $\gsic{2}$ & $(11,11,14,10,8,6)$ & $(13,11,14,6,10,8)$ \\  \hline
%\label{eqn:g7a1g7f1}
	\ref{fig:dmin_LinConst} a) & linear & $\gsia{1}$ & $\gsib{1}$ & $(8,10,11,8,7,6)$ & $(6,6,6,7,6,5)$ \\  \hline
%\label{eqn:g7a1g7c1}
	\ref{fig:dmin_LinConst} a) & linear & $\gsib{1}$ & $\gsie{1}$ & $(7,7,7,7,7,7)$ & $(9,9,9,9,9,9)$ \\  \hline
	\ref{fig:dmin_LinConst} b) & constant & $\gseb{1}$ & $\gseb{2}$ & $(6,6,6,7,6,5)$ & $(6,6,5,6,7,6)$ \\  \hline
%\label{eqn:g7a1g7c2}
	\ref{fig:dmin_LinConst} b) & constant & $\gsie{1}$ & $\gsie{2}$ & $(9,9,9,9,9,9)$ & $(9,9,9,9,9,9)$ \\  \hline
%\label{eqn:g5a1g5b1}
	\ref{fig:dmin_max} a) 	& maximum & $\gfa{1}$ & $\gfb{1}$  & $(9,7,6,6,4,10)$ &$(5,5,7,5,3,5)$ \\  \hline
%\label{eqn:g5a1g5c1}
	\ref{fig:dmin_max} a) 	& maximum & $\gfa{1}$ & $\gfc{1}$  &$(9,7,6,6,4,10)$ &$(5,5,5,5,5,5)$ \\  \hline
	\ref{fig:dmin_max} b) 	& maximum (h) & $\gsia{1}$ & $\gsif{1}$ &$(11,11,14,10,8,6)$ &$(13,9,9,9,9,13)$ \\
%\label{eqn:g7a1g7f2}
	\ref{fig:dmin_max} b) 	& maximum (l) & $\gsia{1}$ & $\gsif{2}$ &$(11,11,14,10,8,6)$ &$(9,13,13,9,9,9)$ \\  \hline
%%\label{eqn:g7c1g7f1}
	\ref{fig:dmin_switching} a) & maximum & $\gsic{1}$ & $\gsif{1}$ &$(11,13,6,10,8,14)$ &$(13,9,9,9,9,13)$ \\
%%\label{eqn:g7c1g7f2}
	\ref{fig:dmin_switching} a) & monotonic & $\gsic{1}$ & $\gsif{2}$ &$(11,13,6,10,8,14)$ &$(9,13,13,9,9,9)$ \\  \hline%\\[-2ex] \cline{1-6} 
%\label{eqn:g7a1g7d1}
	\ref{fig:dmin_switching} b) & maximum & $\gsia{1}$ & $\gsid{1}$ &$(11,11,14,10,8,6)$ &$(11,11,9,11,9,11)$ \\
%%\label{eqn:g7a1g7d2}
	\ref{fig:dmin_switching} b) & monotonic & $\gsia{1}$ & $\gsid{2}$ &$(11,11,14,10,8,6)$ &$(11,11,11,9,11,9)$ \\  \hline
%\label{eqn:g4a1g4a3}
	\ref{fig:dmin_switching} c)	& maximum & $\gva{2}$ & $\gva{3}$ &$(6, 6, 5, 4, 7, 4)$ &$(6, 6, 4, 5, 4, 7)$ \\
%\label{eqn:g4a2g4a3}
	\ref{fig:dmin_switching} c) & constant &$\gva{1}$ & $\gva{3}$ &$(6, 6, 4, 7, 4, 5)$ &$(6, 6, 4, 5, 4, 7)$ %\\[-2ex] \cline{1-6} 
\end{tabular}
	\caption{\label{tab:pathLength} % 
	The path length $\pathVec{1}{}$ given for all combinations shown in Fig.~\ref{fig:dmin_monotonic} - \ref{fig:dmin_switching}. 
	The case \ref{fig:dmin_max} b) represents scenario one, where one rotation combination has a higher (h) maximum value than the other (l).}
\end{table}
%VERSION2: %TABLE%%TABLE%%TABLE%%TABLE%%TABLE%%TABLE%%TABLE%%TABLE%%TABLE%%TABLE%

We start with the discussion on how the rotated versions of the individual patterns influence 
the results for randomised mixed labyrinth fractals.
We observe two different scenarios.
Scenario one: 
The rotation of patterns lead to two distinct behaviours of $\dminMean$ over $p$ within the statistical error.
Scenario two: 
The rotation has no influence within statistical error and 
only one behaviour of $\dminMean$ over $p$ 
{is found for all pattern combinations.

In scenario one, one type of behaviour is found for $|i-j| = \{0,2\}$ for two given rotation indices $i,j$ of the patterns.
	Thus, the rotation angle difference between $i$ and $j$ is either $0^\circ$ or $\pm 180^\circ$.
	The other one is found for $|i-j| = \{1,3\}$, i.e.,
	the rotation angle difference equals $\pm 90^\circ$ or $\pm 270^\circ$. 
One can switch between the two scenarios by rotating one of the given patterns by $\pm 90^\circ$ or $\pm 270^\circ$.}{}
The given examples are 
\gva{} \& \gva{} (Fig.~\ref{fig:dmin_switching} c), 
\gsia{} \& \gsic{} (Fig.~\ref{fig:dmin_monotonic} b), 
\gsia{} \& \gsid{} (Fig.~\ref{fig:dmin_switching} a), 
\gsia{} \& \gsif{} (Fig.~\ref{fig:dmin_max} b), and
\gsic{} \& \gsif{} (Fig.~\ref{fig:dmin_switching} b). 

Scenario two occurs
if at least one of the patterns is non-blocked or is a fully-path-length-symmetric pattern.
The non-blocked or fully-path-length-symmetric pattern do not alter the shortest path scaling factor due to rotation 
and thus all combinations with the second pattern are same due to rotational symmetry arguments.
Given examples are 
\gva{} \& \gvb{} (Fig.~\ref{fig:dmin_monotonic} a), 
\gfa{} \& \gfb{} (Fig.~\ref{fig:dmin_max} a), 
\gfa{} \& \gfc{} (Fig.~\ref{fig:dmin_max} a),
\gsea{} \& \gseb{} (Fig.~\ref{fig:dmin_monotonic} a), 
\gseb{} \& \gseb{} (Fig.~\ref{fig:dmin_LinConst} b),
\gsia{} \& \gsie{} (Fig.~\ref{fig:dmin_LinConst} a),
\gsib{} \& \gsie{} (Fig.~\ref{fig:dmin_LinConst} a), and
\gsie{} \& \gsie{} (Fig.~\ref{fig:dmin_LinConst} b).

The complete list of possible combinations for all patterns with 
their $\dminMean$-$p$-diagrams is given in the Supplementary material,
as well as the comparison of the rotational symmetry.

%FIGURE%%FIGURE%%FIGURE%%FIGURE%%FIGURE%%FIGURE%%FIGURE%%FIGURE%%FIGURE%%FIGURE%%FIGURE%%FIGURE%%FIGURE%%FIGURE%%FIGURE%%FIGURE%%FIGURE%%FIGURE%
%FIGURE%%FIGURE%%FIGURE%%FIGURE%%FIGURE%%FIGURE%%FIGURE%%FIGURE%%FIGURE%
%FIGURE%%FIGURE%%FIGURE%%FIGURE%%FIGURE%%FIGURE%%FIGURE%%FIGURE%%FIGURE%

Subsequently, we discuss why the different behaviours of $\dminMean$ over $p$ arise and how they can be explained.

In Fig.~\ref{fig:dmin_monotonic}, the pattern combinations show nonlinear monotonic behaviours of $\dmin$ 
over the selection probability $p$.
The deviations from the straight line (shown in grey) are clearly visible.
In Fig.~\ref{fig:dmin_monotonic} a), the combinations $\W_\iter(\gva{1}:p\,,\,\gvb{1}:1-p)$ and 
$\W_\iter(\gsea{1}:p\,,\,\gseb{1}:1-p)$ are examples of scenario two, 
where a non-blocked pattern is part of the combinations
and thus rotation does not alter the behaviour.
In Fig.~\ref{fig:dmin_monotonic} b), $\W_\iter(\gsia{1}:p\,,\,\gsic{1}:1-p)$ and $\W_\iter(\gsia{1}:p\,,\,\gsic{2}:1-p)$ 
show different behaviours for both combinations of rotation.
One finds that $\dminApprox$ fits very well with ${\dminMean}$ 
	within the min-max-error of ${\dminMean}$.
Our analysis of the patterns reveals a rather 
``soft'' criterion 
to predict the occurrence of a monotonic behaviour.
This is typically observed
for pattern combinations 
where 
``nearly'' all path lengths $\path{1}{}(\ep)$ of one pattern are larger than the ones of the other pattern.
The patterns are either 
both
totally blocked 
or 
one is totally blocked  and 
 one is non-blocked.
The 
corresponding
path lengths 
are listed in Tab.~\ref{tab:pathLength}.
 
In the following, we  
present two 
special cases for this behaviour 
as well as extensions and exceptions of the soft criterion 
leading to different behaviours.

%FIGURE%%FIGURE%%FIGURE%%FIGURE%%FIGURE%%FIGURE%%FIGURE%%FIGURE%%FIGURE%%FIGURE%%FIGURE%%FIGURE%%FIGURE%%FIGURE%%FIGURE%%FIGURE%%FIGURE%%FIGURE%
%FIGURE%%FIGURE%%FIGURE%%FIGURE%%FIGURE%%FIGURE%%FIGURE%%FIGURE%%FIGURE%
%FIGURE%%FIGURE%%FIGURE%%FIGURE%%FIGURE%%FIGURE%%FIGURE%%FIGURE%%FIGURE%

The special cases of the monotonic behaviour are the 
linear and the constant behaviour of $\dminMean$ as shown in Fig.~\ref{fig:dmin_LinConst}.
In the linear cases, 
two totally blocked patterns with different individual $\dmin$ are combined, where 
all path lengths $\path{1}{}(\ep)$ of one pattern are smaller 
than the path lengths $\path{1}{}(\ep)$ of the other pattern.
Such examples are $\W_\iter(\gsib{1}:p\,,\,\gsie{1}:1-p)$ and $\W_\iter(\gsia{1}:p\,,\,\gsie{1}:1-p)$ 
shown in Fig.~\ref{fig:dmin_LinConst} a), 
with
the corresponding path lengths  $\pathVec{1}{}$  given in Tab.~\ref{tab:pathLength}.

In the case of constant ${\dminMean}$ over $p$ 
there are either two non-blocked patterns or
two totally blocked patterns with equal $\dmin$ combined, 
where the path lengths of both patterns are identical or mirror symmetric.
The constant  ${\dminMean}$ in the case of non-blocked patterns
results from the proof  
that $\darc{}{}=1$ for non-blocked mixed labyrinth patterns
\cite{cristea.l.18.on.575,cristea.l.20.supermixed.183}.
In the case of totally blocked patterns, 
it is convenient that mixing patterns with identical path length does not alter the shortest path length while combining them.
In the case of mirror symmetric pattern the rotation of the pattern by $180^\circ$ does not alter the path length combinations either.
One example for each case is depicted in Fig.~\ref{fig:dmin_LinConst} b) 
$\W_\iter(\gseb{1}:p\,,\,\gseb{2}:1-p)$ and $\W_\iter(\gsie{1}:p\,,\,\gsie{2}:1-p)$ (cf.~Tab.~\ref{tab:pathLength}).
Note that in the case of constant behaviour there is no variation in the scaling of the path lengths, 
as mentioned above,
and thus $\dminVar=0$.

%FIGURE%%FIGURE%%FIGURE%%FIGURE%%FIGURE%%FIGURE%%FIGURE%%FIGURE%%FIGURE%%FIGURE%%FIGURE%%FIGURE%%FIGURE%%FIGURE%%FIGURE%%FIGURE%%FIGURE%%FIGURE%
%FIGURE%%FIGURE%%FIGURE%%FIGURE%%FIGURE%%FIGURE%%FIGURE%%FIGURE%%FIGURE%
%FIGURE%%FIGURE%%FIGURE%%FIGURE%%FIGURE%%FIGURE%%FIGURE%%FIGURE%%FIGURE%

%FIGURE%%FIGURE%%FIGURE%%FIGURE%%FIGURE%%FIGURE%%FIGURE%%FIGURE%%FIGURE%%FIGURE%%FIGURE%%FIGURE%%FIGURE%%FIGURE%%FIGURE%%FIGURE%%FIGURE%%FIGURE%
%FIGURE%%FIGURE%%FIGURE%%FIGURE%%FIGURE%%FIGURE%%FIGURE%%FIGURE%%FIGURE%
%FIGURE%%FIGURE%%FIGURE%%FIGURE%%FIGURE%%FIGURE%%FIGURE%%FIGURE%%FIGURE%

After analysing the nonlinear monotonic, linear and constant behaviour 
we will now investigate the cases 
where we observe either a maximum behaviour or 
two different characteristic behaviours (switching behaviour) depending on the rotation combination.

Examples for the maximum behaviour are shown in Fig.~\ref{fig:dmin_max}.
In Fig.~\ref{fig:dmin_max} a), the maximum is observed for the combinations
$\W_\iter(\gfa{1}:p\,,\,\gfb{1}:1-p)$ and  $\W_\iter(\gfa{1}:p\,,\,\gfc{1}:1-p)$ at
$p\approx 0.9$. 
We observe only one type of behaviour, as the pattern \gfa{} is non-blocked.
In Fig.~\ref{fig:dmin_max} b) 
, we
find different maximum values 
for mixing patterns depending on the combination of rotations,
as both patterns are totally blocked and not mirror symmetric.
For $\W_\iter(\gsia{1}:p\,,\,\gsif{1}:1-p)$ the maximum is reached at $p\approx 0.35$,
whereas for $\W_\iter(\gsia{1}:p\,,\,\gsif{2}:1-p)$, at $p\approx0.4$.
The positions of the maxima are indicated by the horizontal and vertical lines.

Furthermore, there are also
cases where the behaviour of $\dminMean$ over $p$  changes,
as shown in Fig.~\ref{fig:dmin_switching}.
The combinations \ref{fig:dmin_switching} a) $\W_\iter(\gsia{1}:p\,,\,\gsid{1}:1-p)$ and $\W_\iter(\gsia{1}:p\,,\,\gsid{2}:1-p)$,
as well as \ref{fig:dmin_switching} b) $\W_\iter(\gsic{1}:p\,,\,\gsif{1}:1-p)$ and $\W_\iter(\gsic{1}:p\,,\,\gsif{2}:1-p)$ switch from 
a nonlinear monotonic to a maximum behaviour with maximum at $p=0.2$ and $p=0.85$.
In the third case c) $\W_\iter(\gva{1}:p\,,\,\gva{3}:1-p)$ and $\W_\iter(\gva{2}:p\,,\,\gva{3}:1-p)$
the behaviour switches from a constant to a maximum behaviour with maximum at $p=0.5$.

We find   that the maximum and switching behaviours   
are very well captured 
by $\dminApprox$.
It thus provides a convenient tool to predict the behaviour \emph{a priori}.
By just looking at the path lengths of the individual patterns itself,
as given in Tab.~\ref{tab:pathLength}
these types of behaviour are hard to understand.

For instance, in the case of $\W_\iter(\gfa{1}:p\,,\,\gfb{1}:1-p)$, 
where a maximum behaviour is observed,
the path lengths of both patterns look quite similar. 
There is no path that is particularly long or short compared to the others
which would explain the maximum peak directly.

%FIGURE%%FIGURE%%FIGURE%%FIGURE%%FIGURE%%FIGURE%%FIGURE%%FIGURE%%FIGURE%%FIGURE%
%FIGURE%%FIGURE%%FIGURE%%FIGURE%%FIGURE%%FIGURE%%FIGURE%%FIGURE%%FIGURE%%FIGURE%
\begin{figure}[p]
	\small
	\begin{tabular}{cccc}
		a)  $W_1(\gfb{2})\hspace{8ex}$%, $\F$-path 
		&
		b)  $W_1(\gfa{1})\hspace{8ex}$%, $\F$-path 
		&
		&
		\\ %[-.5ex]
		\includegraphics[width=.2275\textwidth]{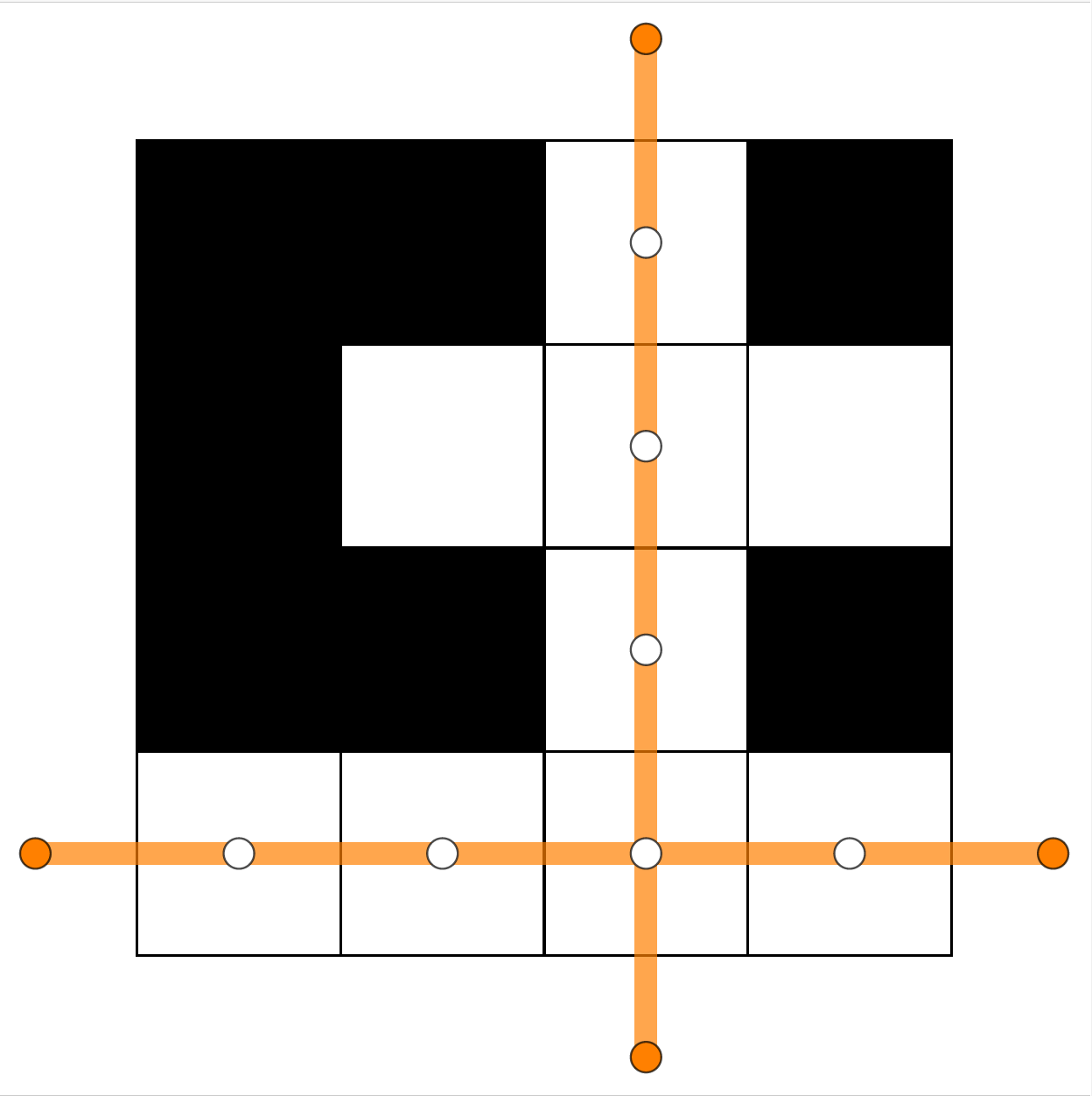} 	
		& \includegraphics[width=.2275\textwidth]{pattern_g5a1.pdf} 
		\\[1ex]
		c) $\W_2(\gfb{2},\gfb{2})$ & 
		d) $\W_2(\gfa{1},\gfb{2})$ &
		e) $\W_2(\gfb{2},\gfa{1})$ &
		f) $\W_2(\gfa{1},\gfa{1})$ \\
		\includegraphics[width=.2275\textwidth]{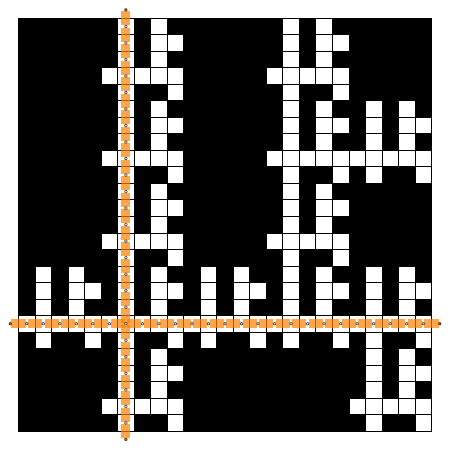} %\\[2ex] %\hline
		& \includegraphics[width=.2275\textwidth]{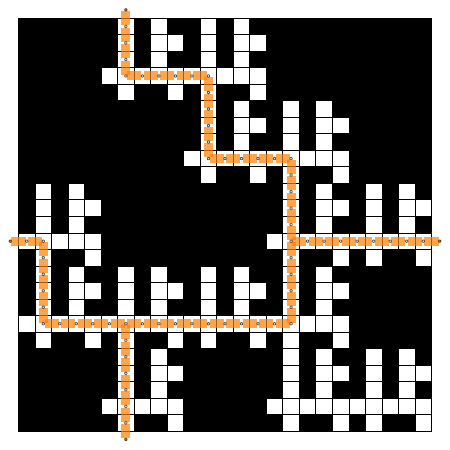} 
		& \includegraphics[width=.2275\textwidth]{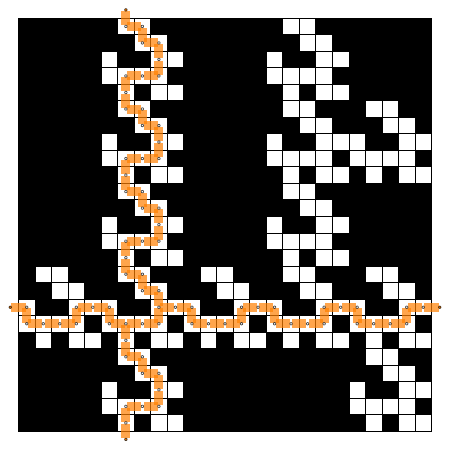} %,cfbox=blue 1pt 1pt
		& \includegraphics[width=.2275\textwidth]{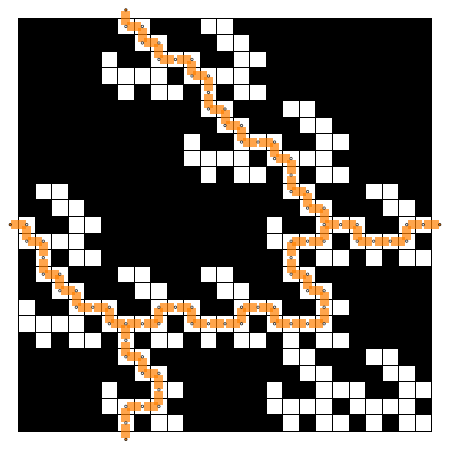} %,cfbox=gray 1pt 1pt 
		\\ [2ex]
		\multicolumn{2}{l}{g) $\W_3(\gfb{2},\gfb{2},\gfb{2})$} & 
		\multicolumn{2}{l}{h) $\W_3(\gfa{1},\gfb{2},\gfa{1})$} \\
		\multicolumn{2}{c}{\includegraphics[width=0.49\textwidth]{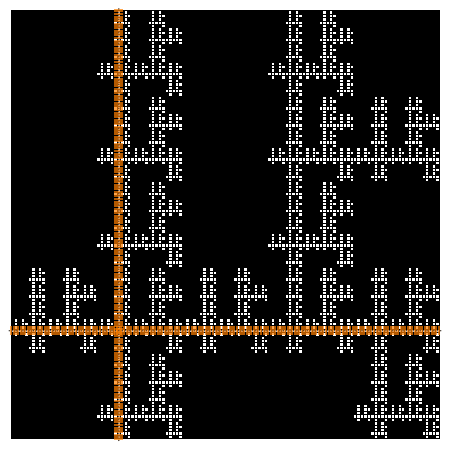}} & 
		\multicolumn{2}{c}{\includegraphics[width=0.49\textwidth]{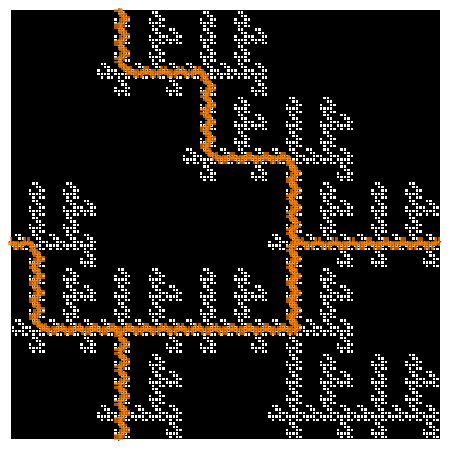}} \\
		
	\end{tabular}
	\caption{ \label{fig:g5ag5bPathLengths}%
		Randomised mixed labyrinth sets $\W_k$ for $k=1,2,3$ for $\W_k(\gfa{1}:p\,,\,\gfb{2}:1-p)$ are given. 
		The notation $\W_\iter(\Ac, \dots)$
		gives the specific labyrinth set realisations of $\W_k(\Ac:p\,,\,\Ac':1-p)$.
		The sets in a) - f) represent all possible realisations for $k=1,2$.
		The realisation shown in g) represents $W_3$ with the smallest $\dmin^3$ 
		and h)  $\W_3$ with the largest $\dmin^3$ are given (cmp.~Tab.~\ref{tab:g5ag5bPathLengthsPo1}).
	}
\end{figure}
%FIGURE%%FIGURE%%FIGURE%%FIGURE%%FIGURE%%FIGURE%%FIGURE%%FIGURE%%FIGURE%%FIGURE%
%FIGURE%%FIGURE%%FIGURE%%FIGURE%%FIGURE%%FIGURE%%FIGURE%%FIGURE%%FIGURE%%FIGURE%

%%TABLE%%TABLE%%TABLE%%TABLE%%TABLE%%TABLE%%TABLE%%TABLE%%TABLE%%TABLE%%TABLE%%
\begin{table}[tb!]
	\small
	
	\begin{tabular}{c|c|c|c}
		level $k$ & labyrinth set realisation &
		$\minPathScale{k}{1}(\ep), \ep=\{\A,\B,\C,\D,\E,\F\}$
		& $\dmin^{k,1}$  \\ \hline \hline
		\multirow{2}{*}{$k=1$}
		& $\W_1(\gfb{2})$ &
		$\{5.0, 5.0, 7.0, 5.0, 3.0, 5.0\}$ & $\mathbf{0.68}$
		\\ \cline{2-4} 
		& $\W_1(\gfa{1})$ & 
		$\{9.0, 7.0, 6.0, 6.0, 4.0, 10.0\}$ & $\mathbf{0.86}$ 
		\\ \hline \hline
		\multirow{4}{*}{$k=2$} 
		& $\W_2(\gfb{2}{},\gfb{2}{})$ &  
		$\{5.0, 5.0, 5.3, 5.0, 4.3, 5.0\}$ & $\mathbf{0.91}$ 
		\\ \cline{2-4} 
		& $\W_2(\gfa{1}{},\gfb{2}{})$ &
		$\{5.0, 5.0, 5.3, 5.0, 4.5, 5.0\}$ & $\mathbf{0.93}$  
		\\ 
		& $\W_2(\gfb{2}{},\gfa{1}{})$ &
		$\{9.0, 7.0, 7.7, 7.2, 6.7, 8.8\}$ 
		& $\mathbf{1.18}$ 
		\\ \cline{2-4}
		& $\W_2(\gfa{1}{},\gfa{1}{})$ &  
		$\{6.8, 6.7, 5.5, 7.5, 5.8, 6.3\}$ & $\mathbf{1.06}$ 
		\\ \hline \hline 
		\multirow{8}{*}{$k=3$} 
		& $\W_3(\gfb{2}{},\gfb{2}{},\gfb{2}{})$ & 
		$\{5.0, 5.0, 5.1, 5.0, 4.8, 5.0\}$ & $\mathbf{0.98}$
		\\ \cline{2-4}
		& $\W_3(\gfa{1}{},\gfb{2}{},\gfb{2}{})$ &  
		$\{5.0, 5.0, 5.1, 5.0, 4.9, 5.0\}$ & $\mathbf{0.99}$ \\ 
		& $\W_3(\gfb{2}{},\gfa{1}{},\gfb{2}{})$ & 
		$\{5.0, 5.0, 5.0, 5.0, 4.9, 5.0\}$ 
		& $\mathbf{0.99}$
		\\ 
		& $\W_3(\gfa{1}{},\gfa{1}{},\gfb{2}{})$ &
		$\{5.0, 5.0, 5.1, 5.0, 4.9, 5.0\}$ & $\mathbf{0.99}$ \\ \cline{2-4}
		& $\W_3(\gfb{2}{},\gfb{2}{},\gfa{1}{})$ &
		$\{9.0, 7.0, 7.9, 7.4, 7.7, 8.6\}$ & $\mathbf{1.21}$ \\ 
		&$\W_3(\gfa{1}{},\gfb{2}{},\gfa{1}{})$ &
		$\{7.8, 7.4, 7.4, 7.7, 7.7, 7.6\}$ & $\mathbf{1.24}$ \\ 
		& $\W_3(\gfb{2}{},\gfa{1}{},\gfa{1}{})$ &
		$\{6.8, 6.7, 6.6, 6.9, 6.6, 6.7\}$ & $\mathbf{1.17}$
		\\
		\cline{2-4}
		& $\W_3(\gfa{1}{},\gfa{1}{},\gfa{1}{})$ &
		$\{6.4, 6.5, 5.8, 6.8, 6.1, 6.1\}$ & $\mathbf{1.09}$ 
		\\ \hline \hline
	\end{tabular}
	\caption{\label{tab:g5ag5bPathLengthsPo1}%
		For the pattern combination $\W(\gfa{1}{} : p\,, \gfb{2}{} : 1-p\,)$
		all possible realisations of the first three iterations $k$ are listed 
		with their path scaling factors $\minPathScale{k}{1}(\ep)$ 
		and the shortest path dimensions $\dmin^{k,1}$. 
	}
\end{table}
%%TABLE%%TABLE%%TABLE%%TABLE%%TABLE%%TABLE%%TABLE%%TABLE%%TABLE%%TABLE%%TABLE%%

So we investigated the iteration process of the path lengths itself
\emph{a posteriori}, 
in order to get an idea of occurring effects.
In Tab.~\ref{tab:g5ag5bPathLengthsPo1} for the first three iteration level $k$
the shortest path scaling factors $\lambda_k(\ep)$ of all paths 
and the corresponding $\dmin^k$ are given 
for all possible randomised mixed labyrinth sets $\W_k$.
In Fig.~\ref{fig:g5ag5bPathLengths} all possible sets $\W_k$ with $k=1,2$ are shown
and for $\W_3$ the set with the largest and the smallest $\dmin^k$ is given.
From Tab.~\ref{tab:g5ag5bPathLengthsPo1} one already recognises the effect of restoration of isotropy for the path scaling factors.
Furthermore, at level $k=3$ there are distinct values $\dmin^{k,1}$ for each realisation $\W_k$, indicating the occurrence of the maximum for $\dminMean$ over selection probability $p$. 
Fig.~\ref{fig:g5ag5bPathLengths} also shows different behaviours of the path lengths for different combinations.
Nevertheless, no indicator based on the patterns' shape could be identified for the observed effects.  

Similarly to the maximum behaviour of $\W_\iter(\gfa{1}:p\,,\,\gfb{2}:1-p)$,
also the switching behaviour from a monotonic to a maximum behaviour 
by combining \gsic{} and \gsif{} is quite surprising from the perspective of the path lengths and symmetries.
The corresponding figures of labyrinth sets and the tables with the path scaling factors and shortest path dimensions are given in the supplementary material.
But there no indicator could be found either.

As the approximated shortest path dimension $\dminApprox$,
that we obtain from the approximated path matrix captures all unexpected behaviours, 
we recommend their use this for future investigations.

%SECTION%%SECTION%%SECTION%%SECTION%%SECTION%%SECTION%%SECTION%%SECTION%%SECTION%%SECTION%%SECTION%
\section{\label{sec:Sum}Summary}
%SECTION%%SECTION%%SECTION%%SECTION%%SECTION%%SECTION%%SECTION%%SECTION%%SECTION%%SECTION%%SECTION%

In the present work we introduce the concept of randomised mixed labyrinth fractals, 
where a labyrinth pattern is chosen randomly  from a given set with selection probability $p$
 for each iteration level.
Here, we restrict ourselves to mixing labyrinth patterns of the same width and of equal fractal dimension.
For the obtained randomised mixed labyrinth fractals 
we determine
the approximated shortest path dimension, 
the approximated arc dimension and 
the spectral radius of the corresponding path matrix 
for an ensemble of different realisations for different values of $p$.
The main focus is to investigate how 
the individual path lengths, the blocking property of the patterns, and
their symmetry 
influence the behaviour of the determined quantities for varying $p$.

First, we found 
that randomised mixed labyrinth fractals
show the effect of restoration of isotropy of the shortest path scaling factors during the iteration process.
However, due to the mixing of the dominant path length, 
determining the overall shortest path dimension is hard.
Second, we found a broad variety of behaviour types for the shortest path dimension over $p$,
from a constant, linear or nonlinear monotonic transition to a transition exhibiting a maximum. 
We find that the cases of constant and linear behaviour can be predicted 
by analysing the individual path lengths.
In order to understand the maximum behaviour,
 the iteration process of the path lengths still needs to be investigated.
 This exceeds the scope of this work.
Let us note, that the finding of the changing behaviour of the shortest path dimension directly influences the dynamics on such structures due to the direct connection to the random walk dimension based on the Einstein relation.
This has also been observed previously for more general \Sierpinski carpets 
\cite{anh.d.07.anomalous.11453}.

Additionally, we introduced the concept of an approximated path matrix 
to directly determine the corresponding shortest path dimension of 
 randomised mixed labyrinth fractals, 
 instead of performing thousands of realisations.
 We can show that
the shortest path dimension of the approximated path matrix 
gives the same results as the statistical analysis of random realisations within statistical errors.
Thus, the approximated path matrix is a convenient tool for an a priori analysis 
of randomised mixed labyrinth fractals for further investigations.

Furthermore, we find that in general for combinations of labyrinth patterns of equal width 
the arc dimension of a randomised mixed labyrinth fractal is 
the averaged sum of the shortest path dimension over all iteration levels.

Open issues are 
a proof that Gelfand's theorem does apply for the here used path matrices and
to what extend
it can be applied to pattern combinations with different widths and fractal dimensions 
as well as what happens while mixing half-blocked and blocked patterns.

\section*{Acknowledgments}:
We thank Sascha Troscheit for valuable comments on Gelfand's theorem.

\section*{Supplementary material}

Here, a full list of all pattern combinations, the resulting data of the shortest path dimensions $\dminMean$ over selection probability $p$. 
Additionally a discussion of shortest path scaling factors over the first three iteration level is given for the pattern combination $\gsic{}$ and $\gsif{}$ that might give  some insights to the observed behaviours. 

\section*{Conflict of Interest}

The authors have no conflicts to disclose.

\section*{Data Availability Statement}

The data that support the findings of this study are available within the article and its supplementary material.
If there are any further requests, please contact the corresponding author upon reasonable request.

\nocite{*}
\bibliographystyle{unsrt}
\bibliography{article_RandLabFrac}% Produces the bibliography via BibTeX.

@string{A	={Axioms}}

@string{AdP	={Adv. Phys.}}

@string{AN	={Astron. Nachr.}}

@string{ARXV	={arXiv}}

@string{AS	={Am. Sci.}}

@string{C	={Computing}}

@string{CMP	={Comm. Math. Phys.}}

@string{CPC	={Comp. Phys. Comm.}}

@string{EL	={Electron. Lett.}}

@string{EPL	={Europhys. Lett.}}

@string{FF	={Fracal Fract.}}

@string{Fractals	={Fractals}}

@string{H	={Helyion}}

@string{JFG     ={J. Fractal Geom.}}

@string{JPA	={J. Phys. A: Math. Gen.}}

@string{MHfM	={Monatsh Math}}

@string{ND	={Nonlinear Dyn.}}

@string{Nonlin	={Nonlinearity}}

@string{PLA	={Phys. Lett. A}}

@string{PRE	={Phys. Rev. E}}

@string{PRL	={Phys. Rev. Lett.}}

@string{PTRF	={Probab. Theory Relat. Fields}}

@string{SCM	={Sci. China Math.}}

@string{SIGSAM	={SIGSAM Bulletin}}

@string{SJ	={IEEE Sens. J.}}

@string{SREP	={Sci. Rep.}}

@string{TA	={Topol. Appl.}}

@ARTICLE{anguera.j.20.fractal.4,

	RELEVANCE       = { adf },
	LOCATION        = {file},

	AUTHOR          = {Anguera, Jaume and And\'ujar, Aurora and Jayasinghe, Jeevani and Chakravarthy, V.~V.~S.~S.~Sameer: and Chowadary, P.~S.~R. and Pijoan, Joan L. and Ali, Ranweer and Cattani, Carlo},
	TITLE           = {Fractal Antennas: An Historical Perspective},
	JOURNAL         = FF,
	YEAR            = {2020},
	VOLUME          = {4},
	NUMBER          = {1},
	PAGES           = {3},
	ISSN_ISBN       = {},
	DOI		= {10.3390/fractalfract4010003},
	NOTE            = {},
	CLASSIFICATION  = {},
}

@ARTICLE{anh.d.05.diffusion.109,

	RELEVANCE       = { adf },
	LOCATION        = {own, file},

	AUTHOR          = {Anh, Do Hoang Ngoc  and Hoffmann, Karl Heinz and Seeger, Steffen and Tarafdar, Sujata},
	TITLE           = {Diffusion in disordered Fractals},
	JOURNAL         = EPL,
	YEAR            = {2005},
	VOLUME          = {70},
	NUMBER          = {1},
	PAGES           = {109--115},
	ISSN_ISBN       = {},
	DOI		= {10.1209/epl/i2005-10002-x},
	NOTE            = {},
	CLASSIFICATION  = {PACS: 61.43.Hv 66.30.-h},
}

@ARTICLE{anh.d.07.anomalous.11453,

	RELEVANCE       = { adf },
	LOCATION        = {own,file},

	AUTHOR          = {Anh, Do Hoang Ngoc and Blaudeck, Peter and Hoffmann, Karl Heinz and Prehl, Janett and Tarafdar, Sujata},
	TITLE           = {Anomalous diffusion on random fractal composites},
	JOURNAL         = JPA,
	YEAR            = {2007},
	VOLUME          = {40},
	NUMBER          = {38},
	PAGES           = {11453-11465},
	ISSN_ISBN       = {ISSN 1751-8113/07/3811453+13},
	DOI		= {10.1088/1751-8113/40/38/002},
	NOTE            = {},
	CLASSIFICATION  = {},
}

@ARTICLE{barlow.m.95.restoration.3042,

	RELEVANCE          = { adf },
	AUTHOR		   = {Martin T. Barlow and Kumiko Hattori and Tetsuya Hattori and Hiroshi Watanabe},
	JOURNAL		   = PRL,
        NUMBER             = {17},
	PAGES		   = {3042-3045},
	TITLE		   = {Restoration of Isotropy on Fractals},
	VOLUME		   = {75},
	YEAR		   = {1995},
        LOCATION           = {copy},
}

@ARTICLE{barlow.m.97.weak.1,

	RELEVANCE       = { adf },
	LOCATION        = {file},

	AUTHOR          = {Barlow, Martin T. and Hhattori, K. and Hattori, T. and Watanabe, H.},
	TITLE           = {Weak homogenization of anisotropic diffusion on pre-Sierpi\'nski carpet},
	JOURNAL         = CMP,
	YEAR            = {1997},
	VOLUME          = {188},
	NUMBER          = {1},
	PAGES           = {1--27},
	ISSN_ISBN       = {},
	DOI		= {},
	NOTE            = {},
	CLASSIFICATION  = {},
}

@ARTICLE{blaudeck.p.06.coastline.1609,

	RELEVANCE       = { adf },
	LOCATION        = {own},

	AUTHOR          = {Blaudeck, Peter and Seeger, Steffen and Schulzky, Christian and Hoffmann, Karl Heinz and Dutta, Tapati and Tarafdar, Sujata},
	TITLE           = {The coastline and lake shores on a fractal island},
        JOURNAL         = JPA,
        VOLUME          = {39},
	YEAR            = {2006},
        PAGES           = {1609--1618},
	ISSN_ISBN       = {ISSN: 0305-4470},
	DOI		= {10.1088/0305-4470/39/7/006},
	NOTE            = {},
	CLASSIFICATION  = {},
}

@BOOK{bunde.a.96.fractals.book,

	RELEVANCE	= { adf },
        LOCATION	= {"lib:S4/493176"},
	ISSN_ISBN	= {ISBN: 3-540-56219-2},
	AUTHOR		= {},
	EDITOR		= {Bunde, A. and Havlin, S.},
	TITLE		= {Fractals and Disordered Systems},
	PUBLISHER	= {Springer},
	YEAR		= {1996},
	VOLUME		= {},
	NUMBER		= {},
	SERIES		= {},
	ADDRESS		= {Berlin, Heidelberg, New-York},
	EDITION		= {2nd},
	NOTE		= {},
	PACS		= {},
}

@ARTICLE{cristea.l.09.curves.1,

	RELEVANCE       = { adf },
	LOCATION        = {copy},

	AUTHOR          = {Cristea, Ligia Loreta and Steinstky, Bertran},
	TITLE           = {Curves of infinite length in 4$\times$4-labyrinth fractals},
	JOURNAL         = {Geom Dedicata},
	YEAR            = {2009},
	VOLUME          = {141},
	NUMBER          = {},
	PAGES           = {1--17},
	ISSN_ISBN       = {},
	DOI		= {10.1007/s10711-008-9340-3},
	NOTE            = {},
	CLASSIFICATION  = {},
}

@ARTICLE{cristea.l.11.curves.329,

	RELEVANCE       = { adf },
	LOCATION        = {copy},

	AUTHOR          = {Cristea, Ligia L. and Steinsky, Bertran},
	TITLE           = {Curves of infinite length in labyrinth fractals},
	JOURNAL         = {Proc. Edinburgh Math. Soc.},
	YEAR            = {2011},
	VOLUME          = {54},
	NUMBER          = {},
	PAGES           = {329--344},
	ISSN_ISBN       = {},
	DOI		= {},
	NOTE            = {},
	CLASSIFICATION  = {MSC: 14Q05, 26B05, 28A75, 28A80, 51M25, 52A38},
}

@ARTICLE{cristea.l.17.mixed.112,

	RELEVANCE       = { adf },
	LOCATION        = {},

	AUTHOR          = {Cristea, Ligia L. and Steinsky, Bertran},
	TITLE           = {Mixed labyrinth fractals},
	JOURNAL         = TA,
	YEAR            = {2017},
	VOLUME          = {229},
	NUMBER          = {},
	PAGES           = {112--125},
	ISSN_ISBN       = {},
	DOI		= {10.1016/j.topol.2017.06.022},
	NOTE            = {},
	CLASSIFICATION  = {MSC: 14Q05 , 26B05 , 28A75 , 28A80 , 51M25 , 52A38},
}

@ARTICLE{cristea.l.18.on.575,

	RELEVANCE       = { adf },
	LOCATION        = {file},

	AUTHOR          = {Cristea, Ligia L. and Leobacher, Gunther},
	TITLE           = {On the length of arcs in labyrinth fractals},
	JOURNAL         = MHfM,
	YEAR            = {2018},
	VOLUME          = {185},
	NUMBER          = {},
	PAGES           = {575--590},
	ISSN_ISBN       = {},
	DOI		= {10.1007/s00605-017-1056-8},
	NOTE            = {},
	CLASSIFICATION  = {MSC: 28A80 , 05C38 , 28A75 , 51M25 , 52A38},
}

@ARTICLE{cristea.l.20.supermixed.183,

	RELEVANCE       = { adf },
	LOCATION        = {file},

	AUTHOR          = {Cristea, Ligia L. and Leobacher, Gunther},
	TITLE           = {Supermixed labyrinth fractals},
	JOURNAL         = JFG,
	YEAR            = {2020},
	VOLUME          = {7},
	NUMBER          = {2},
	PAGES           = {183--218},
	ISSN_ISBN       = {},
	DOI		= {10.4171/JFG/88},
	NOTE            = {arXiv:1802.05461v1 [math.GT]},
	CLASSIFICATION  = {},
}

@ARTICLE{cristea.l.21.on.07468v1,

	RELEVANCE       = { adf },
	LOCATION        = {file},

	AUTHOR          = {Cristea, Ligia~L. and Vuki\^{c}evi\'c, Damir},
	TITLE           = {On the dimension of arcs in mixed labyrinth fractals},
	JOURNAL         = ARXV,
	YEAR            = {2020},
	VOLUME          = {preprint arXiv:2103.07468},
	NUMBER          = {},
	PAGES           = {[math.DS]},
	ISSN_ISBN       = {},
	DOI		= {},
	NOTE            = {},
	CLASSIFICATION  = {},
}

@INPROCEEDINGS{cristea.l.20.on.1,

	RELEVANCE       = { adf },
	LOCATION        = {file},

	AUTHOR          = {Cristea, Ligia~L. and Vuki\^{c}evi\'c, Damir},
	TITLE           = {On the dimension of arcs in mixed labyrinth fractals},
	EDITOR          = {Tomislav Do\u{s}li\'c, Snje\u{z}ana Majstorovi\'c},
	BOOKTITLE       = {Proceedings of the $3$rd Croatian Combinatorial Days Zagreb 21-22 Sept. 2020},
	ORGANIZATION    = {},
	SERIES          = {},
	PUBLISHER       = {Faculty of Civil Engineering, University of Zagreb},
	ADDRESS         = {},
	YEAR            = {2020},
	VOLUME          = {},
	NUMBER          = {},
	PAGES           = {1--15},
	ISSN_ISBN       = {978-953-8168-49-9},
	DOI		= {10.5592/CO/CCD.2020.01},
	NOTE            = {},
	CLASSIFICATION  = {AMS: 28A80, 05C38, 28A75, 05C05},
}

@BOOK{falconer.k.14.fractal.book,

	RELEVANCE       = { adf },
	LOCATION        = {},

	AUTHOR          = {Falconer, Kenneth},
	EDITOR          = {},
	TITLE           = {Fractal geometry: Mathematical foundations and applications},
	PUBLISHER       = {John Wiley \& Sons},
	ADDRESS         = {Chichester, UK},
	YEAR            = {2014},
	VOLUME          = {},
	NUMBER          = {},
	SERIES          = {},
	EDITION         = {3rd},
	ISSN_ISBN       = {},
	NOTE            = {},
	CLASSIFICATION  = {},
}

@ARTICLE{franz.a.00.efficient.155,

	RELEVANCE       = { adf },
	LOCATION        = {own},

	AUTHOR          = {Franz, Astrid and Schulzky, Christian and Seeger, Steffen and Hoffmann, Karl Heinz},
	TITLE           = {An Efficient Implementation of the Exact Enumeration Method for Random Walks on Sierpinski Carpets},
	JOURNAL         = Fractals,
	YEAR            = {2000},
	VOLUME          = {8},
	NUMBER          = {2},
	PAGES           = {155-161},
	ISSN_ISBN       = {ISSN: 0218-348X(00)},
	DOI		= {10.1142/S0218348X00000172},
	PACS            = {},
}

@ARTICLE{franz.a.01.einstein.1411,

	RELEVANCE       = { adf },
	LOCATION        = {own},

	AUTHOR          = {Franz, Astrid and Schulzky, Christian and Hoffmann, Karl Heinz},
	TITLE           = {The Einstein relation for finitely ramified Sierpinski carpets},
	JOURNAL         = Nonlin,
	YEAR            = {2001},
	VOLUME          = {14},
	NUMBER          = {5},
	PAGES           = {1411--1418},
	ISSN_ISBN       = {PII: S0951-7715(01)16979-3},
	DOI		= {10.1088/0951-7715/14/5/324},
	NOTE            = {},
	CLASSIFICATION  = {PACS: 05.45.D 05.40.F 61.43.H 66.10.C},
}

@ARTICLE{franz.a.01.pore.8751,

	RELEVANCE	= { adf },
	LOCATION	= {own},

	AUTHOR		= {Franz, Astrid and Schulzky, Christian and Tarafdar, Sujata and Hoffmann, Karl Heinz},
	TITLE		= {The pore structure of Sierpinski carpets},
	JOURNAL		= JPA,
	YEAR		= {2001},
	VOLUME		= {34},
	NUMBER		= {42},
	PAGES		= {8751--8765},
	ISSN_ISBN	= {PII: S0305-4470(01)19486-2},
	DOI		= {10.1088/0305-4470/34/42/303},
	NOTE		= {},
	CLASSIFICATION	= {PACS: 05.45.Df 61.43.Hv 61.43.Gt 81.05.Rm 05.40.Fb},
}

@INCOLLECTION{franz.a.02.diffusion.52,

	RELEVANCE       = { adf },
	LOCATION        = {own},

	AUTHOR          = {Franz, Astrid and Schulzky, Christian and Seeger, Steffen and Hoffmann, Karl Heinz},
	EDITOR          = {Blackledge, Jonathan~M. and Evans, Allan~K. and Turner, Martin~J.},
	TITLE           = {Diffusion on Fractals -- Efficient algorithms to compute the random walk dimension},
	BOOKTITLE       = {Fractal Geometry: Mathematical Methods, Algorithms, Applications},
	EDITION         = {},
	SERIES          = {IMA Conference Proceedings},
	TYPE            = {},
	YEAR            = {2002},
	VOLUME          = {},
	NUMBER          = {},
	CHAPTER         = {},
	PAGES           = {52--67},
	PUBLISHER       = {Horwood Publishing Ltd., Chichester, West Sussex},
	ADDRESS         = {},
	ISSN_ISBN       = {1-904275-00-1},
	DOI		= {10.1533/9780857099594.52},
	NOTE            = {},
	CLASSIFICATION  = {},
}

@ARTICLE{franz.a.02.using.18,

	RELEVANCE       = { adf },
	LOCATION        = {own},

	AUTHOR          = {Franz, Astrid and Schulzky, Christian and Hoffmann, Karl Heinz},
	TITLE           = {Using Computer Algebra Methods to Determine the Chemical Dimension of Finitely Ramified Sierpinski Carpets},
	JOURNAL         = SIGSAM,
	YEAR            = {2002},
	VOLUME          = {36},
	NUMBER          = {2},
	PAGES           = {18--30},
	ISSN_ISBN       = {ISSN: 0163-5824},
	DOI		= {10.1145/581316.581318},
	NOTE            = {},
	CLASSIFICATION  = {},
}

@ARTICLE{golmankhaneh.a.18.sub.960,

	RELEVANCE       = { adf },
	LOCATION        = {},

	AUTHOR          = {Golmankhanehh, Alireza K. and Balankin, Alexander S.},
	TITLE           = {Sub- and super-diffusion on {Cantor sets}: Beyond the paradox},
	JOURNAL         = PLA,
	YEAR            = {2018},
	VOLUME          = {382},
	NUMBER          = {14},
	PAGES           = {960--967},
	ISSN_ISBN       = {},
	DOI		= {10.1016/j.physleta.2018.02.009},
	NOTE            = {},
	CLASSIFICATION  = {},
}

@ARTICLE{haber.r.13.diffusion.2840,

	RELEVANCE       = { adf },
	LOCATION        = {own},

	AUTHOR          = {Haber, Ren\'e and Prehl, Janett and Hermann, Heiko and Hoffmann, Karl Heinz},
	TITLE           = {Diffusion of oriented particles in porous media},
	JOURNAL         = PLA,
	YEAR            = {2013},
	VOLUME          = {377},
	NUMBER          = {},
	PAGES           = {2840--2845},
	ISSN_ISBN       = {},
	DOI		= {j.physleta2013.08.036},
	NOTE            = {},
	CLASSIFICATION  = {},
}

@ARTICLE{havlin.s.02.diffusion.187,

	RELEVANCE          = { adf },
        AUTHOR             = {Havlin, S. and Ben-Avraham, D.},
        JOURNAL            = AdP,
        NUMBER             = {1},
        PAGES              = {187-292},
        TITLE              = {Diffusion in Disordered Media},
        VOLUME             = {51},
        YEAR               = {2002},
        LOCATION           = {},
}

@ARTICLE{husain.a.22.fractals.379,

	RELEVANCE       = { adf },
	LOCATION        = {file},

	AUTHOR          = {Husain, Akhlaq and Nanda, Manikyala Navaneeth and Chowdary, Movva Sitaram and Sajid, Mohommad},
	TITLE           = {Fractals: An Eclectic Survey, Part II},
	JOURNAL         = FF,
	YEAR            = {2022},
	VOLUME          = {6},
	NUMBER          = {},
	PAGES           = {379},
	ISSN_ISBN       = {},
	DOI		= {10.3390/fractalfract6070379},
	NOTE            = {},
	CLASSIFICATION  = {},
}

@ARTICLE{juhasz.r.08.superdiffusion.066106,

	RELEVANCE       = { adf },
	LOCATION        = {file},

	AUTHOR          = {R\'obert Juh\'asz},
	TITLE           = {Superdiffusion in a class of networks with marginal long-range connections},
	JOURNAL         = PRE,
	YEAR            = {2008},
	VOLUME          = {78},
	NUMBER          = {},
	PAGES           = {066106},
	ISSN_ISBN       = {},
	DOI		= {10.1103/PhysRevE.78.066106},
	NOTE            = {},
	CLASSIFICATION  = {PACS: 89.75.Hc , 64.60.aq , 05.40.Fb},
}

@ARTICLE{tian.f.21.application.14587,

	RELEVANCE       = { adf },
	LOCATION        = {file},

	AUTHOR          = {Tian, Fengchun and Jiang, Anyan and Yang, Taicong and Qian, Junhui and Lui, Ran and Jiang, Maogang},
	TITLE           = {Application of Fractal Geometry in GasA Sensor:
A Review},
	JOURNAL         = SJ,
	YEAR            = {2021},
	VOLUME          = {21},
	NUMBER          = {13},
	PAGES           = {14587--14600},
	ISSN_ISBN       = {},
	DOI		= {10.1109/JSEN.2021.3072621},
	NOTE            = {},
	CLASSIFICATION  = {},
}

@ARTICLE{kessebohmer.m.20.sierpinski.113,

	RELEVANCE       = { adf },
	LOCATION        = {file},

	AUTHOR          = {Kesseb\"ohmer, Marc and Samuel, Tony and  Sender, Karenina},
	TITLE           = {The {Sierpi\'nski} gasket as the {Martin} boundary of a non-isotropic {Markov} chain},
	JOURNAL         = JFG,
	YEAR            = {2020},
	VOLUME          = {7},
	NUMBER          = {},
	PAGES           = {113--136},
	ISSN_ISBN       = {},
	DOI		= {10.4171/JFG/86},
	NOTE            = {},
	CLASSIFICATION  = {MSC (2020): 31C35, 28A80, 60J10, 60J50},
}

@ARTICLE{kumagai.t.96.homogenization.375,

	RELEVANCE       = { adf },
	LOCATION        = {file},

	AUTHOR          = {Kumagai, T. and Kusuoka, S.},
	TITLE           = {Homogenization on nested fractals},
	JOURNAL         = PTRF,
	YEAR            = {1996},
	VOLUME          = {104},
	NUMBER          = {},
	PAGES           = {375--398},
	ISSN_ISBN       = {},
	DOI		= {},
	NOTE            = {},
	CLASSIFICATION  = {MSC (1991): 31C25 , 47A35 , 60J60},
}

@ARTICLE{lau.k.12.martin.475,

	RELEVANCE       = { adf },
	LOCATION        = {file},

	AUTHOR          = {Lau, Ka-Sing and Ngai, Sze-Man},
	TITLE           = {Martin boundary and exit space on the {Sierpinski} gasket},
	JOURNAL         = SCM,
	YEAR            = {2012},
	VOLUME          = {55},
	NUMBER          = {3},
	PAGES           = {475--494},
	ISSN_ISBN       = {},
	DOI		= {10.1007/s11425-011-4339-x},
	NOTE            = {},
	CLASSIFICATION  = {MSC (2020): 31C35, 60J50, 28A80, 60J10},
}

@INPROCEEDINGS{potapov.a.13.nano.941,

	RELEVANCE       = { adf },
	LOCATION        = {file},

	AUTHOR          = {Potapov, A.~A. and German, V.~A. and Grachev, V.~I.},
	TITLE           = {"Nano-" and radar signal processing: Fractal reconstruction complicated images, signals and radar backgrounds based on fractal labyrinths},
	EDITOR          = {},
	BOOKTITLE       = {2013 14th International Radar Symposium (IRS)},
	ORGANIZATION    = {},
	SERIES          = {},
	PUBLISHER       = {IEEE},
	ADDRESS         = {},
	YEAR            = {2013},
	VOLUME          = {2},
	NUMBER          = {},
	PAGES           = {941--946},
	ISSN_ISBN       = {},
	DOI		= {},
	NOTE            = {},
	CLASSIFICATION  = {},
}

@INPROCEEDINGS{potapov.a.16.simulation.1,

	RELEVANCE       = { adf },
	LOCATION        = {file},

	AUTHOR          = {Potapov, A.~A. and Zhang, W.},
	TITLE           = {Simulation of new ultra-wide band fractal antennas based on fractal labyrinths},
	EDITOR          = {},
	BOOKTITLE       = {2016 CIE International Conference on Radar (RADAR)},
	ORGANIZATION    = {},
	SERIES          = {},
	PUBLISHER       = {IEEE},
	ADDRESS         = {},
	YEAR            = {2016},
	VOLUME          = {},
	NUMBER          = {},
	PAGES           = {1--5},
	ISSN_ISBN       = {},
	DOI		= {10.1109/RADAR.2016.8059186},
	NOTE            = {},
	CLASSIFICATION  = {},
}

@INPROCEEDINGS{potapov.a.17.fractal.499,

	RELEVANCE       = { adf },
	LOCATION        = {file},

	AUTHOR          = {Potapov, A.~A. and Potapov, Alexey~A. and Potapov, V.~A.},
	TITLE           = {Fractal Radioelement's, Devices and Fractal Systems for Radar and Telecommunications},
	EDITOR          = {},
	BOOKTITLE       = {Proc.~14th Sino-Russia Symp.~Advanced Materials and Technologies, Sanya, China},
	ORGANIZATION    = {},
	SERIES          = {},
	PUBLISHER       = {},
	ADDRESS         = {},
	YEAR            = {2017},
	VOLUME          = {},
	NUMBER          = {},
	PAGES           = {499--506},
	ISSN_ISBN       = {},
	DOI		= {},
	NOTE            = {},
	CLASSIFICATION  = {},
}

@ARTICLE{puente.c.96.fractal.1,

	RELEVANCE       = { adf },
	LOCATION        = {file},

	AUTHOR          = {Puente, C. and Romeu, J. and Pous, R. and Garcia X. and Benitez, F.},
	TITLE           = {Fractal multiband antenna based on the
Sierpinski gasket},
	JOURNAL         = EL,
	YEAR            = {1996},
	VOLUME          = {32},
	NUMBER          = {1},
	PAGES           = {1--2},
	ISSN_ISBN       = {},
	DOI		= {},
	NOTE            = {},
	CLASSIFICATION  = {},
}

@ARTICLE{schulzky.c.00.resistance.1,

	RELEVANCE       = { adf },
	LOCATION        = {own},

	AUTHOR          = {Schulzky, Christian and Franz, Astrid and Hoffmann, Karl Heinz},
	TITLE           = {Resistance Scaling and Random Walk Dimensions for Finitely Ramified Sierpinski Carpets},
	JOURNAL         = SIGSAM,
	YEAR            = {2000},
	VOLUME          = {34},
	NUMBER          = {3},
	PAGES           = {1--8},
	ISSN_ISBN       = {ISSN: 0163-5824},
	DOI		= {10.1145/377604.377608},
	NOTE            = {},
	CLASSIFICATION  = {},
}

@ARTICLE{seeger.s.01.random.307,

	RELEVANCE       = { adf },
	LOCATION        = {own},

	AUTHOR          = {Seeger, Steffen and Franz, Astrid and Schulzky, Christian and Hoffmann, Karl Heinz},
	TITLE           = {Random Walks on Finitely Ramified Sierpinski Carpets},
	JOURNAL         = CPC,
	YEAR            = {2001},
	VOLUME          = {134},
	NUMBER          = {3},
	PAGES           = {307--316},
	ISSN_ISBN       = {PII: S0010-4655(00)00208-3},
	DOI		= {10.1016/S0010-4655(00)00208-3},
	NOTE            = {},
	CLASSIFICATION  = {PACS: 05.40.Fb 05.45.Df 61.43.Hv 66.10.Cb},
}

@ARTICLE{seeger.s.09.random.225002,

	RELEVANCE       = { ADF },
	LOCATION        = { own,copy },

	AUTHOR          = {Seeger, Steffen and Hoffmann, Karl Heinz and Essex, Christopher},
	TITLE           = {Random Walks on random {Koch} curves},
	JOURNAL         = JPA,
	YEAR            = {2009},
	VOLUME          = {42},
	NUMBER          = {22},
	PAGES           = {225002-1--11},
	ISSN_ISBN       = {},
	DOI		= {10.1088/1751-8113/42/22/225002},
	NOTE            = {},
	CLASSIFICATION  = {PACS: 05.45.Df, 66.10.Cb, 05.40.Fb},
}

@ARTICLE{troscheit.s.17.on.257,

        RELEVANCE       = { adf },
        LOCATION        = {file},

        AUTHOR          = {Sascha Troscheit},
        TITLE           = {On the dimensions of attractors of random self-similar graph directed iterated function systems}, 
        JOURNAL         = JFG,
        YEAR            = {2017},
        VOLUME          = {4},
        NUMBER          = {},
        PAGES           = {257--303},
	ISSN_ISBN       = {},
	DOI		= {10.10.4171/JFG/51},
	NOTE            = {},
	CLASSIFICATION  = {MSC(2010): 28A80, 60J80, 37C45},
}

@ARTICLE{yang.t.21.optimizing.16675,

        RELEVANCE       = { adf },
        LOCATION        = {file},

        AUTHOR          = {Taicong Yang},
        TITLE           = {Optimizing electrode structure of carbon nanotube gas sensors for sensitivity improvement based on electric field enhancement effect of fractal geometry}, 
        JOURNAL         = SREP,
        YEAR            = {2021},
        VOLUME          = {11},
        NUMBER          = {},
        PAGES           = {16675},
	ISSN_ISBN       = {},
	DOI		= {10.1038/s41598-021-96239-1},
	NOTE            = {},
	CLASSIFICATION  = {},
}

\end{document}